\input harvmac.tex
\input epsf
\noblackbox
\def\figin{\epsfcheck\figin}\def\figins{\epsfcheck\figins}
\def\epsfcheck{\ifx\epsfbox\UnDeFiNeD
\message{(NO epsf.tex, FIGURES WILL BE IGNORED)}
\gdef\figin##1{\vskip2in}\gdef\figins##1{\hskip.5in}
\else\message{(FIGURES WILL BE INCLUDED)}%
\gdef\figin##1{##1}\gdef\figins##1{##1}\fi}
\def\DefWarn#1{}
\def\figinsert{\goodbreak\midinsert}
\def\ifig#1#2#3{\DefWarn#1\xdef#1{fig.~\the\figno}
\writedef{#1\leftbracket fig.\noexpand~\the\figno}%
\figinsert\figin{\centerline{#3}}\medskip\centerline{\vbox{\baselineskip12pt
\advance\hsize by -1truein\noindent\footnotefont{\bf Fig.~\the\figno } \it#2}}
\bigskip\endinsert\global\advance\figno by1}


\def\encadremath#1{\vbox{\hrule\hbox{\vrule\kern8pt\vbox{\kern8pt
 \hbox{$\displaystyle #1$}\kern8pt}
 \kern8pt\vrule}\hrule}}
 %
 %
 

 \font\cmss=cmss10
 \font\cmsss=cmss10 at 7pt
 \def\rlx{\relax\leavevmode}
 \def\inbar{\vrule height1.5ex width.4pt depth0pt}
 \def\IC{\relax\,\hbox{$\inbar\kern-.3em{\rm C}$}}
 \def\IN{\relax{\rm I\kern-.18em N}}
 \def\IP{\relax{\rm I\kern-.18em P}}

\def\ZZ{\rlx\leavevmode\ifmmode\mathchoice{\hbox{\cmss Z\kern-.4em Z}}
  {\hbox{\cmss Z\kern-.4em Z}}{\lower.9pt\hbox{\cmsss Z\kern-.36em Z}}
  {\lower1.2pt\hbox{\cmsss Z\kern-.36em Z}}\else{\cmss Z\kern-.4em Z}\fi}
 \def\IZ{\relax\ifmmode\mathchoice
 {\hbox{\cmss Z\kern-.4em Z}}{\hbox{\cmss Z\kern-.4em Z}}
 {\lower.9pt\hbox{\cmsss Z\kern-.4em Z}}
 {\lower1.2pt\hbox{\cmsss Z\kern-.4em Z}}\else{\cmss Z\kern-.4em Z}\fi}
 \def\IZ{\relax\ifmmode\mathchoice
 {\hbox{\cmss Z\kern-.4em Z}}{\hbox{\cmss Z\kern-.4em Z}}
 {\lower.9pt\hbox{\cmsss Z\kern-.4em Z}}
 {\lower1.2pt\hbox{\cmsss Z\kern-.4em Z}}\else{\cmss Z\kern-.4em Z}\fi}

 \def\narrowplus{\kern -.04truein + \kern -.03truein}
 \def\narrowminus{- \kern -.04truein}
 \def\narrowminussub{\kern -.02truein - \kern -.01truein}

 \def\a{{\alpha}}
 
 \def\frac#1#2{{#1\over #2}}

 \def\IZ{\relax\ifmmode\mathchoice
 {\hbox{\cmss Z\kern-.4em Z}}{\hbox{\cmss Z\kern-.4em Z}}
 {\lower.9pt\hbox{\cmsss Z\kern-.4em Z}}
 {\lower1.2pt\hbox{\cmsss Z\kern-.4em Z}}\else{\cmss Z\kern-.4em Z}\fi}
 \def\IB{\relax{\rm I\kern-.18em B}}
 \def\IC{{\relax\hbox{$\inbar\kern-.3em{\rm C}$}}}
 \def\Ic{{\relax\hbox{$\inbar\kern-.22em{\rm c}$}}}
 \def\ID{\relax{\rm I\kern-.18em D}}
 \def\IE{\relax{\rm I\kern-.18em E}}
 \def\IF{\relax{\rm I\kern-.18em F}}
 \def\IG{\relax\hbox{$\inbar\kern-.3em{\rm G}$}}
 \def\IGa{\relax\hbox{${\rm I}\kern-.18em\Gamma$}}
 \def\IH{\relax{\rm I\kern-.18em H}}
 \def\II{\relax{\rm I\kern-.18em I}}
 \def\IK{\relax{\rm I\kern-.18em K}}
 \def\IP{\relax{\rm I\kern-.18em P}}
\def\Tr{{\rm Tr}}
 \font\cmss=cmss10 \font\cmsss=cmss10 at 7pt
 \def\IR{\relax{\rm I\kern-.18em R}}

 %

 %
 %
 \def\eqnn#1{\xdef
#1{(\secsym\the\meqno)}\writedef{#1\leftbracket#1}%
 \global\advance\meqno by1\wrlabeL#1}
 \def\eqna#1{\xdef
#1##1{\hbox{$(\secsym\the\meqno##1)$}}

\writedef{#1\numbersign1\leftbracket#1{\numbersign1}}%
 \global\advance\meqno by1\wrlabeL{#1$\{\}$}}
 \def\eqn#1#2{\xdef
#1{(\secsym\the\meqno)}\writedef{#1\leftbracket#1}%
 \global\advance\meqno by1$$#2\eqno#1\eqlabeL#1$$}

\newdimen\tableauside\tableauside=1.0ex
\newdimen\tableaurule\tableaurule=0.4pt
\newdimen\tableaustep
\def\phantomhrule#1{\hbox{\vbox to0pt{\hrule height\tableaurule width#1\vss}}}
\def\phantomvrule#1{\vbox{\hbox to0pt{\vrule width\tableaurule height#1\hss}}}
\def\sqr{\vbox{%
  \phantomhrule\tableaustep
  \hbox{\phantomvrule\tableaustep\kern\tableaustep\phantomvrule\tableaustep}%
  \hbox{\vbox{\phantomhrule\tableauside}\kern-\tableaurule}}}
\def\squares#1{\hbox{\count0=#1\noindent\loop\sqr
  \advance\count0 by-1 \ifnum\count0>0\repeat}}
\def\tableau#1{\vcenter{\offinterlineskip
  \tableaustep=\tableauside\advance\tableaustep by-\tableaurule
  \kern\normallineskip\hbox
    {\kern\normallineskip\vbox
      {\gettableau#1 0 }%
     \kern\normallineskip\kern\tableaurule}%
  \kern\normallineskip\kern\tableaurule}}
\def\gettableau#1 {\ifnum#1=0\let\next=\null\else
  \squares{#1}\let\next=\gettableau\fi\next}

\tableauside=1.0ex
\tableaurule=0.4pt

\def\IE{\relax{\rm I\kern-.18em E}}
\def\IP{\relax{\rm I\kern-.18em P}}

 \def\cM{{\cal M}}

\lref\nikitab{
A.~S.~Losev, A.~Marshakov and N.~A.~Nekrasov,
``Small instantons, little strings and free fermions,''
arXiv:hep-th/0302191.
}
\lref\nikitaa{
N.~A.~Nekrasov,
``Seiberg-Witten prepotential from instanton counting,''
arXiv:hep-th/0206161.
}
\lref\gvmone{R. Gopakumar and C. Vafa, ``M-theory and topological
strings, I," hep-th/9809187. }
\lref\gz{T. Graber and E. Zaslow, ``Open-string Gromov-Witten invariants:
calculations and a mirror `theorem','' hep-th/0109075.}
\lref\wittcs{E. Witten, ``Chern-Simons gauge theory as
a string theory,'' hep-th/9207094, in {\it The Floer memorial volume},
H. Hofer, C.H. Taubes, A. Weinstein and E. Zehner, eds.,
Birkh\"auser 1995, p. 637.}
\lref\gv{R. Gopakumar and C. Vafa, ``On the gauge theory/geometry
correspondence," hep-th/9811131, Adv. Theor. Math. Phys. {\bf 3} (1999)
1415.}
\lref\lm{J.M.F. Labastida and M. Mari\~no, ``Polynomial invariants
for torus knots and topological strings,''  hep-th/0004196,
Commun. Math. Phys. {\bf 217} (2001) 423.}
\lref\gvtwo{R. Gopakumar and C. Vafa, ``M-theory and topological
strings, II," hep-th/9812127. }
\lref\ov{H. Ooguri and C. Vafa, ``Knot invariants and topological
strings," hep-th/9912123, Nucl. Phys. {\bf B 577} (2000) 419.}
\lref\jones{E. Witten, ``Quantum field theory and the Jones polynomial,"
Commun. Math. Phys. {\bf 121} (1989) 351.}
\lref\lmv{J.M.F. Labastida, M. Mari\~no and C. Vafa, ``Knots, links and
branes at large $N$,'' hep-th/0010102, JHEP {\bf 0011} (2000) 007.}
\lref\mv{M. Mari\~no and C. Vafa, ``Framed knots at large $N$,''
hep-th/0108064.}
\lref\macdonald{I.G. Macdonald, {\it Symmetric functions and Hall
polynomials}, 2nd edition, Oxford University Press, 1995.}
\lref\ml{H.R. Morton and S.G. Lukac, ``The HOMFLY polynomial of the
decorated Hopf link,'' math.GT/0108011.}
\lref\lukac{S.G. Lukac, ``HOMFLY skeins and the Hopf link,'' Ph.D. Thesis,
June 2001, in http://www.liv.ac.uk/~su14/knotgroup.html}
\lref\ckyz{T.~M.~Chiang, A.~Klemm, S.~T.~Yau and E.~Zaslow,
``Local mirror symmetry: Calculations and interpretations,''
hep-th/9903053,
Adv.\ Theor.\ Math.\ Phys.\  {\bf 3} (1999) 495. }
\lref\proof{H. Ooguri and C. Vafa, ``Worldsheet derivation of a large $N$
duality,'' hep-th/0205297, Nucl.\ Phys.\ {\bf B 641}, 3 (2002).}
\lref\AV{M. Aganagic and C. Vafa, ``Mirror symmetry, D-branes and
counting holomorphic discs,'' hep-th/0012041.}
\lref\AKV{M. Aganagic,
A. Klemm and C. Vafa, ``Disk instantons, mirror symmetry and the duality
web,'' hep-th/0105045, Z.\ Naturforsch.\ {\bf A 57} (2002) 1.}
\lref\ahk{O.~Aharony and A.~Hanany,
``Branes, superpotentials and superconformal fixed points,''
hep-th/9704170, Nucl.\ Phys.\  {\bf B 504} (1997) 239.
O.~Aharony, A.~Hanany and B.~Kol,
``Webs of (p,q) 5-branes, five dimensional field theories and grid
diagrams,'' hep-th/9710116,
JHEP {\bf 9801} (1998) 002.}
\lref\lv{N.~C.~Leung and C.~Vafa,
``Branes and toric geometry,'' hep-th/9711013,
Adv.\ Theor.\ Math.\ Phys.\  {\bf 2} (1998) 91.}
\lref\HV{K.~Hori and C.~Vafa,``Mirror symmetry,'' hep-th/0002222.}
\lref\phases{E.~Witten,``Phases of N = 2 theories in two dimensions,'' hep-th/9301042,
Nucl.\ Phys.\ B {\bf 403}, 159 (1993).}
\lref\kon{M.~Kontsevich, ``Enumeration of rational curves via torus
actions,'' hep-th/9405035, in {\it The moduli space of curves}, p. 335,
Birkh\"auser, 1995.}
\lref\verlinde{E. Verlinde,
``Fusion rules and modular transformations
in 2-D conformal field theory,''
Nucl.\ Phys.\ {\bf B 300} (1988) 360.}
\lref\BCOV{
M.~Bershadsky, S.~Cecotti, H.~Ooguri and C.~Vafa,
``Kodaira-Spencer theory of gravity and exact results for quantum string
amplitudes,'' hep-th/9309140,
Commun.\ Math.\ Phys.\  {\bf 165} (1994) 311.}
\lref\vaaug{C. Vafa, ``Superstrings and topological strings at large $N$,''
hep-th/0008142, J.\ Math.\ Phys.\  {\bf 42} (2001) 2798.}
\lref\newp{J.M.F.~Labastida and M.~Mari\~no,
``A new point of view in the theory of
knot and link invariants,'' math.QA/0104180, J. Knot Theory
Ramifications {\bf 11} (2002) 173.}
\lref\amv{M.~Aganagic, M.~Mari\~no and C.~Vafa,
``All loop topological string amplitudes from Chern-Simons theory,''
hep-th/0206164.}
\lref\dfg{D.~E.~Diaconescu, B.~Florea and A.~Grassi,
``Geometric transitions and open string instantons,'' hep-th/0205234.
``Geometric transitions, del Pezzo surfaces and open string instantons,''
hep-th/0206163.}
\lref\amer{A. Iqbal, ``All genus topological string amplitudes and 5-brane
webs as Feynman  diagrams,'' hep-th/0207114; A. Iqbal and
A.K. Kashani-Poor, to appear.}
\lref\topstrings{E. Witten, `Topological sigma models,''
Commun.\ Math.\ Phys.\  {\bf 118}, 411 (1988).
``On the structure of the topological phase of two-dimensional gravity,''
Nucl.\ Phys.\ {\bf B 340}, 281 (1990).}
\lref\antibr{C. Vafa, ``Brane/anti-brane systems and $U(N|M)$ supergroup,''
hep-th/0101218.}
\lref\hl{R. Harvey and H.B. Lawson Jr., ``Calibrated geometries,'' Acta Math. {\bf 148} (1982) 47.}
\lref\mayr{P. Mayr, ``Summing up
open string instantons and ${\cal N} = 1$
string amplitudes,'' hep-th/0203237.}
\lref\ik{A.~Iqbal and A.~K.~Kashani-Poor,
``Instanton counting and Chern-Simons theory,'' hep-th/0212279.}
\lref\rama{
P.~Ramadevi and T.~Sarkar,
``On link invariants and topological string amplitudes,''
Nucl.\ Phys.\ B {\bf 600}, 487 (2001)
[arXiv:hep-th/0009188].}

\lref\kon{M.~Kontsevich, ``Enumeration of rational curves via torus
actions,'' hep-th/9405035, in {\it The moduli space of curves}, p. 335,
Birkh\"auser, 1995.}

\lref\kz{A. Klemm and E. Zaslow,
``Local mirror symmetry at higher genus,'' hep-th/9906046, in {\it Winter
School on Mirror Symmetry, Vector bundles and Lagrangian
Submanifolds}, p. 183, American Mathematical Society 2001.}

\lref\kl{S. Katz and C-C. Liu, ``Enumerative geometry of stable
maps with Lagrangian boundary conditions and multiple covers of the disc,''
math.AG/0103074, Adv.\ Theor.\ Math.\ Phys.\  {\bf 5} (2002) 1.}

\lref\faber{C. Faber, ``Algorithms for computing intersection numbers
of curves, with an application to the class of the locus of Jacobians,"
alg-geom/9706006, in {\it New trends in algebraic geometry},
Cambridge Univ. Press, 1999.}
\lref\greene{
P.~S.~Aspinwall, B.~R.~Greene and D.~R.~Morrison,
``Calabi-Yau moduli space, mirror manifolds and spacetime topology
  change in string theory,''
Nucl.\ Phys.\ B {\bf 416}, 414 (1994)
[arXiv:hep-th/9309097].
}
\lref\bmodel{
W.~Lerche and P.~Mayr, ``On ${\cal N} = 1$
mirror symmetry for open type II strings,'' hep-th/0111113.
S.~Govindarajan, T.~Jayaraman and T.~Sarkar,
``Disc instantons in linear sigma models,'' hep-th/0108234,
Nucl.\ Phys.\ B {\bf 646}, 498 (2002).}
%
\lref\Hosono{S.~Hosono,
``Counting BPS states via holomorphic anomaly equations,''
hep-th/0206206.}

\Title
{\vbox{
 \baselineskip12pt
\hbox{hep-th/0305132}
\hbox{CALT-68-2439}\hbox{HUTP-03/A032}
\hbox{HU-EP-03/24 }\hbox{CERN-TH/2003-111}
}}
 {\vbox{
 \centerline{The Topological Vertex}
 }}
\centerline{Mina Aganagic,$^{a}$ Albrecht Klemm,$^{b}$ Marcos Mari\~no,$^{c}$
and Cumrun Vafa$^{a,d}$}
\bigskip\centerline{$^{a}$ Jefferson Physical Laboratory, Harvard University}
\centerline{Cambridge, MA 02138, USA}\smallskip
\centerline{$^b$ Humboldt-Universit\"at zu Berlin, Institut f\"ur
Physik}
\centerline{D-10115 Berlin, Germany} \smallskip
\centerline{$^c$ Theory Division, CERN, Geneva 23, CH-1211 Switzerland}
\smallskip
\centerline{$^d$ California Institute of Technology, 452-48, Pasadena, CA 91125, USA}
\smallskip
 \vskip .2in \centerline{\bf Abstract}
 {We construct a cubic field theory which provides
all genus amplitudes of the topological A-model
for all non-compact toric Calabi-Yau
threefolds.  The topology of a given Feynman diagram
encodes the topology of a fixed Calabi-Yau, with
Schwinger parameters playing the role of K\"ahler classes
of the threefold.  We interpret this result
as an operatorial computation of the amplitudes in the B-model
mirror which is the quantum Kodaira-Spencer theory.
The only degree of freedom of this theory
is an unconventional chiral scalar on a Riemann surface.
In this setup we identify the B-branes on the mirror Riemann
surface as fermions related
to the chiral boson by bosonization.}
 \smallskip \Date{}

\newsec{Introduction}
Topological strings have been a focus of much interest since
they were proposed more than a decade ago \topstrings.
A central question has been how to compute the corresponding amplitudes.
There have been two natural approaches available: i) using mirror symmetry
to transform the problem to an easier one; ii) mathematical idea of
localization.  Both approaches can in principle yield
answers to all genus amplitudes (at least in the non-compact case).
However the computations get more and more involved as one goes to higher
genera and neither method becomes very practical.

Ever since the discovery of large $N$ Chern-Simons/topological string
duality \gv\ another approach has opened
up:  Chern-Simons amplitudes seem to give an efficient way
to sum up all genus amplitudes.  This idea was developed
recently \refs{\amv ,\dfg}
where it was shown that one can compute all genus A-model
amplitudes on local toric 3-folds
 from its relation to Chern-Simons amplitudes.  However
in trying to obtain amplitudes in this way, one had often
to take certain limits.  The main aim of the present paper is
to bring this line of thought to a natural conclusion by
giving the direct answer for the topological string amplitudes, without
any need to take any limits.

Toric 3-folds are characterized by a graph which encodes
where the cycles of a $T^2$ fibration degenerates.  The vertices of this graph
are generically trivalent.
The computations in \amv\ and \dfg\
were more natural in the context of tetravalent vertices of the
toric graph.  To obtain the generic situation of trivalent graph, one
had to take particular limits in the Calabi-Yau moduli space.
Thus the basic goal is to
directly capture the structure of the trivalent vertex.
That there should be such a vertex has already been noted
\refs{\amer,\ik}\foot{This had also been
noted in our discussions with D.-E. Diaconescu and A. Grassi.  In particular,
progress towards formulation of the vertex in terms of mathematical
 localization techniques
has been made \ref\digr{D.-E. Diaconescu and A. Grassi,
 unpublished manuscript.}.}.
In this paper we show how this can be achieved.  The
idea can be summarized, roughly, as putting many brane/anti-brane
pairs which effectively chop off the Calabi-Yau to patches with
trivial topology of ${\bf C}^3$.  Computing open topological
string on ${\bf C}^3$ defines the cubic topological
vertex.  Gluing these together yields the closed topological
string results (with or without additional D-branes).
Thus the full amplitude can be obtained from a cubic field theory,
where each Calabi-Yau corresponds
to a Feynman graph with some fixed Schwinger times (determined
by the K\"ahler class of the Calabi-Yau).

This result can best be understood in the mirror picture as
computation of the quantum Kodaira-Spencer theory \BCOV . The
Kodaira-Spencer theory is, in this context of non-compact
Calabi-Yau, captured locally by a chiral boson on a Riemann
surface. The degrees of freedom on the brane get mapped, in this
setup, to coherent states of the chiral boson, and the trivalent
vertex gets identified with the quantum correlations of the chiral
boson. Moreover, the brane in the B-model gets identified with the
fermions of this chiral boson.  Thus the fact that knowing
amplitudes involving branes leads to closed string results
translates to the statement that knowing amplitudes involving
fermions leads via bosonization to the full answer for the chiral
boson.  The topological vertex gets mapped, in this setup, to a
state in the three-fold tensor product of the Fock space of a
single bosonic string oscillator.  To leading order in string
coupling and oscillator numbers this is a squeezed state as in the
conventional approaches to the operator formulation in the Riemann
surface.  However the full topological vertex is far more
complicated;  the chiral scalar is not a conventional field.  The
full vertex involves infinitely many oscillator terms together
with highly non-trivial $g_s$ dependence. Nevertheless, we find
the following closed formula for this highly non-trivial vertex
$|C\rangle$:
\eqn\anothn{\langle t_n^1,t_m^2,t_p^3|C\rangle=
\sum_{Q_1,Q_3} N_{Q_1Q_3^t}^{~~~R_1R_3^t}
q^{\kappa_{R_2}/2+\kappa_{R_3}/2}
{W_{R_2^t Q_1}W_{R_2 Q_3^t}\over W_{R_2 0}} tr_{R_1}V_1
\ tr_{R_2}V_2\ tr_{R_3} V_3.}
where
$$N_{Q_1Q_3^t}^{~~~R_1R_3^t}=\sum_R N_{Q_1 R}^{~~R_1
}N_{Q_3^t R}^{~~R_3^{t}}$$
Here $R_i,Q_i$ are representation of $U(N)$, $N_{R_i R_j}^{~~R_k}$
is the number of times the representation $R_k$ appears in the
tensor product of
 representations $R_i$ and $R_j$, $R^t$
denotes the representation whose Young Tableau is the transpose of
that of $R$ and $W_{RQ}=S_{R{\overline Q}}/S_{00}$ where $S$ is
the S-matrix of the modular transformation of the characters of
$U(N)_k$ WZW for fixed $k+N=2\pi i/g_s$ and $N\rightarrow \infty$.
The $t_n^i$ are the coherent states of a single bosonic string
oscillator and they are related to $V_i$ by $t_n^i=tr (V_i)^n$ in the
fundamental representation, $\kappa_R$ is related to the quadratic
Casimir of the representation $R$ and $q=exp(g_s)$.  This is
obtained by considering certain amplitudes in the context of large
$N$ topological duality \gv .

The organization of this paper is as follows:  In section 2 we
review the relevant facts about local toric Calabi-Yau threefolds
including their $T^2$ fibration structure and its relation to
$(p,q)$ 5-branes. We also review mirror symmetry of these manifolds,
where mirror geometry reduces, in appropriate sense, to a Riemann surface.
In section 3 we discuss how the knowledge of A-model
open topological string amplitudes on ${\bf C}^3$ with 3 sets of
Lagrangian D-branes (defining a trivalent vertex)
 can be used to compute the A-model
amplitudes for all toric Calabi-Yau threefolds with or without
D-branes.  In section 4 we formulate the vertex
in terms of a chiral bosonic oscillator in 1+1 dimension.
In section 5 we formulate the mirror B-model and discuss
the interpretation of the vertex in this setup.
In section 6 we derive the complete expression for the cubic vertex
using the large $N$ topological duality in terms
of certain Chern-Simons
amplitudes.  In section 7 we explain how the vertex
 can be evaluated explicitly. In section 8 we evaluate
the vertex for low excitations and show that it passes some
highly non-trivial tests.
 In section 9 we apply our
formalism to a number of examples.

\newsec{Toric Geometry and Mirror Symmetry}

A smooth Calabi-Yau three-fold can be obtained by gluing together
${\bf C}^3$ patches in a way that is consistent with Ricci-flatness.
For toric Calabi-Yau threefolds the gluing data
and the resulting manifold are simple to describe.

The toric Calabi-Yau 3-folds are special Lagrangian $T^2\times
{\bf R}$ fibrations over the base ${\bf R}^3$ (they are also
Lagrangian $T^3$ fibrations, but this will not be relevant for
us). The geometry of the manifold is encoded in the one
dimensional planar graph $\Gamma$ in the base that corresponds to the
degeneration locus of the fibration. The edges of the graph are oriented
straight lines labeled by vectors $(p,q)\in {\bf Z}^2$, where the
label corresponds to the generator of $H_1(T^2)$ which is the
shrinking cycle. Changing the orientation on each edge replaces
$(p,q)\rightarrow (-p,-q)$ which does not change the Calabi-Yau
geometry.
 The condition
of being a Calabi-Yau is equivalent to the condition that on each vertex,
if we choose the
edges to be incoming with charges ${v}_i = (p_i,q_i)$, one must have
\eqn\cnda{ \sum_{i} {v}_i =0.}
If the local geometry of the threefold near the vertex is ${\bf C}^3$,
then the vertex is trivalent.
Moreover, for any pair of incoming edges
one has that
\eqn\cndb{|{v}_i \wedge {v}_j |=1,}
where $\wedge $ denotes the symplectic product on $H^1(T^2)$.  This
condition ensures smoothness.

\ifig\graph{The degenerate locus
of the $T^2\times {\bf R}$ fibration
of ${\bf C}^3$ in the base ${\bf R^3}=(r_\alpha,r_\beta,r_\gamma)$. This
locus is a graph $\Gamma$. The labels $(-p_i,q_i)$ correspond to the cycles of
$T^2$ which vanish
over the corresponding edge.}
{\epsfxsize 2.0truein\epsfbox{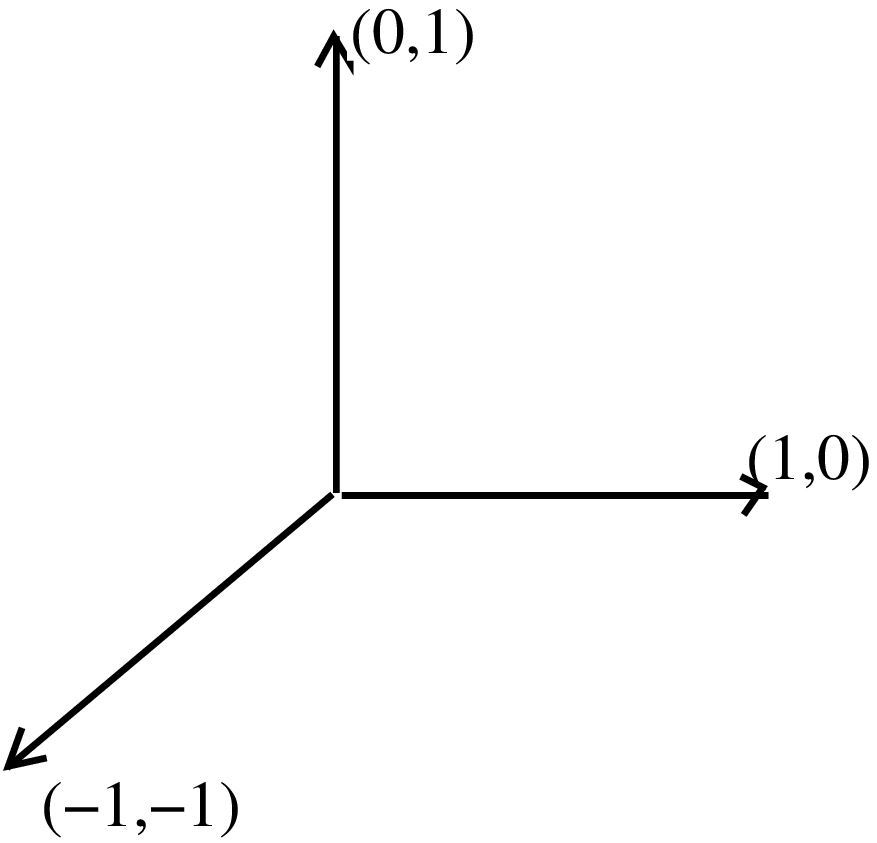}}

The graph corresponding to ${\bf C^3}$ can be obtained as follows.
Let $z_i$ be complex coordinates on ${\bf C}^3$, $i=1,2,3$. The
base of the $T^2\times {\bf R }$ fibration is the
image of moment maps $r_{\alpha}(z)=|z_1|^2-|z_3|^2$,
$r_{\beta} (z)=|z_2|^2-|z_3|^2$, and $r_{\gamma} (z)
={\rm Im}(z_1 z_2 z_3)$. The
special Lagrangian fibers are then generated by the action of the three
``Hamiltonians'' $r_{\alpha,\beta,\gamma}$ on ${\bf C}^3$ via the standard
symplectic form $\omega = i\sum_i dz_i \wedge d\overline z_i$
on ${\bf C}^3$ and Poisson brackets, $\del_{\epsilon} z_i = \{
\epsilon\cdot  r,
z_i\}_{\omega}$. In particular,
the $T^2$ fiber is generated by circle actions
\eqn\act{\exp(i \alpha r_\alpha + i \beta r_\beta):\;\;
(z_1,z_2,z_3)\;\rightarrow \; (e^{i\alpha} z_1,e^{i\beta} z_2,
e^{-i(\alpha+\beta)}z_3),}
and $r_\gamma$ generates the real line ${\bf R}$.
We will call the cycle generated by $r_{\alpha}$
the $(0,1)$ cycle, the $(1,0)$ cycle is generated by $r_{\beta}$.

We have that the $(0,1)$ cycle
degenerates over $z_1=0=z_3$.
This subspace of ${\bf C^3}$ projects to
the $r_\alpha$ and $r_\gamma$ vanishing in the base and $r_\beta\geq 0$,
by their definition.
Similarly over $z_2=0=z_3$, where
$(1,0)$-cycle degenerates, $r_\beta$ and $r_\gamma$ vanish and
$r_\alpha\geq 0$, and  1-cycle parameterized by
$\alpha+\beta$ degenerates over $z_1=0=z_2$ where
$r_\alpha-r_\beta =0 =r_\gamma$ and $r_\alpha\leq 0$ degenerate.
To correlate the cycles unambiguously with the lines in the graph (up
to $(q,p)\rightarrow (-q,-p)$) we
will let a $(-q,p)$ cycle of the $T^2$ degenerate
over an edge that corresponds to $p r_\alpha+ q r_\beta = 0$.
The places in the base where $T^2$ fibers degenerate are
correlated with the zero's of the corresponding Hamiltonians.
 This yields the
graph in \graph\ (drawn in the $r_\gamma=0$ plane).

Above we have made a choice for generators of $H_1(T^2)$
to be the 1-cycles generated by $r_\alpha$ and $r_\beta$.
Other choices will differ from this one by an $SL(2,{\bf Z})$
transformation that acts on the $T^2$. We can have
$r_\alpha$ generate a $(p,q)$ 1-cycle and $r_\beta$ the $(t,s)$
1-cycle where $ps-qt=1$. This of course is a symmetry of ${\bf C}^3$.
However, when gluing
different ${\bf C}^3$'s together, as we will discuss below,
the relative choices will matter and will give rise to different
geometries.

\subsec{More general geometries}

Other toric Calabi-Yau threefolds
can be obtained by gluing together ${\bf C}^3$'s. First, one adds
more coordinates e.g. $z_4, \ldots, z_{N+3}$, so that flat patches
are described by certain triples of the coordinates.
Gluing different patches corresponds, in terms of the base, to
identifying some of the
coordinates by $N$ linear relations:
\eqn\Dt{\sum_i Q^A_i |z_i|^2 = t^A}
where $Q_A$, $A=1,\ldots N$ are integral charges satisfying
\eqn\flat{\sum_{i} Q_i^A=0}
which is the Calabi-Yau constraint. Finally, one  divides the
space of solutions to \flat\ by $U(1)^N$ action on $z's$ where the
$A$-th $U(1)$ acts on $z_i$ by
$$z_i \, \rightarrow \,\exp(i \; Q_{i}^A\; \theta_A) z_i.$$
The $N$ parameters $t^A$ are K\"ahler moduli of the
Calabi-Yau.
The mathematical construction above arises in the
physical context of two-dimensional
linear sigma model with ${\cal N}=(2,2)$ supersymmetry
on the Higgs branch \phases.
The theory has $N+3$ chiral fields, whose lowest components
are $z$'s, which are charged under
$N$ vector multiplets with charges $Q_i^A$. The
equations \Dt\ give minima of the D-term potential as solutions.
Dividing by the $U(1)^N$ gauge group, the Higgs
branch is a K\"ahler manifold, and when \flat\ holds, the theory flows to
a two dimensional conformal sigma model in the IR.

{}From the linear sigma model data described above, i.e. the set
of $N+3$ coordinates $z_i$'s and the D-term equations one
can construct the graph $\Gamma$ corresponding to
the toric Calabi-Yau manifold.
First, we must find a decomposition of the set of all coordinates
$\{z_i\}_{i=1}^{N+3}$
into triplets $U_{\a} = (z_{i_a}, z_{j_a},{z_k}_a)$ that
correspond to the decomposition of $X$ into ${\bf C^3}$ patches.
We will describe this below in an example, but it should
be clear how to do this in general. We can pick one of the
${\bf C}^3$ patches, say $U_1$ and in this patch we get the Hamiltonians
$r_{\alpha} = |z_{i_1}|^2 - |z_{i_3}|^2$,
$r_{\beta} = |z_{i_2}|^2 - |z_{i_3}|^2$ which generate the $T^2$ fiber in
this patch. As it turns out, these can serve as $global$ coordinates in the
base ${\bf R^3}$. Correspondingly, they generate a globally
defined $T^2$ fiber\foot{The third coordinate in the base
 is $r_{\gamma} = Im(\prod_{i=1}^{N+3}z_i)$
which is manifestly gauge invariant and moreover, patch by patch,
can be identified with the coordinate used in the ${\bf C^3}$
example above.}. We can call the cycle generated by $r_{\alpha}$
the $(1,0)$ cycle, and that generated by $r_{\beta}$ the $(0,1)$
cycle. The \Dt\ equation then can be used to find the action of
$r_{\alpha, \beta}$ on the other patches. Namely, in the $U_{a\neq
1}$ patch, we can solve for all the other $z$'s in terms of
$z_{i_a},\ldots, z_{k_a}$ using \Dt , since this is by the
definition what we mean by the $U_a$ patch. The degenerate locus
in this patch is then found analogously to the case of the ${\bf
C^3}$ above, where we use the $r_{\alpha}$ and $r_{\beta}$ as
generators of the fiber globally.

$Example~:~${${\cal O}(-3)\rightarrow {\bf P}^2$}
A familiar example of a Calabi-Yau manifold, $X$, of this type is
the ${\cal O}(-3)$ bundle over ${\bf P^2}$. In this case, there are four
coordinates $z_0,z_1,z_2, z_3$, and the D-term constraint is
\eqn\smb{|z_1|^2+|z_2|^2+|z_3|^2-3|z_0|^2=t,}
There are three patches $U_{i}$ defined by $z_{i}\neq 0$,
for $i=1,2,3$, since at least one of these three coordinates must be
non-zero in $X$. All of these three patches look like ${\bf C}^3$.
For example,
for $z_3\neq 0$, we can ``solve'' for $z_3$ in terms of the other
three unconstrained coordinates which then parameterize ${\bf C^3}:$
$U_3 = (z_0,z_1,z_2)$.  Namely, in this
patch, we can use \smb\ to solve for the absolute value of $z_3$,
in terms of $z_{0,1,2}$, and moreover
its phase can be gauged away by dividing with the $U(1)$ action
of the symplectic quotient construction:
$(z_0,z_1,z_2,z_3)\rightarrow(
e^{-3 i \theta}z_0,e^{i \theta}z_1,e^{i \theta}z_2,e^{i \theta}z_3)$ .
We are left with the space of three unconstrained
coordinates $z_0,z_1,z_2$ as we claimed and
this is of course ${\bf C}^3$. Similar statement holds for the other two
patches.
\ifig\ptwo{The graph of ${\cal
O}(-3)\rightarrow {\bf P^2}$. The manifold is built out of three
${\bf C}^3$ patches with the different orientations as in the
figure. The transition functions correspond to $SL(2,{\bf Z})$
transformations of the $T^2$ fibers as one goes from one patch to
the next.} {\epsfxsize 3.0truein\epsfbox{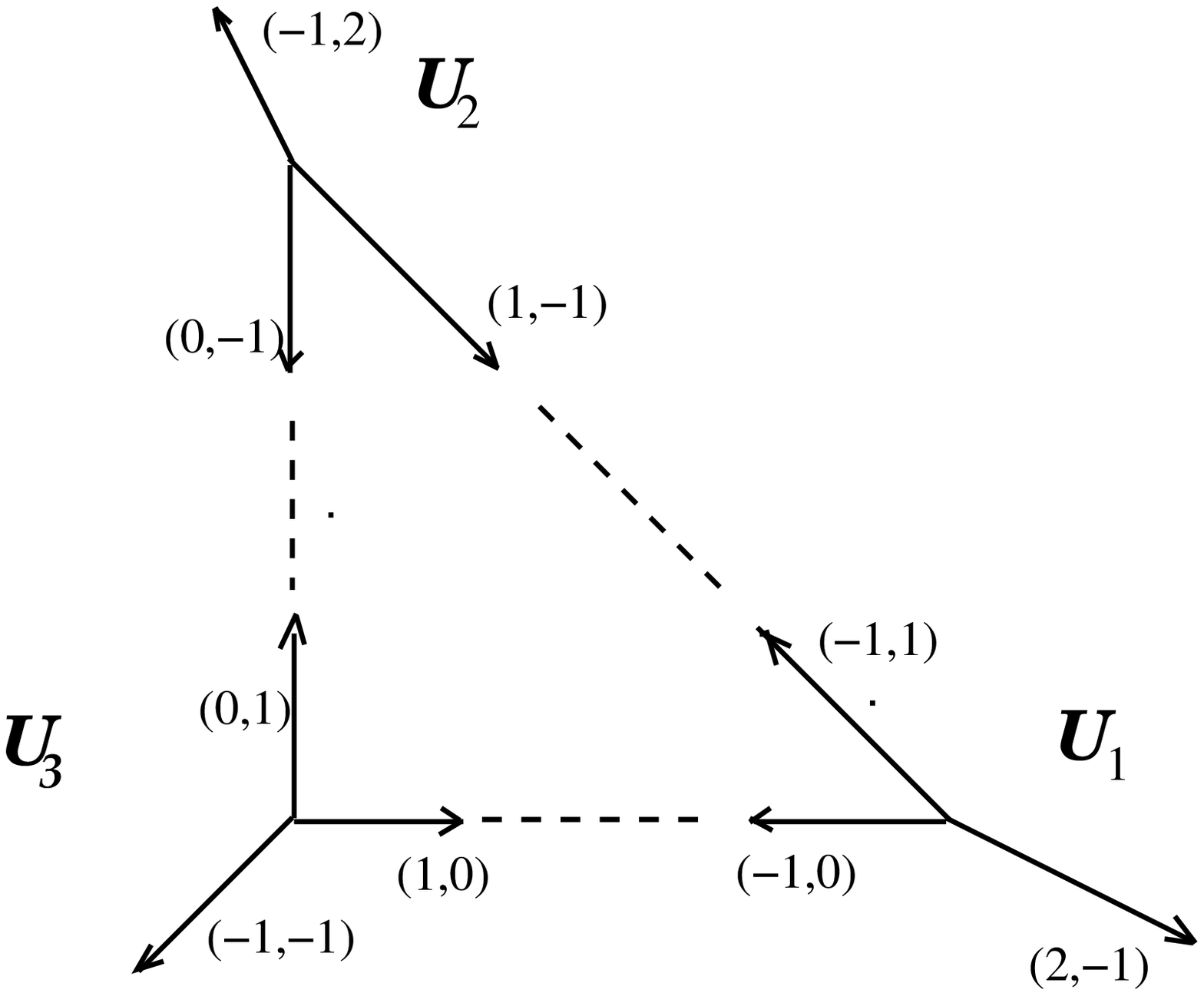}}
Now let us construct the corresponding degeneration graph $\Gamma$.
Let the $T^2$ fiber in the $U_3= (z_0,z_1,z_2)$ patch be generated by
$r_\alpha$ and $r_\beta$ where $r_\alpha = |z_1|^2-|z_0|^2$ and $r_\beta =
|z_2|^2-|z_0|^2$. The graph of the degenerate fibers in the
$r_\alpha-r_\beta$ plane is the same
as in our first ${\bf C}^3$ example, \graph\
(the third direction
in the base, $r_\gamma$ is now given by the gauge invariant product
$r_\gamma= {\rm Im}(z_0 z_1 z_2 z_3)$).
The same two Hamiltonians $r_{\alpha,\beta}$  generate the action
in the $U_{2}=(z_0,z_1,z_3)$ patch, where we use the \smb\
constraint to rewrite them as follows. Since both $z_{0}$ and
$z_1$ are the coordinates of this patch $r_\alpha$ does not
change, $r_\alpha = |z_1|^2 - |z_0|^2$. On the other hand,
$r_\beta$ changes as $z_2$ is not a natural coordinate here, so
instead we have $r_{\beta} = t+ 2 |z_0|^2 - |z_1|^2 - |z_3|^2$,
and hence
$$\exp{(i\alpha~ r_\alpha +i \beta r_\beta)}: (z_0,z_1,z_3)\rightarrow
(e^{i(-\alpha+2\beta)} z_0, e^{i(\alpha-\beta)} z_1, e^{-i\beta}z_3),$$
We see from the above that
the fibers degenerate over three lines: i) $r_\alpha+r_\beta=0,$
and since $z_0=0=z_3$
there, $t\geq r_{\alpha}\geq 0$ where the fact that we have to stop when
$r_{\alpha}=t$ comes from \smb. Over this line $(-1,1)$ cycle degenerates.
ii) There is a line over which a $(-1,2)$ cycle degenerates
where $z_1=0=z_3$, $2r_{\alpha}+r_{\beta}=t$,  and $t\geq r_{\beta}\geq 0$
and
finally, iii) There is a line over which $r_\alpha=0$, $t \geq
r_{\beta}\geq 0$ where $z_0=0=z_1$ and $(0,1)$-cycle  degenerates. The $U_1$ patch is similar, and
we end up with the graph for ${\cal O}(-3)\rightarrow {\bf P^2}$ shown in
\ptwo. Since at least two of the $z$'s have to be zero for the
fiber to degenerate, the graph lies in the $r_{\gamma}=0$ plane.

\subsec{Toric algorithm for general geometries}

The above way of constructing $\Gamma$ becomes cumbersome for
 more complicated geometries.
There is an algorithm which does this efficiently. It is a
standard construction in toric geometry and we will review it here.
This is not meant to be didactical, so for a more thorough exposition
see for example \greene . The algorithm is as follows. To
each coordinate $z_i$ associate a vector $\vec{v}_i$
in ${\bf Z^3}$. The $\vec{v}_i$ are chosen to satisfy an equation analogous to
\Dt , i.e.
$$\sum_i Q_i^{A} {\vec v}_i =0.$$
Since the charges $Q^A$ are integral, the equations can be solved.
The Calabi Yau condition, $\sum_i Q_i^A = 0$ implies in fact that
we can choose all the vectors to lie on a plane $P$, a unit distance
from the origin, e.g. we can choose all the $v_i$'s to be of the form
$\vec{v_i} = (\vec{w}_i,1)$, where ${\vec{w}_i}$ is now a two-vector with
integer entries.
This provides an easy
way to partition the $z$'s into triplets that
parameterize ${\bf C}^3$ patches.
Namely, the $z$'s correspond to
a collection of integral points on a plane $P$ whose
coordinates are $\vec{w}_i$, and this can be triangulated by
considering triangles whose vertices
are triplets of ${\vec w}$'s.
The triangulation that gives a good covering of $X$ is such that
all the triangles in $P$ have unit area.
This is in a sense a maximal
triangulation. For example, for ${\bf C^3}$ we can take: $w_1=(0,0)$,
$w_2 = (1,0)$ and $w_3 =(0,1)$ and the triangulation has a triangle with
these vectors as vertices. For $O(-3) \rightarrow {\bf P^2}$
we can take
$w_0=(0,0)$, $w_1=(-1,0)$, $w_2=(0,-1)$ and $w_3=(1,1)$
with 3 triangles corresponding to $(w_{0} ,w_1,w_2)$, $(w_0,w_1,w_3)$
and $(w_{0},w_2,w_3)$.
In general the choice of the triangulation is not unique. There
is an obvious $SL(2,{\bf Z})$ action of the plane $P$, which is a symmetry
of the closed string theory on $X$. But, in general there are also
different possible triangulations of the same set of points.
and these correspond to different phases in the K\"ahler moduli
space. For a given choice of the K\"ahler parameters in \Dt\ the allowed
triangulation is such that the triplets of coordinates corresponding
to every unit-volume triangle
can all be simultaneously zero in $X$.
We can think about this triangulation as giving rise to a graph
${\hat\Gamma} \in P$.
Given ${\hat \Gamma}$ finding the
graph $\Gamma$ describing the degeneration of the $T^2$ fibers is trivial
-- it is simply the dual graph in the sense that edges of
$\hat{\Gamma}$ are
normals to the edges of $\Gamma$ and vice versa!
\ifig\toric{The graph $\hat{\Gamma}$ of
${\cal O}(-3)\rightarrow {\bf P^2}$.
The black points correspond to vectors ${\vec
w}_i$. Its dual is the graph of degenerate fibers $\Gamma$.}
{\epsfxsize 2.0truein\epsfbox{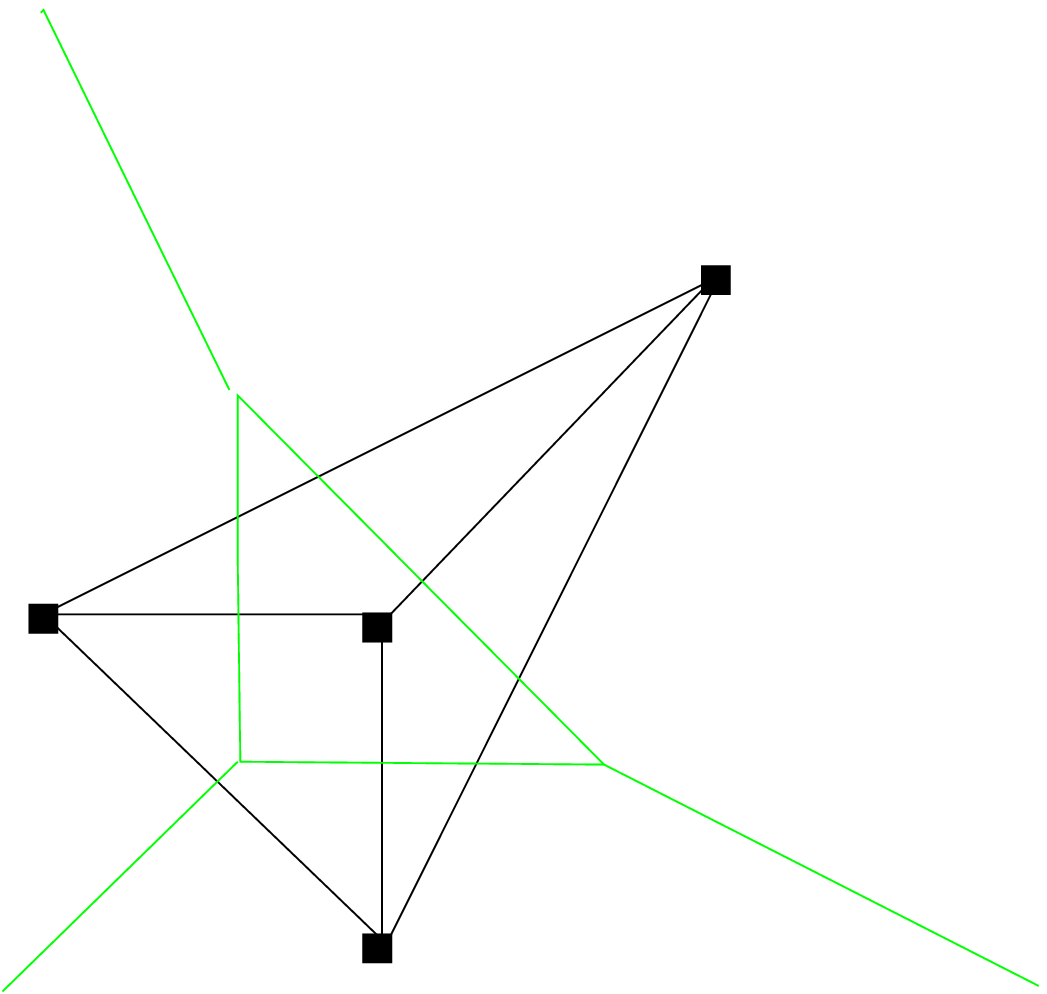}}
In fact, running this algorithm backwards,
provides a fast way to, associate a Calabi-Yau manifold to
a graph $\Gamma$. We first find a dual graph $\hat \Gamma$. The vertices of
this graph are vectors $\hat w_i$, with integer entries. Linear relations
between the vectors ${\hat v}_i=(\hat w_i,1)$ allow us to read off the
charges $Q_A^i$.

To completely specify the geometry we also have to specify the
K\"ahler form $\omega  \in H^{1,1}(X)$. As discussed above, this is
captured by the moduli $t_A$ in the D-term equations \Dt\ of the
linear sigma model. In the formalism we will develop in the
following sections we will need to know the areas of holomorphic
curves in $X$ which are fixed by the torus actions. It is in fact
very easy to determine these directly from the graph $\Gamma$.
This however is easiest to physically motivate in terms of the
$(p,q)$ five-brane picture which we will discuss below.
\subsec{Semi-compact theories}
In the same spirit, we can also consider
certain semi-compact models.
Namely, the geometries discussed so far had only a $T^2$ subspace of the
fiber compact. Here we show that we can also consider some models where
four out of six directions in the geometry are compact.
These geometries can be obtained by
imposing identifications on two of the directions in the base
corresponding to the plane of the graph $\Gamma$.
Clearly, not all toric Calabi-Yau manifolds will admit the
compactifications, but only those with enough symmetry. For those that do,
some of the moduli
that exist in the non-compact geometry are frozen in the compact
one, as they are not consistent with the identifications which we impose.

For example, consider the graph corresponding to $O(-K) \rightarrow
{\bf P^1} \times {\bf P^1}$. When the sizes of the two ${\bf P^1}$'s
are equal we can consider a compactification that corresponds to
identifying points related by:
$$(r_{\alpha},r_{\beta}) \sim
(r_{\alpha} +2 \pi R, r_{\beta} +2 \pi R)
\sim (r_{\alpha} -2 \pi R, r_{\beta} +2 \pi R)$$
The resulting geometry has a single K\"ahler modulus instead of the
two that exist in the non-compact case.
\ifig\compact{The graph of the semi-compact
${\cal O}(-K)\rightarrow {\bf P^1}\times {\bf P^1}$ which arises
by imposing identifications in the base ${\bf R^2}$. The sizes of
the two ${\bf P^1}$'s in the base, that are usually independent,
must be equal here.} {\epsfxsize 3.0truein\epsfbox{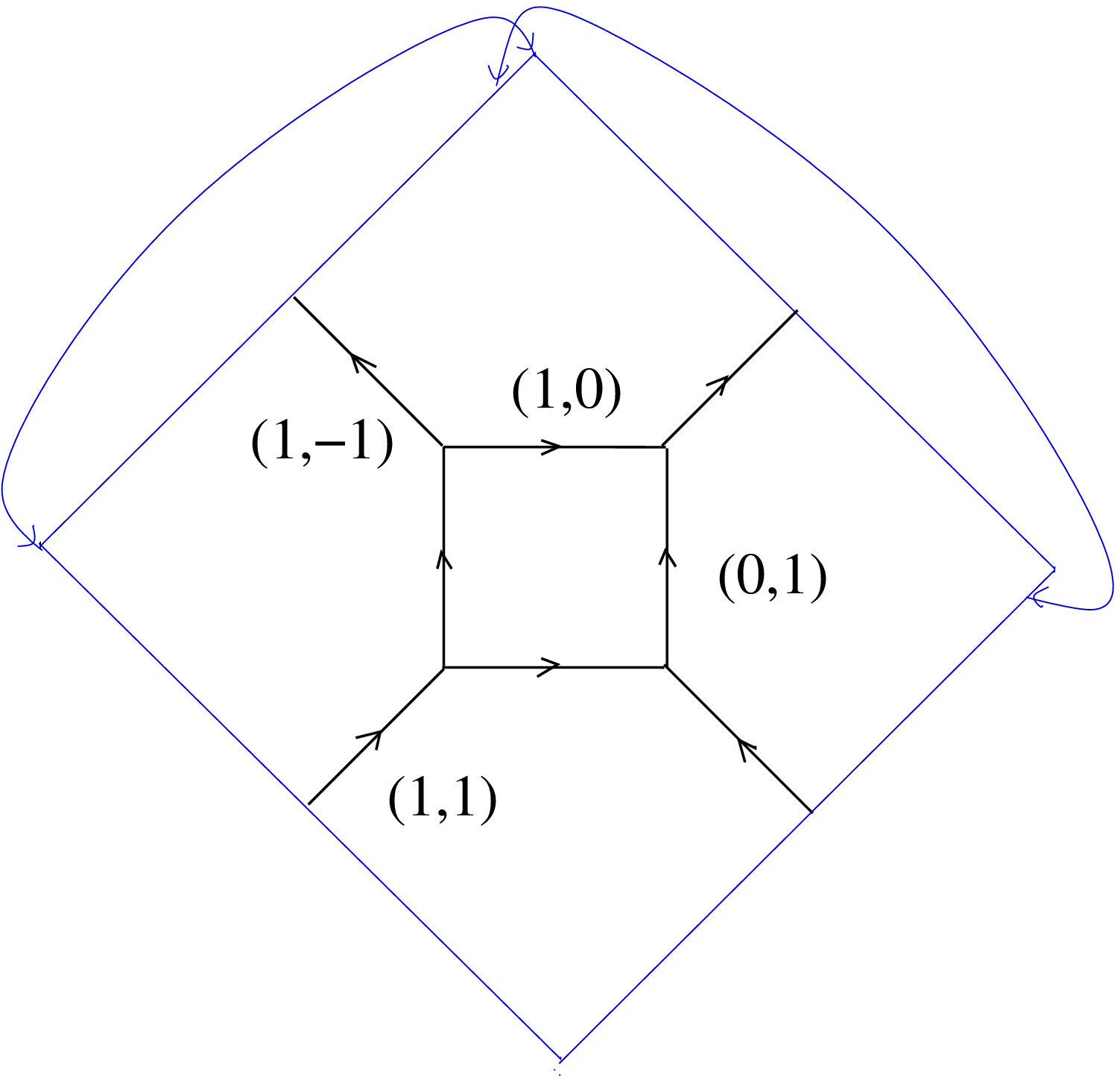}}
As in examples studied in \AKV\ various degenerations of the graph
$\Gamma$ and its different phases allowed by the charge
conservation \cnda\ correspond to geometric transitions of the
full Calabi-Yau geometry. It is easy to see that the semi-compact
models often have obstructions to existence of transitions that
exist in the fully non-compact models. As the simplest example
consider the semi-compact version of $T^* S^3$. This corresponds
to having a $(1,0)$ and $(0,1)$ cycle degenerating over the
corresponding cycles of the ``base'' $T^2$, at different values of
$r_{\gamma}$, i.e. the graph has two components. and corresponding
manifold has $b_3=1$. This geometry, however, does not have a
geometric transition $O(-1) \oplus O(-1) \rightarrow {\bf P^1}$
since the blowup-mode that gives the ${\bf P^1}$ a finite size is
projected out in the semi-compact case. Such an obstruction to a
transition from a single ${\bf S^3}$ is familiar from the fully
compact Calabi-Yau manifolds. Once we discuss the $(p,q)$
five-brane language, it will be manifest that these models have
the same obstructions to resolutions of singularities as the
compact manifolds do. This is related to the fact that the gauge
theories obtained by compactifying string theory on these
geometries are honestly 4-dimensional.

\subsec{Relation to ${(p,q)}$ 5-brane webs}

We can connect the description of Calabi-Yau geometry
by a duality to the web of $(p,q)$ fivebranes \ahk.
This gives an intuitive picture of the geometry.  
The connection was
derived in \lv\ and we will now review it.

Recall that M-theory on $T^2$ is related to type IIB string theory
on ${\bf S}^1$. Since the Calabi-Yau manifolds we have been
considering are $T^2$ fibered over $B={\bf R}^4$, we can relate
geometric M theory compactification on Calabi-Yau manifold $X$ to
type IIB on flat space $B\times {\bf S}^1$. However, due to the fact
that $T^2$ is not fibered trivially, this is not related to the
vacuum type IIB compactification.
The
local type of singularity over a line in the graph
is the Taub-Nut
space, where the $(p,q)$ label denotes which cycle of
the $T^2$ corresponds to the ${\bf S}^1$ of the Taub-Nut geometry. Under the
duality, this local degeneration of $X$ is mapped to the $(p,q)$
five-brane that wraps the discriminant locus in the base space
$B$, and lives on a point on the ${\bf S}^1$. The fact that the $(p,q)$
type of the five brane is correlated with its orientation in
the base is a consequence of the BPS condition. More precisely,
a configuration of five branes that preserves supersymmetry and
$4+1$ dimensional Lorentz invariance is pointlike in a fixed
${\bf R}^2$ subspace of the base.
In the two remaining directions of the base, which we parameterized by
$(r_\alpha,r_\beta)$ above, the five branes are lines
where the equation of the $(p,q)$ five brane is
$pr_{\alpha}+qr_{\beta}={\rm const}$.

The $(p,q)$ five-brane picture provides a simple way to read off
the sizes of various holomorphic curves embedded in the Calabi-Yau
$X$. For this paper, we will only need to know this for the curves
which are invariant under the $T^2$ action. It is clear from the
discussion in the previous subsections of this section that these
are the curves in $X$ which correspond to the edges of the graph
$\Gamma$. The duality of M-theory on $X$ to the IIB with 5-branes
relates the membranes wrapping holomorphic curves in $X$ to
$(p,q)$ string webs ending on the 5-brane web. The masses of
corresponding the BPS states get related by the duality. In the
M-theory picture, the masses of BPS states are the K\"ahler volumes
of the holomorphic curves, and in the IIB language they are the
tensions$\times$lengths of the corresponding strings. The curves
that project to the edges of the graph correspond to strings that
are within the five-branes themselves. These strings are
instantons of the five-brane theory. As discussed in \ahk\ the
instanton of a $(p,q)$ five brane is a string whose tension is
$$T_{p,q}= {Im(\tau)\over |p\tau+q|}T_s$$
where $\tau$ is the type IIB dilaton-axion field $\tau =
{\chi\over 2\pi} + {i\over g_s}$, and the $T_{s}$  is a tension of
a fundamental string which is an instanton in an NS, or $(1,0)$,
five-brane. Note that this is not a conventional free $(r,s)$
string tension for any $r,s$ (which would have been $T_s
\sqrt{r^2+q^2}$), and correspondingly the instanton in general
does not correspond to any free $(p,q)$ string. This is because
the action of an instanton (i.e. tension in this case, as the
instanton is a string) is simply governed by the coefficient of
the $F\wedge * F$ term in the five-brane action and this is $
T_s/g_s$ for a D5 brane and $T_{p,q}$ for a $(p,q)$ five-brane.
Thus, knowing the length $x$ of a $(p,q)$ edge in $\Gamma$, the
area of a holomorphic curve corresponding to this edge is
$x/\sqrt{p^2+q^2}$ (we will take $\tau=i$ which is the square
$T^2$). On the other hand, the slope of the five-brane in the
${\bf R^2}$ is correlated with its $(p,q)$ type as we said above,
and this allows us to read-off the lengths of all the edges in the
graph in terms of the few independent ones which correspond to the
K\"ahler moduli $t^A$ in \Dt . For example, suppose that the length
of the horizontal edge in the graph of $O(-3)\rightarrow {\bf
P^2}$ is $t$. Then the length of the $(1,-1)$ edge is $t
\sqrt{2}$. However the tension of the corresponding instanton in
$1/\sqrt{2}$ so the area of the holomorphic curve corresponding to
this leg is $t$. Similarly, we find that the area of the curve
corresponding to the $(0,1)$ leg is also $t$.

\subsec{Mirror Symmetry and the dual B-model Geometries}

Mirror manifolds of the local toric Calabi-Yau manifolds were
derived in \HV , by using T-duality in the linear sigma model in the
previous section. The result of \HV\ is as follows.
The mirror theory is a theory of variation of complex structure
of a certain hypersurface $Y$ which is given in terms of $n+3$ dual
variables $y^i$ \HV\ with the periodicity $y^i\sim y^i+2\pi i$.
The variables $y^i$
are related to variables of the linear
sigma model \Dt\ as
 \eqn\dual{Re(y^i) = |z^i|^2,}
so in particular, the D-term equation \Dt\  is
mirrored by
\eqn\mdf{\sum_i Q_i^A ~ y^i = t^A.}
Note that \mdf\ has a
three-dimensional family of solutions. One parameter is trivial
and is given by $y^i\rightarrow y^i+c$. Let us parameterize the
two non-trivial family of solutions by $u,v$, and pick an inhomogenous
solution.
The the hypersurface $Y$ is given by
\lref\HoriCK{
K.~Hori, A.~Iqbal and C.~Vafa,
``D-branes and mirror symmetry,''
arXiv:hep-th/0005247.
}
\HoriCK
\eqn\mirr{x\;\tilde x  = e^{y_1(u,v)} + e^{y_2(u,v)}\ldots
+e^{y_{N+3}(u,v)}\equiv P(u,v),} where
$y_i(u,v)$ solve \mdf .
The solutions to \mdf\ are of the form
$$y_i = w_i^u \; u + w_i^v\; v +t_i(t),$$
for some vector $\vec{w}_i=(w_i^u,w_i^v)$ with integer entries.
In fact this is the same vector that we associated to the coordinate $z_i$
in the previous section, when we discussed toric geometry.
The monomials $e^{w_1 u + w_2 v}$ are in one to one correspondence
with points of the graph $\hat \Gamma$.

In the sections to follow a prominent role will be played by the
Riemann surface $\Sigma_X$, obtained by setting $x, \tilde x$ to
zero in \mirr.
\eqn\rs{\Sigma_X\; :\; 0  = P(u,v)}
Note that this Riemann surface is $closely$ related to the graph $\Gamma$
and it is in fact obtained by the fattening of its edges.
For example, for the mirror of ${\bf C^3}$ we get
\eqn\panp{e^{u}+e^{v}+1=0}
and this has three asymptotic regimes corresponding to $u\rightarrow
\infty$ where the equation of the Riemann surface is $v=i\pi$. This is
a long cylinder parameterized by $u$. Similarly, there is a long cylinder
parameterized by $v \rightarrow \infty $ where $u=i\pi$ and there is a
third cylinder where $u=v+i\pi$, and $u,v\rightarrow \infty$, so that
this Riemann surface corresponds to a sphere with three punctures.
{}From ``far away'' the Riemann surface will look like the graph $\Gamma$ of
${\bf C^3}$. Similarly, the Riemann surface $\Sigma_X$ of any $X$
has a degenerate limit where it looks like the graph $\Gamma$.
It is clear that by gluing various patches given by \panp\ dictated
by the graph $\Gamma$
we can obtain the full Riemann surface $\Sigma_X$.

\newsec{Topological A-model and the vertex}

The amplitudes of
topological A-model localize on holomorphic maps from the worldsheet
to the target space \topstrings. In particular the path integral
defining the free energy of the theory reduces to a sum
over the topological type of holomorphic maps from the worldsheet to the
Calabi-Yau space $X$. Each term in the sum involves an integral over
the moduli space $\cM$ of that type of maps, which leads to the so called Gromov-Witten
invariant of that map, weighted by the $e^{-{\rm Area}}$,
where the area is the one of the target space curve.

In this paper we will find a very efficient way to calculate
the A-model amplitudes on local toric Calabi-Yau manifolds described in
the previous section, to all genera, exactly.
The rough idea is to place Lagrangian  D-brane/anti-D-brane
pairs
in appropriate places (one on each edge of the toric
diagram)
 to cut the Calabi-Yau manifold $X$ into patches which
are ${\bf C}^3$.  They do not quite cut the Calabi-Yau in pieces
as their dimension is too low, but all closed string worldsheet
configurations will nevertheless cross them. More precisely
using toric actions the configurations can be made to pass
through the lines of the toric graph \kon.
 Thus if we are interested in the closed
string amplitudes we could use the D-branes as ``tags'' for
when the closed string goes from one patch to another.  Thus
the open string amplitudes on each patch, glued together in
an appropriate way, should have the full information about
the closed string amplitudes.

\ifig\localization{The curves that are invariant under
$T^2$ action
in the ${\cal O}(-3)\rightarrow {\bf P^2}$ geometry. All the invariant curves are
${\bf P^1}$'s.
The maps from $\Sigma_{g>0}$ that give a non-zero
contribution to the A-model amplitudes
are degenerate maps where the genus $g>0$ parts of the curve are mapped
to vertices.}
{\epsfxsize 4.0truein\epsfbox{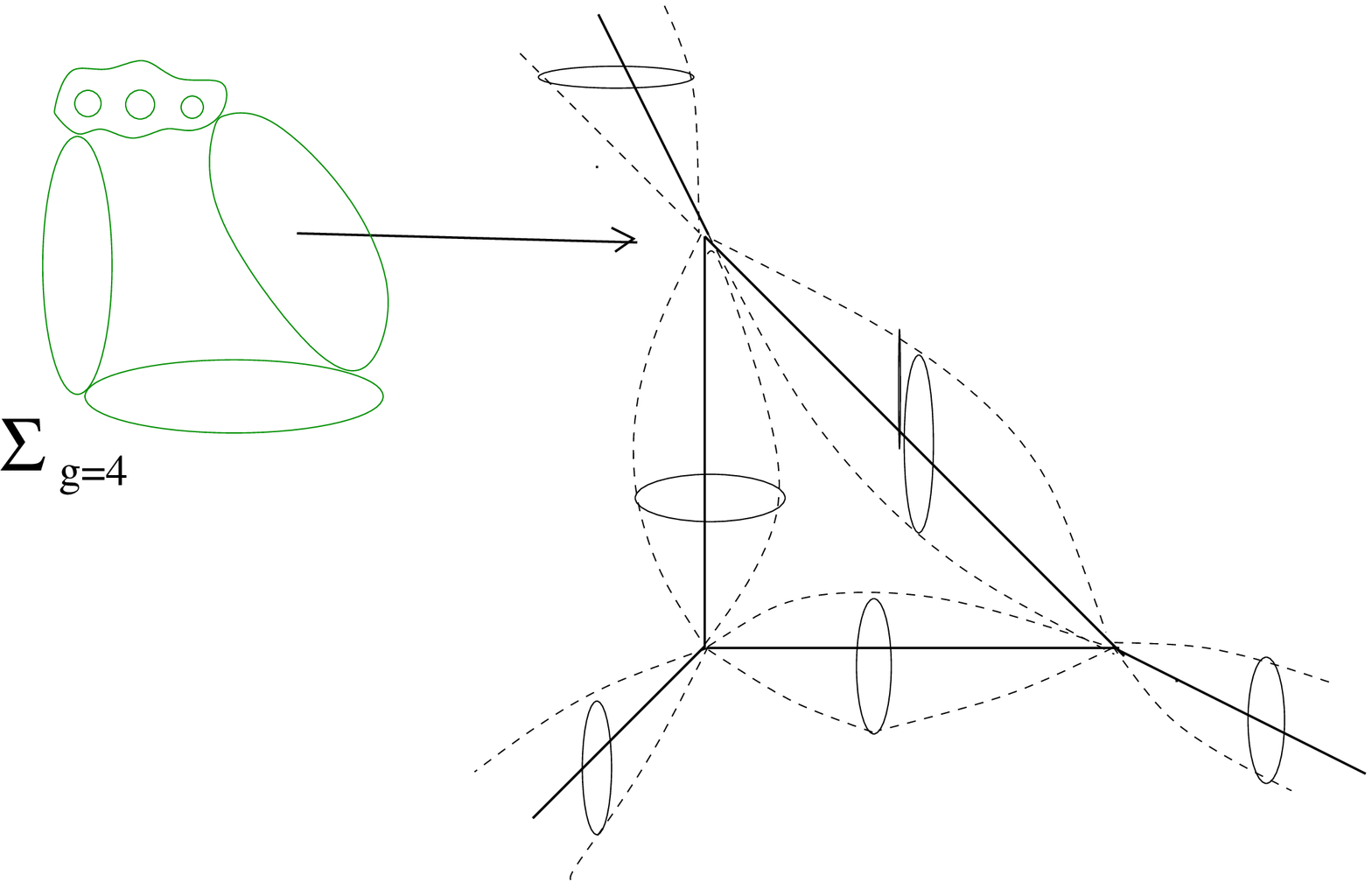}}

The idea is then as follows. Consider chopping the graph $\Gamma$
into tri-valent vertices by cutting each of the legs into two.
Physically we can view this as placing a D-brane/anti-D-brane
pair. Each vertex corresponds to a ${\bf C}^3$ patch, as in
\figs{\ptwo, \localization}. This cuts the ${\bf P^1}$'s which
correspond to compact legs of $\Gamma$ into disks. Consider the
maps $\Sigma_g \rightarrow X$. The maps which contribute to the
A-model amplitudes themselves project to (subgraph of) $\Gamma$,
and cutting the graphs cuts the maps as well, so we get Riemann
surfaces with boundaries. We are led to consider open topological
string on ${\bf C}^3$ with three (stacks of) Lagrangian D-branes
of the appropriate kind, one on each leg.  From these data we
should be able to obtain, by suitable gluing, closed string
amplitudes on arbitrary toric Calabi-Yau threefolds.

\subsec{The vertex as an open string amplitude}

Consider again the description of the ${\bf C}^3$ in section 2.
The Lagrangian D-branes we need are in fact among the original examples of
special Lagrangians of Harvey and Lawson \hl.
The topology of all of the Lagrangians is ${\bf C} \times {\bf S}^1$.
In particular, they project to lines in the base ${\bf R}^3$, and
wrap the $T^2$ fiber.
In the base, the
three Lagrangians $L_{1,2,3}$ are given by\foot{The Lagrangians are pointlike in the
fiber generated by
$r_{\gamma}$. The fiber is parameterized by ${\rm Re}(z_1z_2z_3)$ and the Lagrangians
are where this vanishes.}
\eqn\lag{\eqalign{L_1 \;: \quad & r_{\alpha} = 0,\quad  r_{\beta}=r^*_1,
\quad r_{\gamma}\geq 0 \cr
L_2\;:\quad& r_{\beta} = 0,\quad  r_{\alpha}=r^*_2,\quad r_{\gamma}\geq 0 \cr
L_3\;: \quad &r_{\alpha}-r_{\beta}=0,\quad  r_{\alpha}=r^*_3, \quad r_{\gamma}\geq 0.}}
In order not to have the boundary at
$r_{\gamma}=0$, $L_i$'s are constrained to end on the graph $\Gamma$, where
one of the 1-cycles of the $T^2$ degenerates to ${\bf S}^1$. The
parameters $r^*_i$ correspond to the moduli of $L_i$'s, and the ``no
boundary'' constraint that we just mentioned is what constrains the
number of the moduli to one.
The Lagrangians are easily seen to
intersect the fixed ${\bf P^1}$'s along ${\bf S}^1$'s so the boundaries of
the maps can end on them. For example, a holomorphic disc
ending on $L_1$ is given by $z_1=0 = z_3$, $|z_2|^2 \leq r^*_1.$
\ifig\withbranes{A ${\bf C}^3$ with
three-stacks of Lagrangian D-branes of the type discussed in the
text. The A-model amplitudes localize on holomorphic maps with
boundaries where all the higher genus information is mapped to the
vertex.} {\epsfxsize 4.0truein\epsfbox{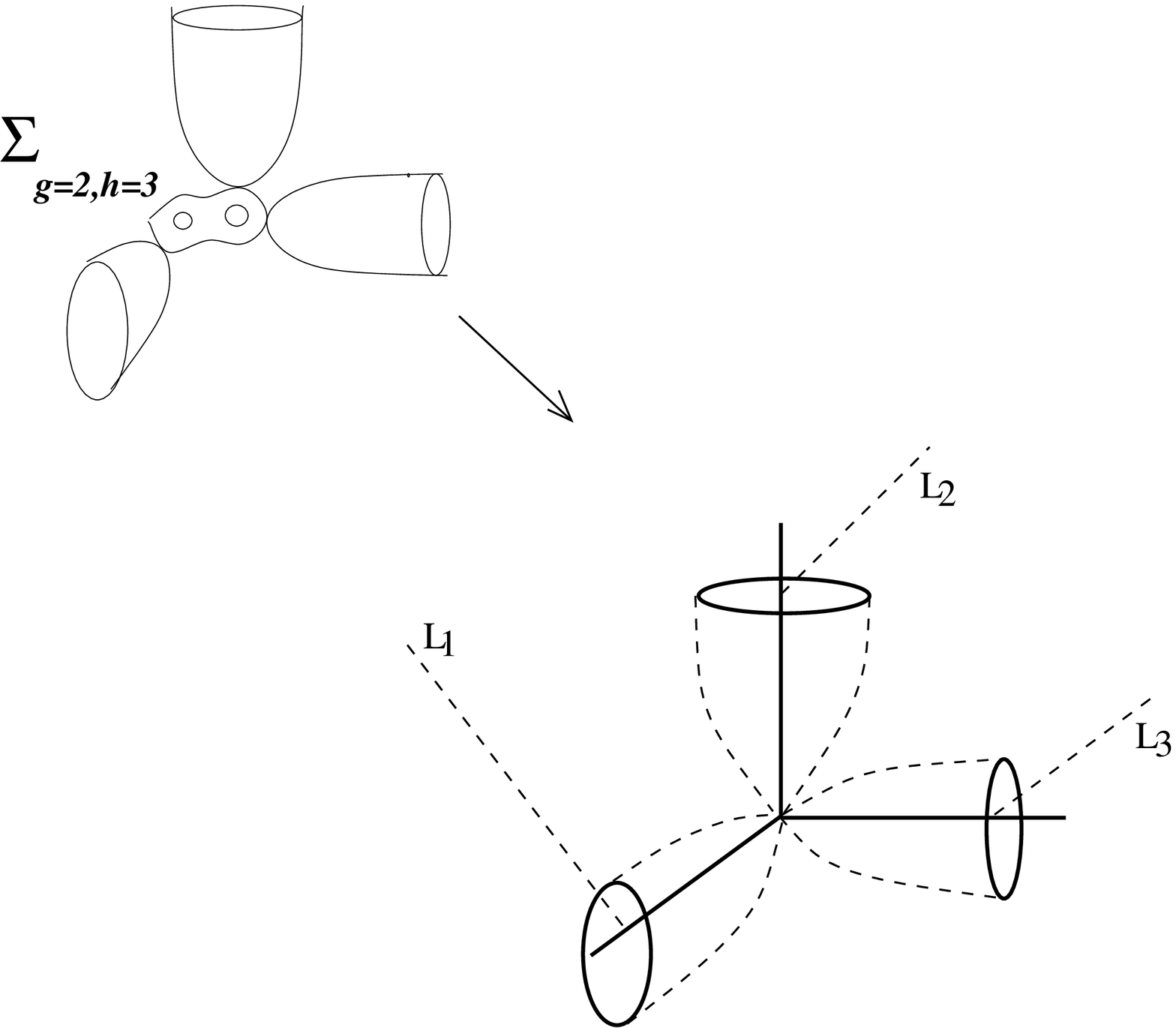}}
Now, consider the topological A-model string amplitude
corresponding to some number of D-branes $N_i$ on the $i$-th
Lagrangian $L_i$ on ${\bf C}^3$. The partition function takes the
form
\eqn\ver{Z = \sum_{\vec{k}^{(1)},\vec{k}^{(2)},\vec{k}^{(3)}}
C_{\vec{k}^{(1)},\vec{k}^{(2)},\vec{k}^{(3)}}\;\prod_{i=1}^3
{1 \over z_{\vec k^{(i)}}}\;
Tr_{\vec{k}^{(i)}}V_{i}}
where $V_i$ is the path ordered exponential of the Wilson-line on
the $i$-th D-brane, $V_i = {\rm P}\exp[\oint A_1]$ around the ${\bf S}^1$,
$$Tr_{\vec{k}}V=\prod_{j=1}^{\infty} ({\rm tr}V^{j})^{k_j},$$
and
$$
z_{\vec k}=\prod_j k_j! \; j^{k_j}.
$$
Note that there are $k_j$ holes of winding number $j$ so the sum
$h=|{\vec k}|=\sum_j k_j$ is the total number of holes on a fixed D-brane,
and $\ell= \sum_{j}j k_j$ is the total winding number. We have
absorbed the modulus of the Lagrangian into the corresponding $V$
which is complexified in string theory.
The vertex amplitude,
$C_{\vec{k}^{(1)},\vec{k}^{(2)},\vec{k}^{(3)}}$ is naturally a
function of the string coupling constant $g_s$ and, in the genus
expansion, it contains information about maps from Riemann
surfaces of arbitrary genera into ${\bf C}^3$ with boundaries on
the D-branes, see \withbranes.

The vertex $C$ is the basic object from which, by gluing, we
should be able to obtain closed string amplitudes on arbitrary toric
geometries. As we will see later, the vertex is naturally used to
calculate general A-model amplitudes with boundaries as well.

\subsec{Framing of the vertices}

Because of the above considerations we are led to consider non-compact
D-branes in ${\bf C}^3$. Due to the non-compactness of the
world-volume of D-branes, to fully specify the quantum theory we must
specify the boundary conditions on the fields on the D-branes at
infinity. This was discovered in \AKV\ and
is the closed string dual to the framing ambiguity of the
Chern-Simons amplitudes \jones.

To keep track of the
boundary condition at infinity, we can
use the following trick \AV. We modify the
geometry in a way that makes the Lagrangian cycles wrapped by the
D-branes compact, while not affecting the topological A-model amplitudes.
We do so by introducing compact ${\bf S}^3$ cycles in the geometry by
allowing the $T^2$ fiber to degenerate at additional locations in
the base ${\bf R^3}$, as in the figure 7. The additional three
lines $F_i$ in the base correspond to degeneration of a
fixed $f_i=(p_i,q_i)$ cycle there.
There are now compact special Lagrangian ${\bf S}^3$
cycles ${\tilde L}_{1,2,3}$
which correspond to paths of the shortest distance between the
graphs $\Gamma$ and $F_i$. For
this cycle to be a non-degenerate
 ${\bf S}^3$ we need the following
condition on the
holonomy
\eqn\fr{f_i \wedge v_i  =1,}
where $v_i$ corresponds to the $H_1(T^2)$ class of the edges
of the graph.  Note that we have chosen a particular
orientation for the framing so that the above product
is always $+1$.
Clearly, if $f_i$ is a solution to \fr\ , so is $f_i-n v_i$ for any
integer $n$. This ${\bf Z}$ valued choice does affect the physics
of the D-brane. To specify the theory on the D-brane fully, we must
specify a choice of framing \AKV, i.e. a choice of the integer
$n$. This is a quantum ambiguity and only the relative values of
$n$ are meaningful. Given an (arbitrary) choice of
framing for the
i-th leg, i.e. a vector $f^{(0)}_i$, the vector $f^{(n)}$
corresponds to a relative framing associated to an integer $n$ if
\eqn\deffr{f^{(n)}\wedge f^{(0)}=n.}
It is crucial for us to keep track of framing. 
The relevant object is a framed vertex, 
$$C^{(f_1,f_2,f_3)}_{\vec{k}^{(1)},\vec{k}^{(2)},\vec{k}^{(3)}},$$
where we specify the framing of the D-branes on the three legs.

Without loss of generality we can take the $v_i$ to be
$v_1= (-1,-1),v_2 = (0,1), v_3 =(1,0)$, since any other choice is related to this
one by an $SL(2,{\bf Z})$ transformation. More generally
we can introduce a vertex which depends on both $v_i$ and $f_i$, but
knowing the vertex for the canonical choice of $v_i$ with arbitrary
framing $f_i$ is enough. Moreover, if we know the vertex in any one framing,
the vertex in any other framing is related to
it in a simple way \refs{\AKV ,\mv}.  In order to describe
this it is most convenient to go to the
``representation basis'' for the vertex which we
will now turn to.

\subsec{The vertex in the representation basis}

Topological open string amplitudes can be written
in terms of products of traces to various powers, as
in \ver. They
can also be rewritten in the representation basis, and
this can be done unambiguously in the limit where we take $N_i\rightarrow
\infty$ branes.  We define the representation basis for the vertex by
$$\sum_{R_1,R_2,R_3}
C^{f_1,f_2,f_3}_{R_1,R_2,R_3} \prod_{i=1}^3 \; {\rm Tr}_{R_i} V_i
= \sum_{\vec{k}^{(1)},\vec{k}^{(2)},\vec{k}^{(3)}}
C^{f_1,f_2,f_3}_{\vec{k}^{(1)},\vec{k}^{(2)},\vec{k}^{(3)}}\prod_{i=1}^3
\;{1 \over z_{\vec k^{(i)}}}\;Tr_{\vec{k}^{(i)}}V_{i}$$
To obtain $C$ in the representation basis defined above, we make
use of Frobenius formula
$$Tr_{\vec{k}} V = \sum_{R} \chi_{R}(C(\vec{k})) {\rm Tr}_R V,$$
where $\chi_{R}(C(\vec{k}))$ is the character of the symmetric
group $S_{\ell}$ of $\ell$ letters for the conjugacy class
$C(\vec k)$, in representation corresponding to the Young tableau of
$R$. Using this we obtain
\eqn\verr{
C^{f_1,f_2,f_3}_{R_1,R_2,R_3} =
 \sum_{\vec{k}^{(1)},\vec{k}^{(2)},\vec{k}^{(3)}}
C^{f_1,f_2,f_3}_{\vec{k}^{(1)},\vec{k}^{(2)},\vec{k}^{(3)}}
\prod_i {\chi_{R_i}(C(\vec{k}^{(i)})) \over z_{\vec k^{(i)}}}}
Now we are ready to describe the framing dependence of the vertex.
We have \mv\
\eqn\framedc{
C^{f_1-n_1v_1,f_2-n_2v_2,f_3-n_3v_3}_{R_1,R_2,R_3} =
(-1)^{\sum_i n_i \ell(R_i)}q^{\sum_i n_i \kappa_{R_i}/2}C^{f_1,f_2,f_3}_{R_1,R_2,R_3},}
where $\kappa_R$ is related to the quadratic Casimir $C_R$ of the
representation $R$ of $U(N)$ as $\kappa_R = C_R - N \ell(R)$, and
$\ell(R)$ is the number of boxes of the representation (which is
the same as the total winding number in the $\vec{k}$-basis). If the
representation $R$ is associated to a Young tableaux whose $i$-th row has
$\ell_i$ boxes, $\ell(R) = \sum_i \ell_i$ one has
\eqn\framfac{
\kappa_R= \sum_i \ell_i (\ell_i-2i+1).
}

\subsec{Symmetries of the vertex}

Consider an $SL(2,{\bf Z})$ transformation that acts on the
$T^2$ fiber of ${\bf C}^3$, in the presence of D-branes. As already
noted the vertex depends on three pairs $(f_i,v_i)$ where
$v_i$ denotes the $(p,q)$ structure of the edge and $f_i$
denotes the framing associated to that edge, and one has
$$f_i\wedge v_i =1$$
which means that $(f_i,v_i)$ forms an oriented
basis for $H_1(T^2)$. Moreover, if we orient the edges
inward towards the vertex, then $\sum_i v_i=0$.
One also has that $v_i\wedge v_j=\pm 1$ for $i\not =j$.
We can choose a cyclic ordering of $v_i$ according
to the embedding of the corresponding vectors in ${\bf R}^2$.
In terms of this cyclic ordering we have
$$v_2 \wedge v_1 =v_1 \wedge v_3 =v_3\wedge v_2 = 1.$$

It is
clear that an element $g \in SL(2,{\bf Z})$ generates a symmetry
of the vertex while replacing
$$(f_i,v_i)\rightarrow (g\cdot f_i,g\cdot v_i)$$
\ifig\canframing{ Three stacks of D-branes on
${\bf C}^3$. We have introduced graphs $F_{1,2,3}$ to help us keep
track of framing. $F_i$ are straight lines in the base,
corresponding to vectors $f_i$ in the text. Different choices of
$f_i$ give different amplitudes. The choice in this figure is
defined to be canonical framing.} {\epsfxsize
3.0truein\epsfbox{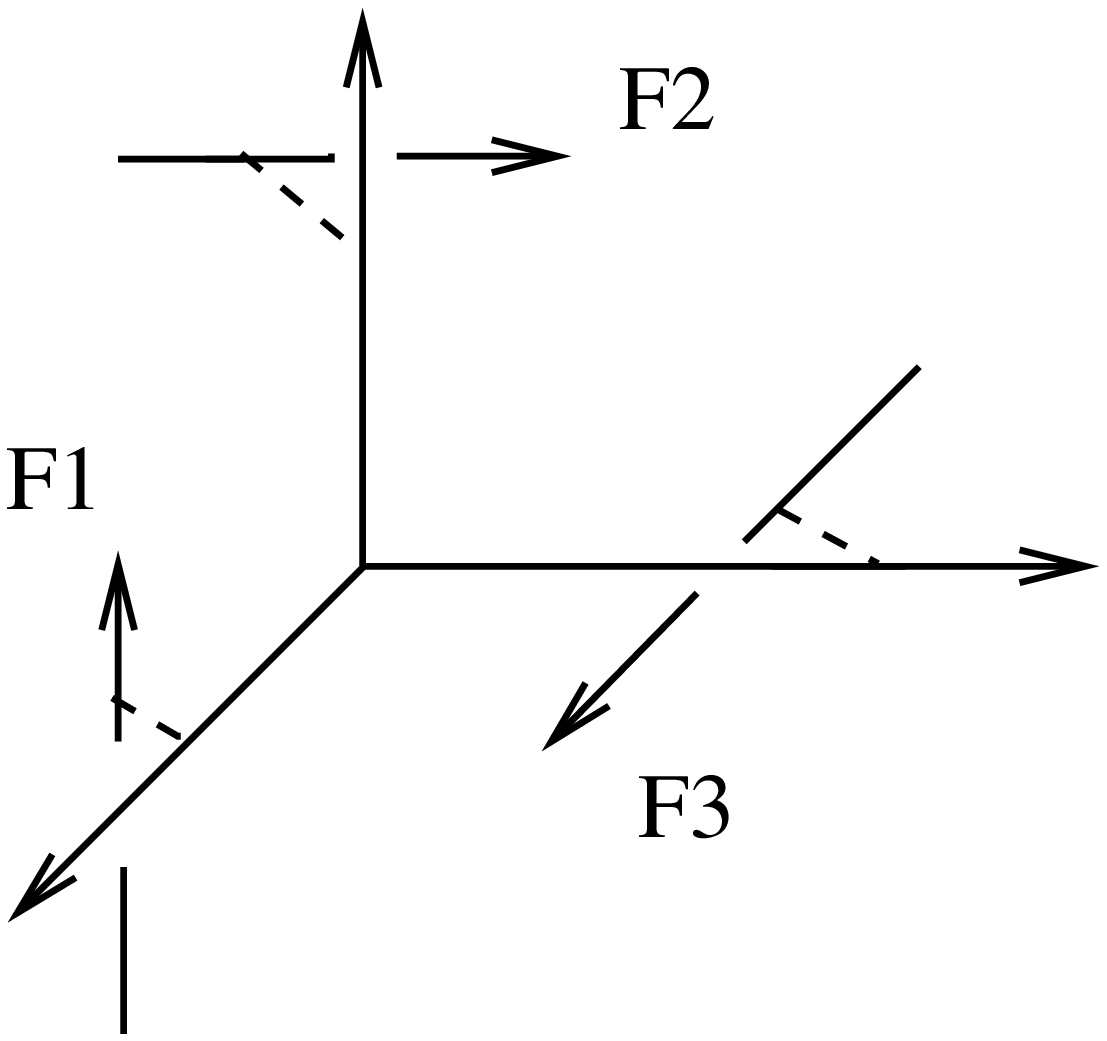}}
There is one particularly natural choice of framing $f_i$
based on symmetry considerations, namely (see \canframing)
$$(f_1,f_2,f_3)=(v_2,v_3,v_1)$$
Note that this has the required property that $f_i\wedge v_i=1$.
For any given choice of $v_i$ cyclically ordered in this way,
we shall call this the canonical
framing and denote the corresponding
vertex by $C$.  Any other choice of framing, relative
to this canonical choice, will be denoted by $C^{n_1,n_2,n_3}$
where $n_i$ denote the amount of change in framing relative
to the canonical choice.  Let $C_{R_1,R_2,R_3}$ denote
the vertex for the canonical framing for $v_i:(-1,-1),(0,1),(1,0)$.
Then it follows that
\eqn\frdep{C^{(f_i,v_i)}_{R_1,R_2,R_3}=(-1)^{\sum_i n_i\;
\ell (R_{i})} q^{{1\over 2}\sum_i n_i\;\kappa_{R_{i}}}
C_{R_1,R_2,R_3}}
where
$$n_i = f_{i}\wedge v_{i+1}$$
and $i$ runs mod $3$.
With three D-branes on the legs of the vertex, the vertex amplitude
$C_{R_1,R_2,R_3}$
is invariant under the ${\bf  Z}_3$ subgroup
of $SL(2,{\bf Z})$ taking
$$v_1\rightarrow v_2, \qquad v_2\rightarrow v_3, \qquad v_3\rightarrow v_1.$$
 Note that the condition
that $v_3\rightarrow v_1$ follows from the first two from
$\sum_i v_i=0$.
Clearly there is such an $SL(2,{\bf Z})$ transformation, because $(v_1,v_2)$
and $(v_2,v_3)$ form an oriented basis for $H_1(T^2)$.  For example
for the simple choice of $v_i: (-1,-1),(0,1),(1,0)$ it is
 generated by $TS^{-1}$ in the standard basis for
generators of $SL(2,{\bf Z})$, so we see
that the vertex amplitude with canonical
choice of framing, which is compatible with this
cyclicity, has a cyclic symmetry,
\eqn\syma{C_{R_1,R_2,R_3}= C_{R_3,R_1,R_2}= C_{R_2,R_3,R_1}.}

So far we have oriented edges of the vertex away from
the vertex.  In gluing vertices together we would need also
to deal with arbitrary orientation of the edges.  Suppose for
example we take $v_1\rightarrow -v_1$.  What this does is to change
the orientation of the circle on the corresponding D-brane.
This is a parity operation on the D-brane, which changes
the action to minus itself.  Thus a genus $g$ topological
string amplitude
with $h$ boundaries on the corresponding D-brane
(in the `t Hooft notation) gets modified by
$$(-1)^{\rm loops}=(-1)^{2g-2+h}=(-1)^h$$
This can also be obtained by viewing the change of
the sign of the action as replacing a topological
brane by a topological anti-brane which replaces $N\rightarrow
-N$ \antibr.
It is convenient to write how this modifies
the vertex in the representation basis. This can be
done using
\eqn\tran{\chi_{Q^t}(C(\vec{k}))=(-1)^{|{\vec{k}}|+\ell(Q)}\chi_Q(C(\vec{k})),}
where $\ell(Q)$ denotes the number of boxes of representation $Q$.
It follows that
$$C_{R_1,R_2,R_3}\rightarrow_{v_1\rightarrow -v_1}
(-1)^{\ell(R_1)} C_{R_1^t,R_2,R_3}.$$
Similarly we can change any of the other $v_i\rightarrow -v_i$.

\ifig\symme{Various symmetries of the
three-point vertex. The figures in the top row are related by a
${\bf Z}_3$ subgroup of $SL(2,{\bf Z})$. The figure in the bottom
row is generated from the top left one by a symmetry of ${\bf
C}^3$ that exchanges $z_1$ and $z_2$. This also maps a D-brane (D)
to an anti D-brane (A).} {\epsfxsize
4.0truein\epsfbox{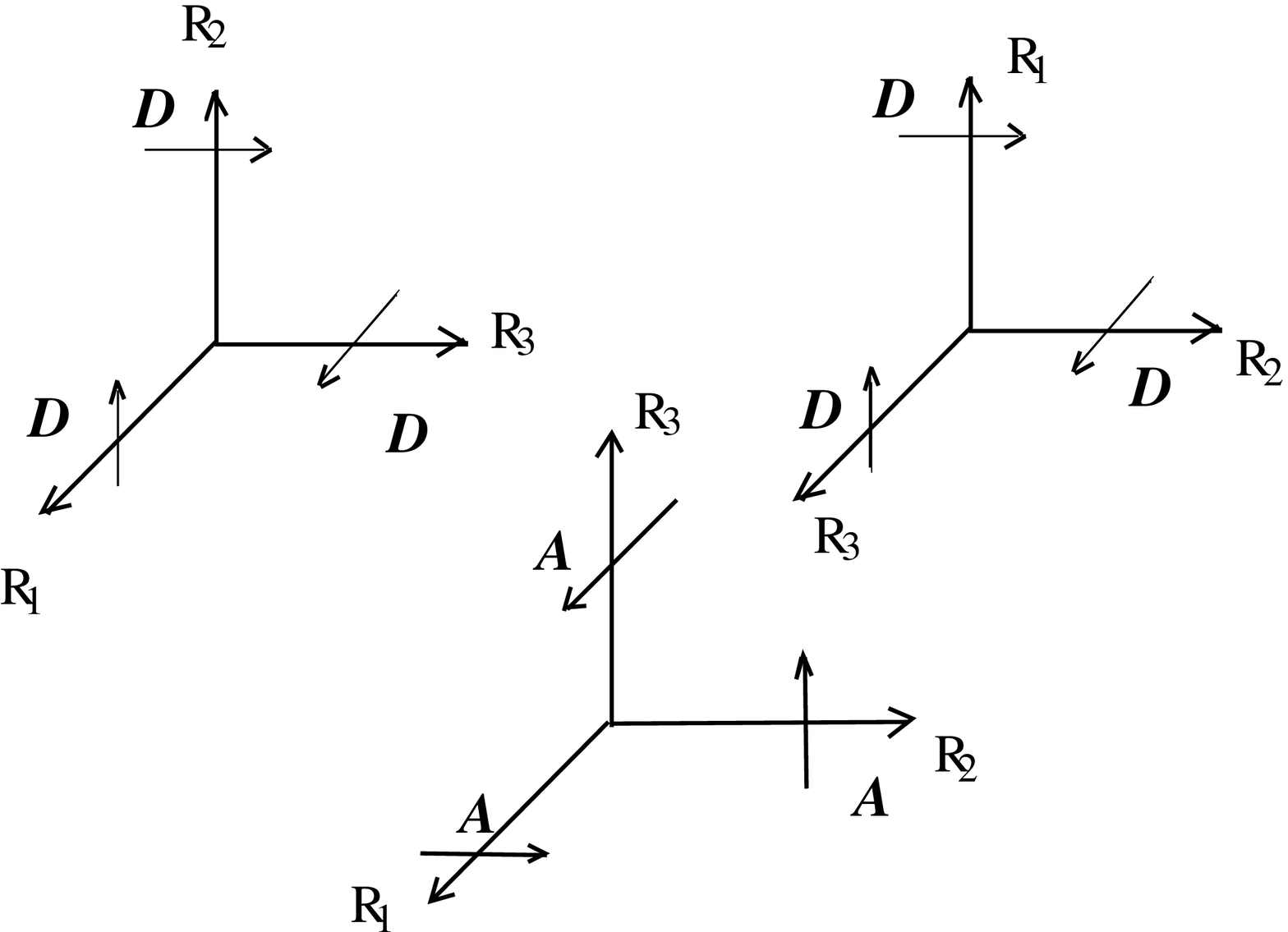}}

We have seen that the vertex has cyclic symmetry
in the canonical framing. It is natural to ask what
symmetry it has under permutation of any of the two representations.
There is a symmetry of ${\bf C}^3$ that exchanges
any pair of its coordinates, say $z_1$, $z_2$. This acts as
orientation reversal on the world-volume of all three D-branes, as
it acts on the $T^2$ fiber by exchanging $(1,0)$ and $(0,1)$
1-cycles, and the $T^2$ is wrapped by all the D-branes.  In addition
the framings are shifted by one unit:  We have
$$((f_1,v_1),(f_2,v_2),(f_3,v_3))\rightarrow ((f_2,v_1),(f_3,v_2),(f_1,v_3))$$
and from \frdep\ it follows that the new framing
is shifted by
$$(f_2\wedge v_2,f_3\wedge v_3 ,f_1 \wedge v_1)=(v_3\wedge
v_2,v_1\wedge v_3,v_2\wedge v_1)=(1,1,1)$$
(see also \symme). From this
it follows that
\eqn\symbf{
C_{R_1,R_2,R_3}= (-1)^{\sum_i \ell_i}\;
C^{-1,-1,-1}_{R_1^t,R_3^t,R_2^t}.}
Since $\kappa_{R^{t}}=-\kappa_R$, we can write this as
\eqn\symb{C_{R_1,R_2,R_3}=
q^{\sum_{i} \kappa_{R_i}/2} C_{R_1^t,R_3^t,R_2^t}.}

\subsec{Gluing the vertices}

In this section, we discuss how to glue open string amplitudes to
obtain closed string amplitudes.
Consider a leg of some graph $\Gamma$, as in the figure 3. The
leg will contribute to closed string amplitudes via holomorphic
curves that map to the corresponding ${\bf P^1}$. By cutting the
curve in the middle of the leg, we obtain a product of open string
amplitudes. Clearly, connected closed string graphs can give open
string graphs that are disconnected, so the gluing must be done at
the level of the partition function, schematically,
$$Z(\Gamma) \sim  Z(\Gamma_L)\times Z(\Gamma_R),$$
where $Z(\Gamma)$ is the amplitude corresponding to the graph $\Gamma$
and, by cutting one of its legs, the graph can be decomposed
into $\Gamma_L$ and $\Gamma_R$.

Moreover, clearly the open string Riemann surfaces one gets in this
way have matching number of holes on the leg over which we glue,
and also the winding numbers. Thus, the
right hand side of the above equation is in fact
$$\sum_{\vec{k}}
Z(\Gamma_L)_{\vec{k}}\,{{\exp(- \ell(\vec{k}) t)}\over{\prod_j k_j!
j^{k_j}}}\,  Z(\Gamma_R)_{\vec{k}}.$$
 Above,
$t$ is the size of the relevant ${\bf P^1}$. In gluing
these we have to be careful that both gluing branes are defined
with respect to the {\it same} framing.
The
combinatorial factor comes about because all holes with the same
winding number are indistinguishable, and the factor of $j$ for
each hole of winding number $j$ comes as the gluing respects the
cyclic ordering of the $j$ windings. In addition, we must remember
that ${X_L}$ and $X_R$ (the manifolds corresponding to the graphs
 $\Gamma_{L,R}$) come equipped with a choice of complex
structures, and this induces natural orientation of boundaries of
the two disks in $X_{L,R}$. In order to glue the two disks into a
${\bf P^1}$ their boundaries must be oriented oppositely, which
can be interpreted as putting branes versus anti-branes.
As was already discussed
 this is equivalent to
multiplying the amplitude by $(-1)^h$ where $h$ is
the number of boundaries of the Riemann surface. This gives the gluing a
nice physical interpretation: we put N D-branes on the
relevant leg in $X_L$ and $N$ anti D-branes in $X_R$. The D-branes
annihilate, so from the corresponding open string amplitudes we
obtain the amplitude for closed strings on $X$. To summarize,
\eqn\glue{Z(\Gamma) =  \sum_{\vec{k}}\;
Z(\Gamma_L)_{\vec{k}}\;{(-1)^{|\vec k|} e^{- \ell(\vec{k}) t}
\over z_{\vec k}}Z(\Gamma_R)_{\vec{k}}}
Obviously, \glue\ holds even in the
presence of D-branes in $X$, where $Z(X),$ etc. refer to amplitudes
with D-branes. At the very least, this is true, as long as the D-branes
are at locations away from the relevant leg, as all the considerations that
led to \glue\ are purely local. We will return to this in the sections below.

Note that in the representation basis the gluing
operation is simply:
\eqn\gluez{Z(X) = \sum_Q Z(X_L)_Q (-1)^{\ell_Q} e^{- \ell(Q)t}Z(X_R)_{Q^t},}
which follows from \tran\ and orthonormality of the characters
$$
\sum_{\vec k} {1 \over z_{\vec k}}\chi_R(C(\vec k)) \chi_{R'}(C(\vec k))=\delta_{RR'}.
$$

\subsec{The gluing algorithm for closed and open strings}

Putting together all we have said so far, we can summarize the rules for
computing closed string amplitudes from the cubic vertex and the
gluing rules as follows: \vskip 0.7 cm i) From the toric data
described in the section 2, we can find the graph $\Gamma$
corresponding to the loci where $T^2$ fibration degenerates. The
edges of the graph are labeled by integral vectors $v_i$ that
encode which cycle of the $T^2$ fiber degenerates over the $i$-th
edge. To each edge associate a representation $R_i$. \vskip 0.7 cm
ii) For smooth Calabi-Yau, the graph can be partitioned to
trivalent vertices and corresponding ${\bf C}^3$ patches $U_a$,
where $a$ labels the vertices $a=1,2,\ldots$ \vskip 0.7 cm iii)
This associates to every vertex an ordered triplet of vectors
$(v_{i},v_j,v_k)$ by reading off the three edges that meet at the
vertex in a counter-clockwise cyclic order --  $(v_{i},v_j,v_k)$
is equivalent to $(v_{j},v_k,v_i)$. \vskip 0.7 cm iv) If all the
edges are incoming, we associate a factor $C_{R_i, R_j, R_k}$ to
the vertex $U_a$, otherwise we replace the corresponding
representation by its transpose times $(-1)^{\ell(R)}$. \vskip 0.7
cm v) Let the  vertex $U_a$ share the $i$-th edge with the vertex
$U_b$ whose corresponding triple is $(v_i, v_j', v_k')$. We can
assume $v_i$ is outgoing at $U_a$ and ingoing at $U_b$. We glue
the amplitudes by summing over the representations on the $i$-th
edge as:
\eqn\gluecs{
 \sum_{R_i} C_{R_jR_k R_i} e^{- \ell(R_i) t_i}(-1)^{(n_i+1) \ell(R_i)}
q^{-n_i \kappa_{R_i}/2}
  C_{R_i^t R_j'R_k'}}
where the integer $n_i$ is defined as
$$n_i = |{v_k}' \wedge v_{k}|$$
and $|{v_k}' \wedge v_{k}|$ equals ${v_k}' \wedge v_{k} $
if both $v_k$ and $v_k'$ are
in(out)going, and $-{v_k}' \wedge v_{k}$, otherwise.
This factor reflects the fact that the framing over
the $i$-th edge should be the same on the two side of gluings. The sign
$(-1)^{(n_i +1)\ell(R_i)}$ in \gluecs\ comes from the
sign associated to the framing, and the one associated to the
gluing in \gluez.
\vskip 0.7 cm
vi) From the D-term equations \Dt , or the
$(p,q)$ 5-brane diagrams read off the lengths $t_i$
of the edges in terms of the K\"ahler moduli $t^A$ of $X$, $t_i=t_i(t^A)$.
%
Note that the edges of the graph $\Gamma$ are straight lines on
the plane, with rational slope.  To the $i$-th edge in the
$(p_i,q_i)$ direction of length $x_i$ in the plane, we associate a
K\"ahler parameter $t_i= x_i/\sqrt{p^2+q^2}$.
\vskip 0.7 cm
vii) For a non-compact edge of the graph $\Gamma$ the
corresponding representation $R$ is necessarily trivial, $R=0$ (we will
sometimes denote this also by $R=\cdot$).
\vskip 0.7 cm

Naturally, the vertex can be used for calculating open string
amplitudes on toric Calabi-Yau manifolds as well as the closed
string ones. When we place the D-branes on the non-compact, outer
edges of the graph $\Gamma$, we simply modify the (vii) above to a
sum over arbitrary representations $R$ on the edge, where we
weight the representations by ${\Tr}_R V$ and $V$ is the holonomy
on the corresponding D-brane. For D-branes on the inner edges,
when we glue the maps on $X$ from the maps on $X_L$ and $X_R$, we
must allow for maps with the boundaries on the D-brane. Suppose
that the we wish to calculate an amplitude corresponding to a
D-brane on the $i-$th edge of the toric graph. This modifies the
gluing rule in (v) above as follows.
\vskip 0.3 cm
v') For a single D-brane on the i-th edge, which is shared by vertices
$U_a$ and $U_b$ in the setup of (v),
we glue the amplitudes by summing over the representations on the
$i$-th edge and representations $Q^L_i, Q^R_i$ which ``stop'' on the D-brane
from left and right as:
\eqn\oneinner{
 \sum_{R_i, Q^L_i,Q^R_i} C_{R_j,R_k ,R_i\otimes Q^L_i}
(-1)^{s(i)} q^{f(i)}
e^{-L(i)}
C_{R_i^t\otimes Q^R_i ,R_j',R_k'}\; {\rm Tr}_{Q^{L}_i} V_i\; {\rm Tr}_{Q^R_{i}} V^{-1}_i }
where we have collected the length, framing and sign factors in
functionals $L(i)$, $f(i)$ and $s(i)$ on this leg:
$$L(i) = \ell(R_i)\; t_i + \ell(Q^{L}_i)\; r_i + \ell(Q^{R}_i)\;(t_i-r_i)$$
$$f(i)= p_i\; \kappa_{R_i \otimes Q^{L}_{i}}/2+
(n+ p_i)\;\kappa_{R_i^t\otimes Q^{R}_{i} }/2.$$
$$s(i)= \ell(R_i) +p_i \;\ell(R_i \otimes Q^{L}_{i}) +
(n+ p_i)\; \ell(R_i^t\otimes
Q^{R}_{i}).$$
The piece of the edge to the left of the brane has length $r_i$,
while the right-hand side of the edge has length $t_i-r_i$. $V_i$
is the holonomy on the D-brane. Note that $e^{-r}$ naturally
complexifies $V_i$: changing the holonomy by $V_i \rightarrow
V_i\; e^{i\theta}$ changes ${\rm Tr}_{R} V_i$ to $e^{i \ell(r)
\theta} {\rm Tr}_R V$. The appearance of both ${\rm Tr}_R V_i$ and
${\rm Tr}_{R} V_i^{-1}$ reflects the fact that, along with open
string instantons of area $r$ and charge $+1$ ending on the
D-brane, there are those of area $t-r$ and charge $-1$, where the
``charge'' refers to how their boundaries couple to the holonomy
on the D-brane world volume. The integer $n_i$ is defined as in
(v) and the choice of an integer $p$ corresponds to a choice of
framing. This way of incorporating framing is natural. Namely,
while for the closed string amplitudes only the relative framing
in the left vs. right patch matters, corresponding to $n_i =
|v_{k}' \wedge v_{k}|$, for the open string the absolute choice of
framing matters: we pick a vector $f_i$, which frames the $i$'t
leg both for the left and the right patch so that $f_i \wedge v_i
=1$. This corresponds to the choice of coordinate on the D-brane
which does affect the open string amplitude. Then the framing of
the left and the right vertex are
$$p_i = |f_i \wedge v_{k}|,\quad\quad n_i+p_i= |f_i \wedge v_{k}'|,$$
where $|f_i \wedge v_{k}|$ is $f_i \wedge v_{k}$ if  $v_k$ and
$v_i$ are both in(out) going in the vertex $U_a$, and equals $-f_i
\wedge v_k$ otherwise,  and similarly with $|f_i \wedge v_k'|$.

If there is more than one stack of D-branes on the edge, say $n$
stacks of them, we also must include contributions of $n (n-1)/2$
massive open strings stretching between the D-branes. As shown in
\ov\ the effect integrating out these strings is
\eqn\annulus{\exp(- \sum_{m=1}^{\infty} \frac{1}{m}
 \;{tr}U_1^m {tr}\; U_2^{m})=\sum_R
(-1)^{\ell_R}\; {\rm Tr}_R U_1\; {\rm Tr}_{R^{t}}U_2.}
The relative minus sign
in the exponent in \annulus\
relative to that of \ov\ arises as follows. In
problem studied in \ov\ one had
two D-branes intersecting on ${\bf S}^1$ (the ${\bf S^1}$ corresponds to the
${\bf S^1}$ factor in the D-brane world-volumes which are
$L={\bf S^1 \times C}$)
and the ground state was a boson.
Here we have two D-branes
whose world-volumes are parallel. There is one normalizable mode of the
stretched string supported along an ${\bf S^1}$, and it
turns out to be a fermion.
One way to see this is that, by changing complex structure which does not
affect the A-model amplitudes, we can bring the branes to intersect on an
${\bf S^1}$ at the expense of turning one D-brane into an anti-D brane. The
ground state of the string stretching
between them is a fermion, as argued in \antibr .

For example, for
stacks of $m$ D-branes, we have
\eqn\glueopent{
\sum_{{R_i,Q^L_{a,i},Q^R_{a,i}}}
C_{R_j,R_k, R_i\otimes_{a=1}^{m} Q^L_{i,a}} (-1)^{s(i)}
e^{-L(i)} q^{f(i)}
C_{R_i^t\otimes_{a=1}^{m} Q^R_{i,a}, R_j',R_k'}
\prod_{a=1}^{m} {\rm Tr}_{Q^{L}_{i,a}} V_a \;{\rm Tr}_{Q^{R}_{i,a}} V_a^{-1}}
where
$$L(i) = t_i \ell(R_i) + \sum_{a=1}^{m}
r_a \;\ell(Q^{L}_{i,a})+(t_i - r_a) \; \ell(Q^{L}_{i,a}).$$
$$f(i)=
\sum_{a=1}^{m} \Bigl( p_i\; \kappa_{R_i \otimes Q^{L}_{i,a}}/2 +
(p_i+n_i) \; \kappa_{R_i^t \otimes
Q^{R}_{i,a}}/2\Bigr).$$
$$s(i)= \ell(R_i) + \sum_{a=1}^{m}
\Bigl(p_i \; \ell(R_i \otimes Q^{L}_{i,a}) +(p_i + n_i) \; \ell(R_i^t\otimes
Q^{R}_{i,a})\Bigr). $$

\newsec{Chiral Bosonic Oscillator and the Vertex}

We have seen that the partition functions of the A-model on local
toric 3-folds are computable from a set of gluing rules involving
a cubic vertex and the propagator. The gluing rules are
reminiscent of the construction of the partition function of
bosons on a Riemann surface from the ``pant diagram'' and the tube
propagators.  We will show that this is not accidental.  In fact,
as we will argue, the vertex operator and the propagator we have
obtained can be viewed as construction of the partition function
of the mirror B-model whose geometry, as is well known, is
captured by a Riemann surface. Towards this aim in this section we
reformulate the vertex and the propagator we have obtained in
terms of a free chiral boson on a Riemann surface.  In particular,
we will show that the winding basis can be identified with the
Fock space of the chiral boson.  In this connection the sewing
rule gets mapped identically to the propagator of the chiral
boson.  Moreover the vertex gets identified with a state in the
triple tensor product of the Hilbert space of the free boson. This
vertex is highly non-trivial.  Even in the classical limit it is
more complicated than the usual vertex states one gets for a free
boson on a Riemann surface (which is always given by a Bogoliubov
transformation and can be represented as exponential of quadratic
monomials in the oscillator creation operators).

In the next section
we  explain how to interpret the free chiral boson
as the relevant field for the Kodaira-Spencer theory
of gravity \BCOV\ in this local context (related ideas have appeared in
\refs{\nikitaa, \nikitab}).

\subsec{Reformulation in terms of a chiral boson}

There is a curious similarity between the winding number ${\vec k}$ basis
and oscillator states of a free chiral boson.
Recall the oscillator expansion of the chiral boson $\phi(u)$,
$$\partial_u\phi(u) = \sum_{m\neq 0 }  j_{m} e^{m u },$$
where
$$ [j_{m} , j_{n}] = m \delta_{m+n,0}$$
so that $j_{m>0}$ is the annihilation operator and $j_{m<0}$ the
creation operator. The Hilbert space ${\cal H}$ of a free chiral
boson on a circle is spanned by states of the form
$$|\vec{k}\rangle = \prod_{m>0} j_{-m}^{k_m}|0\rangle.$$
We simply identify the vector $\vec{k}$ above with the vector of
winding numbers.  With this identification we can interpret the
sewing as picking out an element $P$ in the two-fold tensor
product of the Hilbert space $P\in {\cal H}^{\otimes 2}$, and the
vertex as defining a state in the threefold tensor product Hilbert
space $C\in {\cal H}^{\otimes 3}$.  It is natural to ask what
these states are.

We first turn to the propagator $P$. We will see that $P$ is the $
conventional$ state of the free chiral boson on a cylinder. The
path integral of a free chiral boson on a cylinder of length $t$
is a state in the tensor product Hilbert space $P \in {\cal
H}_1\otimes {\cal H}_2$ (where ${\cal H}_{1,2}$ are associated to
the two boundaries) given by
$$ |P\rangle = \exp( - t \sum_{m>0} {1\over m}\;j_{-m}^{1}\; j_{m}^{2}\;)|0\rangle_1 \otimes
\langle 0 |_2.$$
Expanding the exponential
we get
$$\sum_{\vec{k}} e^{-\ell(\vec k)t}{(-1)^{h} \over z_{\vec k}}\;|\vec{k}\rangle
\otimes \langle \vec{k}|
$$
where
$$ z_{\vec k}=\prod_m k_m ! m^{k_m}.$$
This is precisely the gluing rule we had discussed
for A-model amplitudes!

We now turn to the vertex $C$. As we said above, the vertex amplitude, as
formulated in section 3,
$$Z = \sum_{{\vec k}_i}C_{\vec{k}_1\vec{k}_2\vec{k}_3} {1 \over
\prod_i z_{{k}_i}}\;
Tr_{{\vec k}_1} V_1 \; Tr_{{\vec k}_2}V_2\; Tr_{{\vec k}_3} V_3,$$
can be written in terms of the state $C$ in ${\cal H}^3$
of the chiral boson as
$$Z = \sum_{{\vec k}_i}
Tr_{\vec k} V_1 ~Tr_{\vec k} V_2 ~Tr_{\vec k} V_3~
{1\over \prod_i z_{k_i}}\;\langle \vec{k}_1  |
 \otimes \langle \vec k_2|
\otimes \langle \vec{k}_3 |C\rangle.$$
Note that, at the level of the answer, $|C\rangle$ is given by
$$|C\rangle = exp{(\sum_{{\vec k}_i} F_{\vec{k}_1,\vec{k}_2,\vec{k}_3}(g_s)
\;j_{-\vec{k}_1}\; j_{-\vec{k}_2}\; j_{-\vec{k}_3})}
|0\rangle_1
 \otimes |0\rangle_2
\otimes |0\rangle_3.$$
where $F$ is identified with the free energy of the topological
string, and $j_{\pm \vec{k}} = \prod_{m>0} j_{\pm m}^{k_m}$. since
``evaluating'' the amplitude, i.e. performing contractions,
amounts to replacing the creation operators in the free energy
with $V's$, $tr V^{m} \leftrightarrow j_{-m}.$
It is natural to ask what the meaning of the three point vertex $C\in
{\cal H}^{\otimes 3}$ is.  One may at first think that this
may be related to that of a free boson on trice punctured
sphere.  This is almost true.  Namely it is a state
associated with a sphere with three punctures, but the theory
is that of a free scalar theory only to leading order in oscillator
expansion and in the $g_s\rightarrow 0$ limit.
We will discuss this and its interpretation after we
discuss the B-model interpretation of the chiral boson as describing
the quantum field of the Kodaira-Spencer theory on the Riemann surface.
We will explain
why there are more oscillator terms in $C$, including the existence
of non-trivial $g_s$ corrections.  A full study of the vertex
$C$ from the B-model perspective will be done in \ref\adkmv{Work in progress
with Robbert Dijkgraaf.} .

For now, note that,
most remarkably, the D-branes can be thought of as coherent states in the
chiral boson theory! The vertex amplitude $Z$ is computed
by inserting
$$\langle V| = \langle 0| \; exp( \sum_{m>0} \frac{1}{m}\; tr V^m \;j_m)$$
at each of the three punctures that give rise to the vertex state $C$,
$$Z = \langle V_1| \otimes \langle V_2| \otimes \langle V_3| \; C \rangle.$$
We will explain this below.

\newsec{Local B-model mirror and the quantum Kodaira-Spencer theory}
We have seen that the vertex is naturally captured by the
states of a chiral boson on a sphere with three punctures.  In this
section we will identify the chiral boson on each patch
as the quantum field of the Kodaira-Spencer theory on
the mirror B-model involving a Riemann surface.
 The modes of
the chiral boson are affected, as we will explain,
by the degrees of freedom  ($ tr V^n$) on the B-branes.  In particular
we will explain, from the B-model perspective, why the
open string amplitudes know about the closed string B-model
amplitudes.  Moreover we identify the branes in this setup
as the fermions associated to the chiral bosons $\psi(z)=e^{\phi(z)}$.
We use this picture to compute leading
terms in the oscillator expansion of the vertex.  Extension
to the full vertex, from this perspective will appear elsewhere
\adkmv .
Moreover the gluing rules of the vertex can now be directly
interpreted as computations of the Kodaira-Spencer theory
in the operator formulation on the mirror Riemann surface.

The target space of B-model was interpreted in \BCOV\ as describing
the quantum theory of complex deformation of the Calabi-Yau threefold.
This in particular applies to the local Calabi-Yau case at hand.  In
particular, if we consider the A-model in the local toric case, as already
discussed in section 2, the mirror is given by a hypersurface in
$(x,\tilde x ,u ,v) \in{\bf C}\times {\bf C}\times {\bf C}^*
\times {\bf C}^*$:
$$x \tilde x =F(u,v)$$
Moreover $F(u,v)$ can be obtained by gluing pant diagrams of the form
\eqn\pantr{e^u+e^v+1=0.}
The holomorphic 3-form is given by
$$\Omega =dxdudv/x.$$
As is well known in the local context,  integration of $\Omega$
over the non-trivial class of three cycles gets reduced to
computation of a 1-form on the Riemann surface.  The only
non-trivial 3-cycles are formed by the ${\bf S^1}$ fibration (identified
with $(\tilde x, x)\rightarrow (e^{i\theta} \tilde x ,e^{-i\theta}x )$)
 over a domain in the $u,v$ plane bounded by the
Riemann surface $(\Sigma:\ F(u,v)=0)$.  The integral
of $\Omega$ over such cycles reduces
to integrals of the meromorphic 1-form
$$\lambda=udv$$
 on the 1-cycles of the non-compact
Riemann surface
$\Sigma$.
 Note that $\lambda$
is not globally well defined, and it makes sense only patch
by patch.  In particular if we had considered the $u$-patch,
integration of the 2-cycle fiber would have resulted in a 1-form
$-vdu$.  It is this lack of global definition of $\lambda$
which lead to non-trivial interactions, to an otherwise free theory.

The variations of the complex structure
and the corresponding period integral get mapped to
the variation of the complex structure of $\Sigma$ and the periods
of the
corresponding reduced 1-form $\lambda$ on it.
Note that there is a direct relation between
changing the complex structure of $\Sigma$ and the choice
of $\lambda$.
In particular suppose we are in the $v$-patch defined
by being centered at $e^v=0$;
Consider the complex deformation
$$F\rightarrow F(u,v)+\delta F(u,v).$$
The above derivation for the reduced one form will
still go through without any change, 
$$\lambda=udv$$
but now $u$ is a different function of $v$.

  Solving $F(u,v)=0$
we would in principle get $u=f(v)$ and under the complex deformation
we have
$$u=f(v)+\delta f$$
and so the change in the 1-form $\lambda $ is given by
$$\delta \lambda = \delta f\; dv.$$
Thus, as in \BCOV\ the basic quantum field
gets identified with this variation.
For this to be a good deformation of complex structure
$\delta f$ should be a meromorphic function of $e^v$, i.e.
\eqn\ks{\overline \partial_v \;\delta f=0.}
To get an ordinary quantum field it is natural to write $\delta f
=\partial \phi_v$ in which case \ks\ gets mapped to
\eqn\fsc{\overline \partial \partial \phi_v=0.}
In terms of this scalar the variation of the 1-form is given by
\eqn\onfb{\delta \lambda =\partial \phi_v.}
and the equation \fsc\ is sufficient for the condition of
integrability of the complex structure, unlike
the generic 3-fold complex structure deformation
where the story is more complicated.  Thus we have a free boson
propagating on each patch.

In the classical theory we can of course parameterize the
deformation in any way we want, however for the quantum theory
writing the variation this way is more natural.  This is because
the Kodaira-Spencer theory in the formulation of \BCOV\ has the
kinetic term of the form
$${1\over g_s^2}\int_{Calabi \ Yau}\omega \partial^{-1} \overline
\partial \omega$$
where $\omega$ is a $(2,1)$-form representing the change in the complex
structure of Calabi-Yau.  It is natural to write, at least for a patch,
$\omega =\partial \chi$ where $\chi$ is a $(1,1)$ form.  In terms
of $\chi$ the action would become
$${1\over g_s^2}\int_{Calabi\ Yau} \partial \chi {\overline \partial}
\chi .$$
  In the local context that we are discussing, $\chi$ gets identified
with the $\phi_v$ above which is a scalar on $\Sigma$, and we get the
free scalar theory
$${1\over g_s^2}\int_{\Sigma} \partial \phi {\overline \partial}
\phi .$$
Note that the anti-holomorphic piece of $\phi$ is  a gauge
artifact:  Shifting $\phi$ by an anti-holomorphic function will
not affect $\partial \phi$ and so does not change the 1-form
$\lambda$.  So $\phi$ should be viewed as a {\it chiral} boson. We
can study the Kodaira-Spencer theory patch by patch by chiral
fields $\phi_v$.  We will write the variation of the complex
structure as
$$u=f_0(v) +\partial_v \;\phi_v$$
We can also absorb $f_0(v)$ as a classical vev for $\partial_v
\phi_v$, which we will sometimes do. In the $v-$patch which is
cylindrical we can write
\eqn\exf{\partial_v \phi_v=\sum_{n>0}\;j_n\; e^{-nv}+g_s^{2}\;j_{-n}\;e^{nv}}
where we have included factors of $g_s^{2}$ to account for the kinetic
term of the scalar being $1/g_s^2$.
In the quantum formulation $j_{-n}$ and $j_{n}$ are not
independent, and correspond to creation and annihilation operators.
To better understand this we will consider a coherent
set of states given by replacing
$$j_n\rightarrow t_n.$$
This is natural
in this patch, because changing the complex structure at
the infinity of this patch, corresponds to changing $\partial_v\,\phi_v$
at $e^v\rightarrow 0$, and that is determined by the negative
powers $e^{-nv}$ above.  However now the positive powers
of $e^{nv}$ are determined, quantum mechanically.  Let $\langle \Sigma |$
denote the state created by the rest of the Riemann surface.
Let the coherent states be defined by
$$|t\rangle ={\rm exp}(\sum_{n>0}\; {1\over n}\; j_{-n}\;t_n ) |0 \rangle $$
Let us denote the partition function of the theory including the
$t_n$ deformations by $Z(t)$. Then we have
$$Z(t)={\rm exp}(\,F(g_s, t_n))=\langle \Sigma \; |t\rangle $$
To justify constructing the coherent state in terms of $j_{-m}$ alone,
note that if we consider the expectation value
of $\partial_v \phi_v$ on the cylinder at infinity where $e^{-v}\rightarrow
\infty$ then $e^{-nv}$ terms in \exf\ dominate.
This can also be viewed as changing
the 1-form $\lambda$ at infinity by
$$\delta \lambda =dv\;\sum_n t_n e^{-nv}$$
On the other hand if we consider the expectation value of $
\partial_v \phi_v$ for the positive
powers $e^{nv}$ it will not be zero.
It might at first appear that in the classical limit
$g_s\rightarrow 0$ this would be zero because
of the explicit $g_s^2$ dependence in \exf , but this is not the case.
This is because $F(g_s,t_n)$ has a $1/g_s^2$ term in the genus
zero part given by $F_0(t_n)$, so this survives in the limit.  
This will give us for the expectation
value of  $\partial_v\phi_v$ in the limit $g_s\rightarrow 0$
$${1\over Z}\;
\langle \Sigma |\;\partial_v \phi_v\;|t\rangle =\sum_{n>0} t_ne^{-nv}
+n{ \partial F_0 \over \partial t_n} e^{nv}.$$
which means that classically we have
$$\partial_v \phi_v=\sum_{n>0}t_ne^{-nv}+
n{\partial F_0 \over \partial t_n} e^{nv}.$$

Here we need to clarify one important point:  $F$ is
not an unambiguous function of $t_n$'s.  This depends
on how we choose the coordinates on each patch.  Changing
the coordinates, will give rise to a different function $F(t_i)$.
The difference between these results is the same as the
Virasoro action.  This dependence on the choice
of the local coordinates on the Riemann surface will turn out to be related
to the framing ambiguity.  For us the Riemann surface comes
with almost canonical coordinates involving combinations of $u$ and
$v$ with $du\wedge dv$
making sense in the full Calabi-Yau.
 In a given $v$-patch we will have the situation
where $z=e^v\rightarrow 0$, i.e. $v\rightarrow -\infty$ is on the
patch, and $u\rightarrow const.$ as $z\rightarrow 0$.  This almost
uniquely fixes the coordinates except for an integer choice:  The
$du\wedge dv$ is invariant under $SL(2,{\bf Z})$. There is a
subset of $SL(2,{\bf Z})$, indexed by an integer $n$, which
preserve the conditions we have put on each patch, namely $u'= u$
and $v'= v+nu$.  Note that the coefficient of $v$ in $v'$ is 1
because we want $z'=O(z) $ as $z\rightarrow 0$ in order to have a
good coordinate. This transformation will give a new one form
$\lambda =u' dv'$ and a new coordinate
$$z'=e^{v'}=e^{nu} z$$
where $e^u=\sum_{i=0}^{\infty} a_i z^i$.  For example consider
the
pant Riemann surface
$$e^{-u}+e^{v}+1=0$$
in the $v$-patch, which includes $v\rightarrow -\infty$
(where $u\rightarrow i\pi$).  If we now change
coordinates $v'=v+nu$ we will have
$$z'=z[(-1)^n (1+z)^{-n}]$$
So if we compute $F$ in the $z$ patch, in the new coordinate patch
we will need to exponentiate an appropriate element of Virasoro
algebra which changes the coordinates.  Thus the choices of $F$ is
indexed by an integer $n$ in each patch.

In the interest of
comparison with our A-model vertex
we will now specialize to the case of the pant diagram, which
is mirror to ${\bf C^3}$, given
 by the Riemann surface
$$ e^{u_1}+e^{u_2}+e^{u_3}=0$$
where one variable is eliminated by rescaling the equation.
This way of writing it, exhibits the cyclic symmetry between the
three patches. A choice of coordinates that
corresponds to the ``standard framing of the vertex''
discussed before and which preserves the
${\bf Z}_3$ symmetry is as to let
$u=u_3-u_1$ and $v=u_2-u_3$, and
$w=u_1-u_2$ then we have
\eqn\rsc{\eqalign{&e^{-u}+e^{v}+1=0 \quad \lambda=vdu \quad u-{\rm patch}\cr
&e^{-v}+e^{w}+1=0 \quad \lambda=wdv \quad v-{\rm patch}\cr
&e^{-w}+e^{u}+1=0 \quad \lambda=udw \quad w-{\rm patch}}}
Note that we also have the relation
$$u+v+w=0$$

Changing of the coordinates by framings $(n_1,n_2,n_3)$ is obtained by
the choice of the coordinates
$$u\rightarrow u+n_1v$$
$$v\rightarrow v+n_2w$$
$$w\rightarrow w+n_3 u$$
In each patch we can deform the defining equation by a chiral
scalar, as discussed above.  The corresponding scalars we
call $\phi_u,\phi_v,\phi_w$.  The equation of the surface
gets modified, when $\phi_i\not=0$ by following the deformation
discussed in general above, and we get
$$e^{-u}+e^{v+\partial_u \phi_u} +1=0$$
$$e^{-v}+e^{w+\partial_v \phi_v} +1=0$$
$$e^{-w}+e^{u+\partial_w \phi_w} +1=0$$
where in the classical limit
$$\partial_u \phi_u=\sum_{n>0}t^u_ne^{-nu}+n { \partial F_0 \over
 \partial t^u_n} e^{nu}.$$
$$\partial_v \phi_v=\sum_{n>0}t^v_ne^{-nv}+n{ \partial F_0 \over
 \partial t^v_n} e^{nv}.$$
$$\partial_w \phi_w=\sum_{n>0}t^w_ne^{-nw}+n {\partial F_0 \over
 \partial t^w_n} e^{nw}.$$
If we assume $\del \phi$'s are small\foot{This derivation
of the small $\del \phi$ limit of the vertex was suggested
to us by Robbert Dijkgraaf.}, we get the following
equations:
\eqn\rsd{\eqalign{& e^{-u}+e^{v}+1 +e^{v}\partial_u \phi_u=0 \qquad (i)\cr
&e^{-v}+e^{w}+1+ e^{w}\partial_v \phi_v=0 \qquad (ii)\cr
&e^{-w}+e^{u}+1+ e^{u}\partial_w \phi_w=0 \qquad (iii)}}
If we multiply equation $(i)$ by $e^{-v}$, and use the fact
that $u+v=-w$ we get
$$e^{w}+1+e^{-v}+\partial_u \phi_u=0$$
Comparing this with equation $(ii)$ we learn that
$$\partial_u \phi_u =e^{w}\partial_v \phi_v$$
On the other hand, to leading order we
have from equation (i),
$$dv/du =e^{-u-v}=e^{w}$$
we thus have
\eqn\gls{\partial_u \phi_u ={dv\over du} \partial_v \phi_v}
and by the ${\bf Z_3}$ cyclic symmetry similar equations with
$u\rightarrow v\rightarrow w\rightarrow u$.  This in particular
implies that in the classical limit, and to leading order in
oscillators the three $\phi$'s in the three patches can be viewed
as coming from a global $\phi$.  This means that to leading order
in the classical limit and in oscillators, the vertex operator
should be the standard one coming from the well known techniques
of operator formulation on Riemann surfaces
\lref\IshibashiBD{
N.~Ishibashi, Y.~Matsuo and H.~Ooguri,
``Soliton Equations And Free Fermions On Riemann Surfaces,''
Mod.\ Phys.\ Lett.\ A {\bf 2}, 119 (1987).
}
\lref\VafaES{
C.~Vafa,
``Operator Formulation On Riemann Surfaces,''
Phys.\ Lett.\ B {\bf 190}, 47 (1987).
}
\lref\AlvarezGaumeBG{
L.~\'Alvarez-Gaum\'e, C.~G\'omez, G.~W.~Moore and C.~Vafa,
``Strings In The Operator Formalism,''
Nucl.\ Phys.\ B {\bf 303}, 455 (1988).
}
\refs{\IshibashiBD,\VafaES,\AlvarezGaumeBG}\
which lead to Bogoliubov transformations.

Here we will digress to review this derivation.

\subsec{Bogoliubov transformation}

Consider a chiral boson on a Riemann surface mirror to ${\bf
C^3}$. The Riemann surface is a sphere with three punctures, and
the path integral on this gives a state in the tensor product of
three free Hilbert spaces ${\cal H}_1 \otimes {\cal H}_2 \otimes
{\cal H}_3$ corresponding to the punctures. Moreover, mirror
symmetry provides us with a choice of complex structure on the
punctured Riemann surface, and this gives a canonical choice of
coordinates $z_i = e^{u_i}$ near each puncture at $z_i=0$ and
transition functions relating them. The transition functions
between the patches give rise to Bogoliubov transformations that
relate the three Hilbert spaces and these are sufficient to
determine the ray in the Hilbert space to which the path integral
corresponds to. To do so we follow
\VafaES . Note
that the path integral of the chiral boson $\phi$  has infinite
dimensional group of symmetries corresponding to shifting $\phi \rightarrow
\phi + f$  for any function $f$ which is holomorphic on
the punctured Riemann surface (a meromorphic function whose only poles
are at the punctures).
This gives rise to conserved charge
$$Q(f)  = \sum_{i} \oint_{z_i=0} f(z_i) \partial \phi ,$$
which must annihilate the path integral. In each of the three
patches we have a different expansion for the chiral boson in
terms of the local holomorphic coordinate.
$$ \phi (z_i) = \sum_{m\neq 0} \frac{1}{m} j^{(i)}_{-m} z^m_i,$$
The three patches are related by \rsc , where we put $z_1=e^{u}$,
$z_2=e^{v}$ and $z_3 =e^w$, so that the three patches are related
by
$$z_{i+1}  + {1 \over z_{i}} +1 =0.$$
Then, for example, a meromorphic function $f^{(1)}_{m} = z_1^{m}$ has
expansion
$$
z_1^{m} =
\sum_{n\geq m} {m \over n} O_{n,m} z_2^{-n}=
\sum_{m\geq n} O_{m,n} z_3^{-n}$$
where
$$ O_{m,n} =  (-1)^m {m \choose n} $$
The corresponding charge $Q(f)$ is given by
$$Q(f^{(1)}_m) =
j^{(1)}_m
+ \sum_{n\geq m} O_{m,n} j^{(2)}_{-n}
+ \sum_{n\geq m} {m \over n}O_{n,m} j^{(3)}_{-n},$$
The conditions that $Q(f^{(i)}_m)$ annihilate the path integral
suffice to determine it:
\eqn\twop{ |Z\rangle = exp \Bigl( \sum_{m>0, n\geq m} {O_{n,m}\over n} \;[
j^{(1)}_{-m}\; j^{(2)}_{-n}+
 j^{(2)}_{-m}j^{(3)}_{-n}+ j^{(3)}_{-m}j^{(1)}_{-n}\;]\;\Bigr)
|0\rangle_1
 \otimes |0\rangle_2
\otimes |0\rangle_3.}
Here
we have suppressed the linear term in the $j_{-m}$'s corresponding to the
fact that the vacuum $|0\rangle$ is not the ordinary vacuum,
but $\del \phi$ has a piece corresponding to the classical geometry,
which
we have been shifting away so far.
Let us now restore it. The classical piece of the
chiral boson in the u-patch, for example, is
$\del \phi(u)_0 = v(u) du = log(1+e^{-u})du$.
Shifting this away in the action ${1\over g_s^2}\int \del \phi \overline \del \phi$
gives a surface term
\eqn\onep{{1\over g_s}\oint \phi_0 \del \phi = -\sum_{n>0} {(-1)^{n}\over n^2}~j^{(1)}_{-n},}
(after rescaling $\phi \rightarrow g_s \phi$),
so this shifts the vacuum $|0\rangle_1$ to
$$|0\rangle_1 \rightarrow exp(-\sum_{n>0} ~{(-1)^{-n}\over n^2 g_s}~j^{(1)}_{-n}\;)|0\rangle_1,$$
and similarly in the other two patches.

The state we have computed should be accurate to leading order in
$g_s$ and up to quadratic terms in $j_{-m}=t_m$. The full vertex
will have additional terms both in the $g_s$ corrections as well
as in terms involving more $t_m$'s. This is because the derivation
leading to a global chiral boson was valid only in this limit.
 A full discussion
of this from the B-model perspective will be presented
elsewhere \adkmv .

{}From the B-model perspective if we know what the
path-integral gives for the pant-diagram, then we can
obtain any other amplitude by gluing.  This is because
the Kodaira-Spencer equation implies that in each patch
$\phi$ is a chiral boson with the standard propagator.
However we have to make sure that in the gluing the coordinate
choices match--this is the same as making sure that the
framings are compatible in the A-model computation of the
vertex.
Independently of how one computes this trivalent vertex, the
knowledge of
\eqn\pantpant{F=\sum_g F_g(t^u_n,t^v_n,t^w_n)(g_s)^{2g-2}}
will capture arbitrary B-model local amplitudes, as everything
can be obtained from this by gluing, as discussed in section 2.
We thus appear to have a system involving closed strings on
the pant-diagram, in the operator formalism, capturing
arbitrary local models.  On the other hand we have given,
motivated from the A-model considerations a similar gluing
rule and a vertex $C$ involving amplitudes of the A-branes
on ${\bf C}^3$, which is mirror to the pant diagram. To complete
the circle of ideas we have to connect these two facts.  The mirror
of A-branes are  B-branes, which on the Riemann surface
get mapped to points on the pant diagram \AV .  So the question,
posed in purely B-model context, is the following:  What
is the relation of B-branes with closed string Kodaira-Spencer
amplitudes?   We will explain how this works in
the next section.

\subsec{ B-branes and closed string B-model }
We will now argue how $F$ defined in \pantpant\ in terms
of operator formulation of closed string target theory, i.e.
the quantum Kodaira-Spencer theory, can be equivalently phrased
in terms of open string B-model amplitudes.  To this end
we will have to understand the effect of B-brane on closed
string B-model.  We will show
that the insertion of a B-brane at the point $z$ is equivalent,
for the reduced Kodaira-Spencer theory, to the insertion of the
fermionic field
$\psi (z) =e^\phi$.  In other words, fermions, which are
usually viewed as soliton of a chiral scalar can also be viewed
as the B-branes, i.e., the soliton of the Kodaira-Spencer theory.

Consider B-model in the local patch given
by $v$ where $e^{v}\rightarrow
0$ is part of the patch.
Consider placing  many branes at the circumference
of the cylinder given by $v=v_i\ll 0$.  We ask how putting
B-branes back reacts on the gravity B-model?  In other words,
how does the complex structure get modified by the B-branes?
To answer this question
we consider adding an extra brane, viewing it as a probe,
and place it at $v$ in this patch.  We are interested
in what the effect of many branes are on the probe.  If the
probe is placed at $v\ll 0$ then the effect is simply given
by summing over all the open strings stretching between them.
This is the mirror of the computation of \ov\ and was discussed
in the mirror setup in
\lref\AganagicWV{
M.~Aganagic, A.~Klemm, M.~Mari\~no and C.~Vafa,
``Matrix model as a mirror of Chern-Simons theory,''
arXiv:hep-th/0211098.
}
\AganagicWV.  The effect
on the free energy of the probe at $v$ upon
integrating out the stretched strings between branes at $v_i$
and the probe is
\eqn\namit{\delta F=\partial \phi =\sum_{i,n>0} {1\over n}e^{-n(v-v_i)}=\sum_{n>0}
{1\over n} tr V^n e^{-nv}}
where we have used the fact that $e^{v_i}$ are the eigenvalues
of $V$.
Note that for $v\ll 0$ there is no contribution from the rest of the
surface to the free energy of the probe.
On the other hand we can ask which deformed geometry will give
rise to this free energy probed by the brane.  This is given
by the computation in \AV\ where
\eqn\fenq{g_sF=\int^v \lambda=\int^v \partial_v \phi =\phi(v) +const.}
where $\lambda$ is the deformed 1-form on the surface and we
have absorbed back into $\phi$ the classical piece of the 1-form.
Thus placing a brane at $v$ affects the Kodaira-Spencer action
by the addition of $\phi(v)$.
Let us consider how $F$ changes for $v\ll 0$.  Let
us call this singular part of $\phi$ by $\phi^-$. In particular
the deformed one form $\lambda^-$ which dominates for $
v\ll 0$ gets identified with
$$\phi^-(v) = g_sF$$
We thus see that the free energy felt by the probe brane \namit\ is
reproduced by the deformation
$$\phi^-(v) =\sum_{n>0}{1\over n} tr V^n e^{-nv}$$
which leads to the identification
$$t_n=tr V^n.$$
 This explains the observation we made
before about the role of the chiral boson in the vertex
we had obtained from A-model considerations.

This suggests the following interpretation:  To have a brane at
the point $v$ we add to the KS action the operator $\phi(v)/g_s$,
or to the path integral the operator
$${\rm exp}(\phi (v)/g_s)$$
(the $1/g_s$ there is to remind us that the disk amplitude
is proportional to $1/g_s$).  This leads to the same
response in the free energy.
 We can redefine
$\phi /g_s\rightarrow \phi$, which gets rid of the $1/g_s^2$
in the kinetic term.  We are thus led to identify
the operator inserting the brane at $z$ with the insertion
in the path-integral of
$$\psi(z)=e^{\phi (z)}$$
i.e. the fermion operator! (we have absorbed an $i$ in the definition
of $\phi$ in comparison with the conventional description of
bosonization).  Connection between fermions and
the D-branes was anticipated a while back \ref\shen{S. Shenker,
private communication, 1995.}\ and this result makes this
concrete.
The anti-branes get identified with
$$\psi^{\dagger}(z)=e^{-\phi (z)}.$$
This is because the free energy will change sign for an
anti-brane.  As a check of this statement, note that the coherent
state involving the branes at $v_i$ can be viewed as the state
$$\prod_i{\rm exp}(\phi (v_i))|0\rangle ={\rm exp}(\sum_{i,n>0}{1\over n}e^{nv_i}j_{-n})|0\rangle
=|t_i\rangle$$
with $t_n=\sum_i exp(nv_i)$.  Note that in this expression
we have normal ordered the operator.  Not normal ordering it would
have also led to the effect of the branes on each other; i.e. the result
of integrating out the open string stretched between them.
We will now present an additional argument why
the fermion field is the B-brane operator.

Consider a general Calabi-Yau threefold.  Consider wrapping $N$
 B-branes over a (compact or non-compact)
holomorphic curve $C$ in Calabi-Yau.  This curve is of real
codimension 4 in the Calabi-Yau and is surrounded by a 3-cycle
$F_C$. Consider the integral
$$I_N=\int_{F_C } \Omega_N$$
where $\Omega_N$ is the holomorphic 3-form, corrected
by the fact that there are $N$ B-branes on $C$.
Then we claim
$$I_N-I_0=Ng_s$$
This is related to the mirror of the Chern-Simons/topological
strings duality of \gv.
For example consider $N$ B-branes wrapping the ${\bf P}^1$ in
${\rm O(-1)+O(-1)}\rightarrow {\bf P}^1$.  Then there is an
${\bf S^3}$ surrounding ${\bf P}^1$ and the claim is that
$$\int_{S^3}\Omega_N =Ng_s$$
In particular under the large $N$ duality this is consistent
with the size of the ${\bf S_3}$ being given by $Ng_s$.   This
statement should hold for compact or non-compact branes as it
is a local question and it is our definition of the B-brane
in terms of its coupling to the gravitational Kodaira-Spencer
theory.

In the context of the local model we are considering here, the
B-brane wraps over a non-compact plane, which intersects
the Riemann surface at a point.  The integral of the holomorphic
three-form around it, reduces to the integral of the 1-form
$\lambda$ around this intersection point.  So if
we denote
by $B(z)$ the field creating a D-brane at
point $z$ on the Riemann surface, then

$$\oint_z \lambda(z') \ B(z)=g_s \oint_z \partial_{z'} \phi(z')
 B(z)dz' = g_s.$$
This is indeed the correct OPE defining the fermionic field
and we get the identification
$$B(z)=\psi(z)=e^{\phi(z)}$$
Thus the trivalent topological vertex can be viewed as computing
$$\langle \prod_i \psi(u_i) \prod_j \psi(v_j)
\prod_k \psi (w_k)\rangle$$
in the closed string mirror formulation of the problem.
It is then clear why the amplitudes involving brane
can lead to a full reconstruction of the closed string amplitudes:
This is simply the familiar bosonization! We can also consider
mixed amplitudes with some branes left over in a closed string
background.  In the B-model all we need to do is to add certain
fermion operators.  Note that the framing ambiguity for the B-branes
gets mapped to the fact the $\psi(z)$ is a half-differential
and so the amplitudes will depend on the coordinates chosen.

\newsec{The derivation of the vertex amplitude}
In this section we will provide a derivation of the three-point
vertex in the A-model.  It turns out to be more convenient to switch
back to the representation basis in deriving the cubic vertex
$C_{R_1, R_2, R_3} $ .  We will derive this vertex using the large
$N$ topological duality \gv\ relating
large $N$ Chern-Simons amplitudes with those of closed topological
strings.  We will be able to compute
the {\it full vertex} to all orders in $g_s$ and for arbitrary oscillator
numbers.  This will allow us to confirm, in section 8,
that this agrees, in the $g_s\rightarrow 0$ limit, with the linear
and quadratic pieces of oscillators
which we have computed from the perspective
of the B-model in the previous section.

As was conjectured in \gv\ and proven in \proof , the topological open
string A-model of $N$ D-branes on ${\bf S}^3$ in
$Y= T^*{\bf S}^3$ is the same as
the topological A-model closed string theory on $X = {\cal O}(-1)\oplus
{\cal O}(-1)\rightarrow {\bf P^1}$. The large $N$ duality is a geometric
transition where the ${\bf S}^3$ and the D-branes disappear and get
replaced by the ${\bf P^1}$ \gv . The string coupling constant is
the same in both theories, and the size $t$ of the ${\bf P^1}$ is
identified with the 't Hooft coupling $t=N g_s$. The open string
theory, as was shown in \wittcs, is the same as $U(N)$ Chern-Simons
theory on ${\bf S}^3$ where the level $k$ of the Chern-Simons is related
to the string coupling as $g_s={2 \pi i \over k+N}$. Various aspects of
this duality have been studied in \refs{\ov ,\lm ,\lmv ,\rama},
in particular in \ov\
the duality was studied in the presence of non-compact D-branes.
A variant of this is what we need here.
\ifig\CS{The figure on the left corresponds to Chern-Simons theory
on ${\bf S}^3$ with three source D-branes. $Q_{1,2,3},Q$ denote
the bifundamental strings. In the large $N$ limit, the ${\bf S}^3$
undergoes a geometric transition. The figure on the right depicts
the large N dual geometry, with Lagrangians $L_{1,2,3}$, after the
transition. The local patch where the D-branes are is a ${\bf
C^3}$. } {\epsfxsize 5.0truein\epsfbox{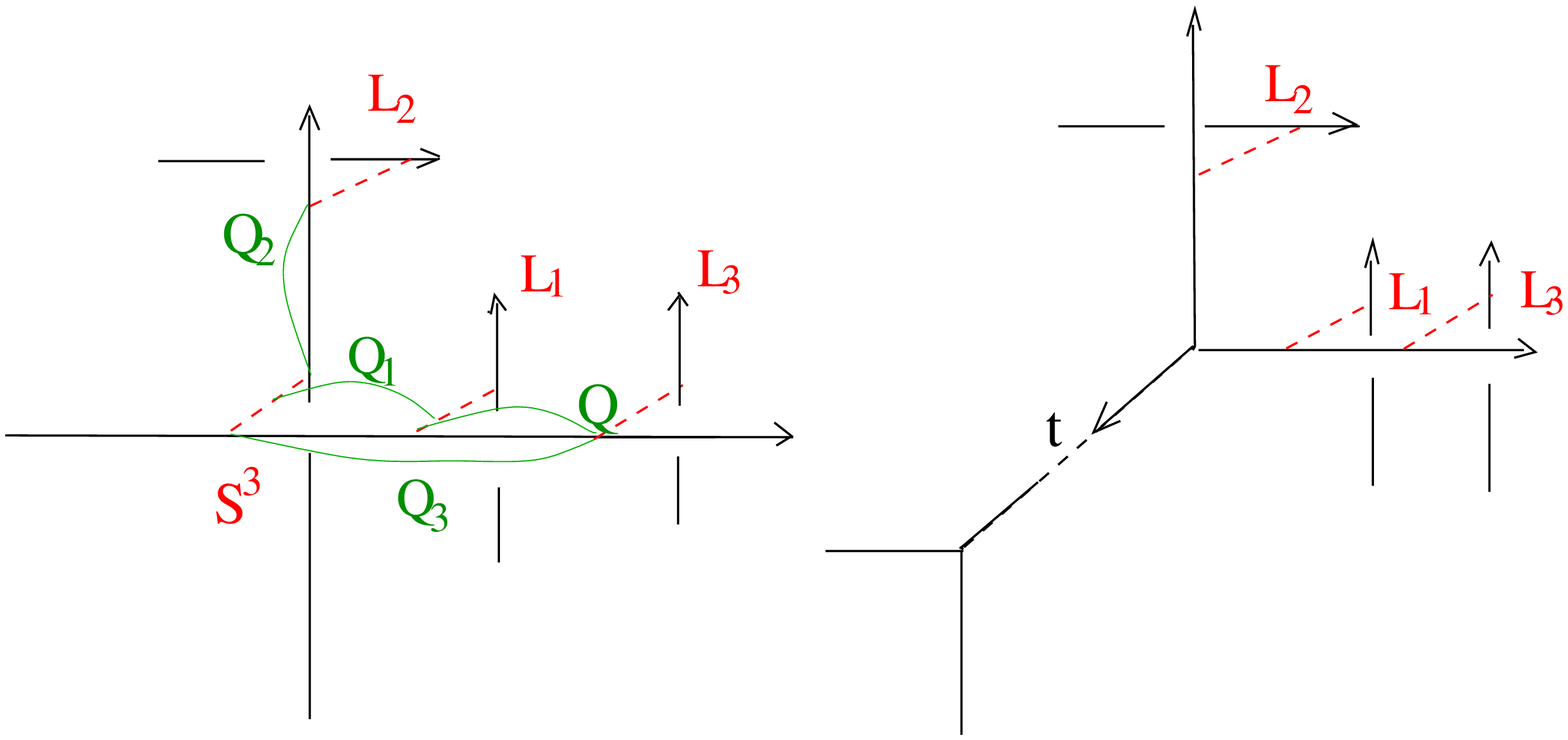}}
Consider then $T^{*} {\bf S}^3$ with $N$ D-branes on the ${\bf
S}^3$, but in addition $N_2$ D-branes on one leg wrapping the
Lagrangian $L_2$, and $N_1$ and $N_3$ D-branes on the other,
wrapping $L_1$ and $L_3$ respectively, as depicted in \CS. In the
dual theory we end up with $Y={\cal O}(-1)\oplus {\cal O}(-1)
\rightarrow {\bf P^1}$. The D-branes wrapping the compact cycle
have disappeared, but the D-branes on the non-compact cycles are
pushed through the transition. The resulting configuration is
shown in the second picture in \CS. The amplitude corresponding to
D-branes on $Y$ can easily be calculated using solvability of
Chern-Simons theory and the large $N$ transition. In the limit
where $N\rightarrow \infty$, the size of the ${\bf P^1}$ in $Y$
grows, and zooming in on the vertex with the D-branes, we are left
with ${\bf C}^3$ and the three D-branes. This is not exactly the
configuration of D-branes that gives the three-point vertex, but
it turns out to be close enough;  we need to move the $L_1$
Lagrangian D-branes through the vertex and put it on the other leg
of ${\bf C}^3$.  We will explain below how this can be achieved.

The open string theory on the ${\bf S}^3$ is $U(N)$ Chern-Simons
theory with some matter fields coming from the three non-compact
Lagrangians $L_{1,2,3}$. As shown in \amv, there are bifundamental
strings stretching between the ${\bf S^3}$ and $L_{1,2,3}$, and is
in addition there are strings between $L_1$ and $L_3$. The ground
state of all of these strings in the topological A-model is a
bifundamental matter field, and integrating it out corresponds to
inserting an annulus operator \annulus.
\ifig\hopftwo{This is a three-component link which is obtained from the
Hopf link by ``doubling'' one of its components, i.e. by replacing it with
two unlinked unknots. The labels denote representations, and the Chern-Simons
invariant associated to the link is $S_{{\overline Q_2}\, Q_1^t\otimes Q_2^t}/S_{00}$.}
{\epsfxsize 1.5truein\epsfbox{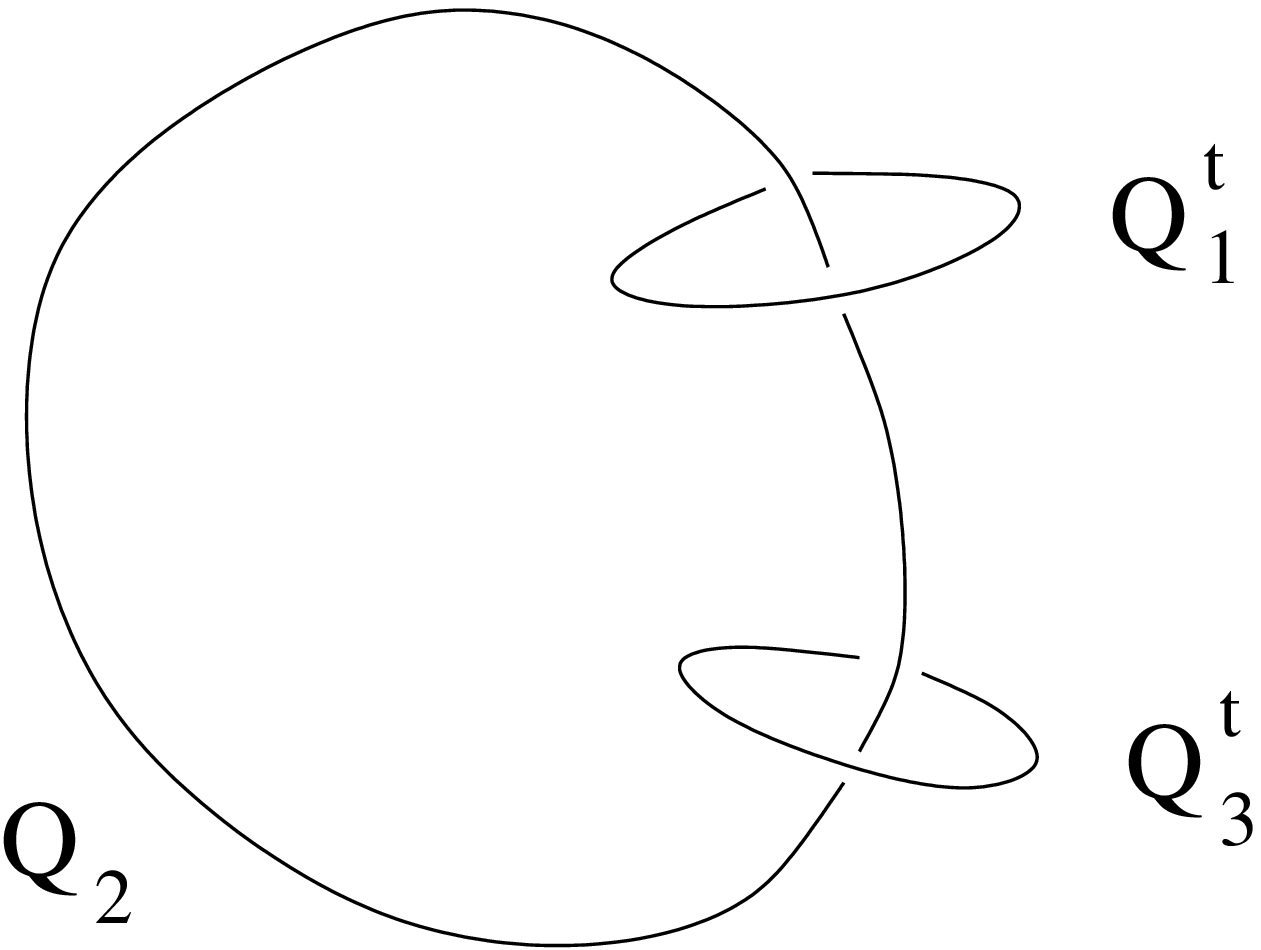}}
Keeping track of the orientations, we have that
\eqn\hopf{\eqalign{Z(V_1, V_2, V_3) ={1\over S_{00}}
\sum_{Q_1,Q_2,Q_3,Q} & (-1)^{\ell(Q_1)}\;
\langle {\rm Tr}_{Q_2}U\; {\rm Tr}_{Q_1^t\otimes Q_3^t} U\rangle \cr
&\times \;
{\rm Tr}_{Q_1} V_{1}\;{\rm Tr}_{Q^t} V_1^{-1}\;{\rm Tr}_{Q_2} V_2\;{\rm
Tr}_{Q\otimes Q_3} V_3.}}
where we have put the $1/S_{00}$ in front to compute
the contribution to the partition function due to the branes, noting
that $S_{00}$ is the partition function of topological string
on ${\cal O}(-1)\oplus
{\cal O}(-1)\rightarrow {\bf P^1}$.
In the equation above the $V_i$ is related
to the holonomy on the $i-$th stack of
non-compact D-branes and $U$ is holonomy on the ${\bf S}^3$.
The vacuum expectation value in \hopf\ corresponds to a Hopf link with
one of its components replaced by two unlinked unknots,
as in \hopftwo, and evaluated on the ${\bf S}^3$.
The unnormalized expectation value is given by \jones\
$$\langle {\rm Tr}_{Q_2}U\; {\rm Tr}_{Q_1^t\otimes Q_3^t} U
\rangle = S_{Q_1^t\otimes Q_3^t{\overline Q_2}},$$
where $S$ is the S-matrix of the $U(N)_k$ WZW model.
 On the other hand, it is well known that $
S_{Q_i\otimes Q_j {\overline Q_l}}= {S_{Q_i {\overline Q_l}}
S_{Q_j {\overline Q_l}}/S_{0 {\overline Q_l}}}$
\refs{\verlinde, \jones}.
Using this, we arrive at the
following expression for \hopf:
\eqn\wf{Z(V_1, V_2, V_3) = \sum_{Q_1,Q_2,Q_3,Q}
 (-1)^{\ell(Q_1)}\,{S_{Q_1^t{\overline Q_2}}S_{Q_3^t {\overline Q_2}}\over
 S_{00} S_{0{\overline Q_2}}}
\;{\rm Tr}_{Q_1} V_1 \;{\rm Tr}_{Q^t} V^{-1}_1
\; {\rm Tr}_{Q_2}V_2
\;{\rm Tr}_{Q \otimes Q_3} V_3.}
By large $N$-duality the above amplitude is computed by
the topological strings on $X$ with three stacks of
D-branes, as in \amv , corresponding to the right figure in fig. 7 .
In
the limit where we send $N$ or
equivalently $t=Ng_s$ to infinity, $Y$ becomes a ${\bf C^3}$, and this is
the limit we are interested in.
All the $N$ dependence in \wf\ is in the $S-$matrices,
and in the following we will use
$W_{Q_i Q_j}$ to mean
the {\it limit} of the matrix $S$-matrix as $N$ goes to infinity,
$W_{Q_i Q_j}= \lim_{t \rightarrow \infty}S_{Q_j{\overline Q_i}}/S_{00}$
 which
was introduced in
\amv. So, in the limit where $Y$ becomes ${\bf C^3}$ the amplitude in \wf\
simply becomes
\eqn\fs{Z(V_1, V_2, V_3) = \sum_{Q_1,Q_2,Q_3,Q}
 \;(-1)^{\ell(Q_1)}\,{W_{Q_2 Q_1^t} W_{Q_2 Q_3^t}\over W_{Q_2 0}}
\;{\rm Tr}_{Q_1} V_1 \;{\rm Tr}_{Q^t} V^{-1}_1
\; {\rm Tr}_{Q_2}V_2
\;{\rm Tr}_{Q \otimes Q_3} V_3.}
\ifig\move{The left and the right hand side of the figure describe
two D-brane configurations in ${\bf C}^3$, the latter obtained by
moving the D-brane $L_1$. The amplitude corresponding to
configuration on the left is $C^{0,0,-1}_{0, Q_2 ,Q_1}$ and on the
right it is $C^{0,0,0}_{Q_1,Q_2,0}$.} {\epsfxsize
4.0truein\epsfbox{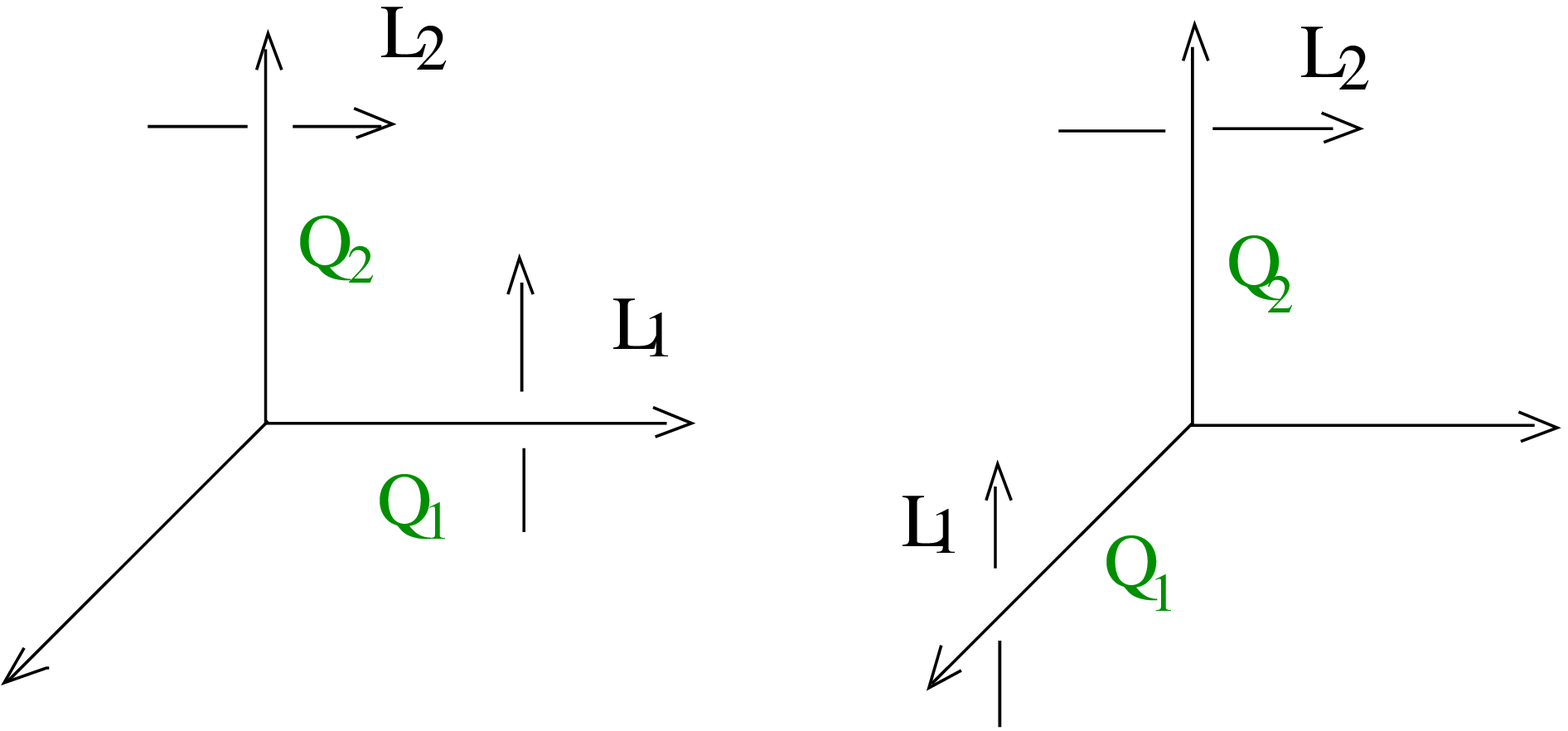}}

Our main interest is in the amplitude where the D-brane on
$L_1$ is on the first leg of ${\bf C}^3$.  This we will
do by a suitable ``analytic continuation''.
 The only part of the
amplitude in \wf\ that can be affected by moving the D-brane there
involves the representation labeled by $Q_1$, as this representation
corresponds to the world-sheet instanton strings which may become
massless in the process. The only other representation that could
have been affected is the one labeled by $Q$, however the action of
the corresponding strings is growing in the process, and as far as
they are concerned, the above expressions are getting more and
more reliable. At any rate, we will provide, in the following
sections, strong evidence that this is correct. With this
assumption, we only need to know what the transition means for the
part of amplitude corresponding to  $W_{Q_1 Q_2}$. To answer this, we
may well study a simpler problem where we have only two stacks of
D-branes, one wrapping $L_{2}$ and one on $L_{1}$, to start with, as
in \move.
Now, notice that this amplitude equals precisely
$$Z(V_1, V_2) = \sum_{Q_1, Q_2} W_{Q_2 Q_1^t}~ (-1)^{\ell(Q_1)}~ {\rm Tr}_{Q_1}
V_1 ~ {\rm Tr}_{Q_2}V_2.$$
By using the definition of the three-point vertex, and keeping track
of framing, $W_{Q_2 Q_1^t} (-1)^{\ell(Q_1)}$ should equal
$$C^{0,0,-1}_{0, Q_2 ,Q_1}
=C_{0, Q_2, Q_1}(-1)^{\ell(Q_2)}q^{-\kappa_{Q_1}/2},$$
from which we conclude that
\eqn\twopoint{C_{0 ,Q_2, Q_1}=W_{Q_2 Q_1^t} q^{\kappa_{Q_1}/2}.}
{}From this we can immediately find that the amplitude
corresponding to the second phase in \move.
{}From the definition of the three-point vertex alone, this is
$C_{Q_1,Q_2 0 }$,
but by cyclic symmetry \syma\ of the vertex and \twopoint , $C_{Q_1,Q_2,0}$
is the same as $W_{Q_2^t Q_1}  q^{\kappa_{Q_2}/2}$ (the
reader
should recall that $S$, and hence $W$ is symmetric).
We conclude that in going from the left to the right hand side of
\move\ we must replace
$$W_{Q_2 Q_1^t}\; {\rm Tr}_{Q_1} V_1\; {\rm Tr}_{Q_2} V_2
\rightarrow W_{Q_2^t Q_1} q^{\kappa_{Q_2}/2}\; {\rm Tr}_{Q_1} V_1^{-1}
\;{\rm Tr}_{Q_2}V_2,$$
in \fs . The strings ending on the $L_1$ D-brane in this new
phase, labeled with $Q_1$ and will naturally have the same charge
on the D-brane as the strings labeled with $Q$ since now moving
the D-brane affects both of their masses in the same way (recall
that $V$ is a phase of a complex field), and this is why we
replace $V_1$ by $V_1^{-1}$ in the above formula. Redefining $V_1
\rightarrow V_1^{-1}$ and collecting the coefficient of ${\rm
Tr}_{R_1} V_1\; {\rm Tr}_{R_2} V_2\; {\rm Tr}_{R_3} V_3$ in the
partition function we compute $C^{0,0,-1}_{R_1,R_2,R_3}$.
Correspondingly, we get the following expression for the
three-point vertex in the canonical framing $C_{R_1,R_2,R_3}$:
\eqn\wfb{C_{R_1,R_2,R_3}=\sum_{Q_1,Q_3}N_{Q_1 Q_3^{t}}^{~~~R_1 R_3^{t}}
q^{\kappa_{R_2}/2+\kappa_{R_3}/2}
{W_{R_2^t Q_1}W_{R_2 Q_3^{t}}\over W_{R_2 0}}.}
where $N_{Q_1 Q_3^{t}}^{~~~R_1R_3^{t}}$
counts the number of ways representations $Q_1$ and
$Q_3^{t}$ go into $R_1$ and $R_3^t$:
$$N_{Q_1 Q_3^{t}}^{~~~R_1 R_3^t} = \sum_{Q} N_{Q Q_1}^{~~R_1} N_{Q
Q_3^t}^{~~R_3^t}$$
 One must be careful to note
that $N_{R_i R_j}^{~~R_k}$ in the
formula above are
the $ordinary$ tensor product coefficients, and not the Verlinde coefficients\foot{The careful reader should note that, in
writing \hopf\ and \wfb, we have used the freedom to scale $V$ to
absorb a factor of $(-1)^{\ell(R)}$, into ${\rm Tr}_R V$.}.

We conjecture that the
above expression \wfb\ is the exact trivalent vertex
amplitude.   Here we have motivated this result based on the large $N$
topological string duality, combined with certain plausible assumptions.
As we discussed in section 3 on general grounds,
the vertex amplitude has a ${\bf Z}_3$ cyclic symmetry \syma\
and transforms simply under exchanges of pairs of indices \symb .
Moreover, in the previous section, we have computed the
leading piece of the vertex amplitude, corresponding to genus zero with
up to two holes. The expression \wfb\ has none of the symmetries
of the vertex amplitude manifest, and checking agreement with \twop\ is
highly involved. Nevertheless, the expression \wfb\ passes all these
checks. We will explicitly demonstrate this in section 8.

\newsec{Review of Chern-Simons and topological string amplitudes}

In order to work out some examples of closed and open
string
amplitudes from the three-point vertex, we need
a more precise definition
of the quantities appearing in \wfb. In this section we review
these ingredients, as well as the integrality properties of
open and closed string amplitudes.

\subsec{Review of necessary Chern-Simons theory ingredients}
In the evaluation of the
amplitudes we will need the
Chern-Simons invariants of the Hopf link in arbitrary representations
of $U(N)$. In this section we collect some formulae for these invariants.
Recall that in terms of Chern-Simons variables
\eqn\csvar{
q=\exp(g_s) = \exp \biggl( { 2\pi i \over k+N} \biggr), \,\,\,\,
\lambda = q^N.}
In the duality with topological string theory \gv, we have that $t=Ng_s$,
so $\lambda={\rm e}^t$.
As a warmup, consider $W_{R} \equiv W_{R0}$, which is related to
the Chern-Simons
invariant of the unknot in an
arbitrary representation $R$. The invariant of the unknot
is given by the quantum dimension of $R$:
\eqn\qdimvev{
{S_{0R} \over S_{00}}={\rm dim}_q R.
}
The explicit expression for ${\rm dim}_q R$ is as
follows. Let $R$ be a representation corresponding to a Young tableau
with row lengths $\{ \mu_i \}_{i=1, \cdots, d(\mu)}$, with
$\mu_1 \ge \mu_2 \ge \cdots$, and where $d(\mu)$
denotes the number
of rows. Define the following $q$-numbers:
\eqn\qnumbers{
\eqalign{
[x] =& q^{{x\over2}} - q^{-{x\over2}},\cr
[x]_{\lambda} =& \lambda^{1\over2} q^{{x\over2}} - \lambda^{-{1\over2}}
q^{-{x\over2}}.\cr}}
Then, the quantum dimension of $R$ is given by
\eqn\qdim{
{\rm dim}_q R=\prod_{1\le i<j\le d(\mu)} { [\mu_i -\mu_j + j-i]
\over [j-i]} \prod_{i=1}^{d(\mu)}  {\prod_{v=-i+1}^{\mu_i-i}
[v]_{\lambda} \over \prod_{v=1}^{\mu_i} [v-i+d(\mu)]}.}
The quantum dimension is a Laurent polynomial in
$\lambda^{\pm {1\over2}}$ whose
coefficients are rational functions of $q^{\pm {1 \over 2}}$.

We are interested in the leading
power of $\lambda$ in \qdim.
As explained in \amv, this power is
$\ell/2$, where $\ell=\sum_i \mu_i$ is the total
number of boxes in the representation $R$, and the coefficient of this
power is the rational function of $q^{\pm  {1 \over 2}}$
\eqn\leadingqdim{
W_R=q^{\kappa_R/4} \prod_{1\le i<j\le d(\mu)} { [\mu_i -\mu_j + j-i]
\over [j-i]} \prod_{i=1}^{d(\mu)}
\prod_{v=1}^{\mu_i} {1 \over [v-i+d(\mu)]},
}
where $\kappa_R$ is the framing factor introduced in

Let us now consider the Hopf link with linking number $1$. Its invariant
for representations
$R_1$, $R_2$, ${\cal W}_{R_1R_2}=S_{R_1 {\overline R_2}}/S_{0 0}$ is given by
\eqn\hopfinv{
({\cal W}_{R_1R_2})_{U(N)}=q^{ \ell_1 \ell_2/N}({\cal
W}_{R_1R_2})_{SU(N)},}
where $\ell_i$ is the total number of boxes in the Young tableau of $R_i$,
$i=1,2$. The prefactor $q^{ \ell_1 \ell_2/N}$ in \hopfinv\ is a
correction which was pointed out in \mv, and is
due to the fact that the vev $W_{R_1, R_2}$ has
to be computed in the theory with gauge group $U(N)$.
The expression we will use for this invariant is the one obtained by
Morton and Lukac in \refs{\ml ,\lukac}.
Their formula is as follows. Let $\mu$
be a Young tableau, and let $\mu^\vee$ denote its transposed
tableau (remember that this tableau is obtained from $\mu$ by
exchanging rows and columns). The Schur polynomial
in the variables $(x_1,\cdots,x_N)$ corresponding to $\mu$
(which is the character of the diagonal
$SU(N)$ matrix $(x_1,\cdots,x_N)$ in the representation
corresponding to $\mu$), will be denoted by  $s_{\mu}$.
They can be written in terms of elementary
symmetric polynomials $e_i(x_1,\cdots,x_N)$, $i\ge 1$, as follows \macdonald:
\eqn\jt{
s_{\mu} = {\rm det} M_\mu}
where
$$M_\mu^{ij}=(e_{\mu^\vee_i +j-i})$$
$M_\mu$ is an $r \times r$ matrix, with
$r= d(\mu^\vee)$. To evaluate $s_{\mu}$ we put $e_0=1$, $e_k=0$
for $k<0$. The expression \jt, known sometimes
as the Jacobi-Trudy identity, can be formally extended to give the
Schur polynomial $s_{\mu} (E(t))$ associated to any formal power series
$E(t)=1 + \sum_{n=1}^{\infty} a_i t^i$. To obtain this, we simply use the
Jacobi-Trudy formula \jt, but where $e_i$ denote now the coefficients
of the series $E(t)$, i.e. $e_i=a_i$. Morton and Lukac define the series
$E_{\emptyset}(t)$ as follows:
\eqn\trive{
E_{\emptyset}(t)=1+ \sum_{n=1}^{\infty} c_n t^n, }
where the coefficients $c_n$ are defined by
\eqn\coeffs{
c_n =\prod_{i=1}^n {1 - \lambda^{-{1}} q^{i-1} \over {q^{i-1}}.}}

They also define a formal power series associated to
a tableau $\mu$, $E_{\mu}(t)$, as follows:
\eqn\eser{
E_{\mu}(t)= E_{\emptyset}(t)
\prod_{j=1}^{d(\mu)} { {1 + q^{\mu_j -j} t}
\over {1 + q^{- j} t}}.} One can then consider the Schur function of the
power series \eser, $s_{\mu}(E_{\mu'}(t))$, for any pair of tableaux
$\mu$, $\mu'$, by expanding
$E_{\mu'}(t)$ and substituting its coefficients in the Jacobi-Trudy
formula \jt. It turns out that this Schur function is essentially the
invariant we were looking for. More precisely, one has
\eqn\mlfor{
{\cal W}_{R_1, R_2}(q, \lambda)= ({\rm dim}_q R_1) (\lambda q)^{\ell_2 \over 2}
s_{\mu_2}(E_{\mu_1}(t)),} where $\mu_{1,2}$ are the tableaux
corresponding to $R_{1,2}$, and
$\ell_2$ is the number of boxes of $\mu_2$. More details and examples
can be found in \ml. It is easy to see from
\mlfor\ that the leading power in $\lambda$ of
${\cal W}_{R_1, R_2}$ is $(\ell_1 +
\ell_2)/2$, and its coefficient is given by the leading coefficient of the
quantum dimension, \leadingqdim, times a rational function of $q^{\pm {1
\over 2}}$ that is given by:
\eqn\leadhopf{
W_{R_1 R_2}(q)= W_{R_1}q^{\ell_2 \over 2} s_{\mu_2}(E^{\rm lead}_{\mu_1}(t)),}
where
\eqn\eserle{
E^{\rm lead}_{\mu}(t)= E^{\rm lead}_{\emptyset}(t)
\prod_{j=1}^{d(\mu)} { {1 + q^{\mu_j -j} t}
\over {1 + q^{-j} t}}}
and
\eqn\trivele{
E^{\rm lead}_{\emptyset}(t)=1+ \sum_{n=1}^{\infty}{ t^n \over \prod_{i=1}^n (q^i-1)}. }

The above results are for knots and links in the standard framing. The
framing can be incorporated as in \mv, by simply multiplying the
Chern-Simons invariant of a link with components
in the representations $R_1, \cdots, R_L$, by the factor
\eqn\framefactor{
(-1)^{\sum_{\alpha =1}^L p_{\alpha}\ell_{\alpha}} q^{{1 \over
2}\sum_{\alpha=1}^L p_{\alpha} \kappa_{R_{\alpha}}},}
where $p_{\alpha}$, $\alpha=1, \cdots, L$ are integers labeling
the choice of framing for each component.

\subsec{Integrality of closed string amplitudes}

In this and the following subsection we recall certain integrality properties that the topological A-model
amplitudes posses on general grounds \gvtwo. This allows one to formulate
our answers in terms of certain integers which capture
BPS degeneracies.

The topological
A-model free energy $F(X)$ has the following structure:
$$
F(X)=\sum_{g=0}^{\infty}g_s^{2g-2}F_g (t). $$ Here, $F_g(t)$ is the free energy at
genus $g$. It can be computed as a sum over two-homology classes
of worldsheet instantons of genus
$g$,
$$
F_g(t) = \sum_{Q}N_{g,Q}e^{- Q \cdot t },$$ where the vector $Q
\in H_2(X,{\bf Z})$ labels the homology class, $t$ is a vector of
K\"ahler parameters, and $N_{g, Q}$ are Gromov-Witten invariants.
The free energy $F(X)$ can be related to counting of certain BPS
states on the Calabi-Yau manifold associated to D2 branes wrapping
holomorphic curves in $X$ \gvtwo. The relation follows from the
embedding of the topological A-model in type IIA string theory on
$X$ and its further embedding in M-theory.  Moreover it relies on
the target string interpretation of topological string amplitudes
\lref\AntoniadisZE{
I.~Antoniadis, E.~Gava, K.~S.~Narain and T.~R.~Taylor,
``Topological amplitudes in string theory,''
Nucl.\ Phys.\ B {\bf 413}, 162 (1994)
[arXiv:hep-th/9307158].
}
\refs{\BCOV ,\AntoniadisZE}.
This implies that the free energy has the following form \gvtwo :
\eqn\closedf{
F(X) = \sum_{n=1}^{\infty}\sum_{Q \in
H_2(X)}\sum_{g=0}^{\infty} n^g_{Q} (2 \sinh (ng_s/2))^{2g-2}
{e^{-n Q \cdot t}\over n}.}
In this formula, $q=e^{g_s}$, $g$
is related to an $SU(2)_L\subset SO(4)$ quantum number denoting
the spin representation of the particle in $4+1$ dimensions , $Q \cdot t$ is
the mass of the BPS state and $n^g_{Q}$ is an integer which counts
the number of BPS states with quantum numbers $Q$ and $g$.

%
%

\subsec{Integrality of open string amplitudes}

The free energy of open strings $F(V)$
is given by the
logarithm of a partition function with the structure
\eqn\partf{
Z_{\rm open} (V_1, \cdots, V_L)=\sum_{R_1, \cdots, R_L}Z_{(R_1, \cdots, R_L)}
\prod_{\alpha=1}^L {\rm Tr}_{R_{\alpha}} V_{\alpha}.}
We define the generating function
$f_{(R_1, \cdots, R_L)} (q, \lambda)$ through the
following equation:
 \eqn\conj{
F(V)= \sum_{n=1}^\infty \sum_{R_1, \cdots, R_L}
{1\over n} f_{(R_1, \cdots, R_L)} (q^n, {\rm e}^{-n t})
\prod_{\alpha=1}^L {\rm Tr}_{R_\alpha}V_{\alpha}^n
 }
where $R_{\alpha}$ denote representations of $U(M)$ and we are considering
the limit $M\rightarrow \infty$. In this limit we can exchange the basis
consisting
of product of traces of powers in the fundamental representation, with
the trace in arbitrary representations.
It was shown in \ov , following
similar ideas in the closed string case \gvtwo, that the open topological
strings compute the partition function of BPS domain walls in a related
superstring theory.  This led to the result that $F(V)$ has an integral
expansion structure. This result was further refined in \lmv\
where it was shown that the corresponding integral expansion
leads to the following formula for
 $f_{(R_1, \cdots, R_L)} (q, \lambda)$:
\eqn\fr{
\eqalign{
&f_{(R_1, \cdots, R_L)}(q, \lambda)=\cr &
(q^{1\over 2}-q^{-{1 \over 2}})^{L-2}
\sum_{g\ge 0} \sum_{Q}
\sum_{R'_1, R_1'' \cdots, R'_L, R_L''}  \prod_{\alpha=1}^L
C_{R_{\alpha}\,R_{\alpha}'\,R_{\alpha}''}S_{R_{\alpha}'}(q)
N_{(R_1'', \cdots, R_L''),g,Q}
 (q^{1\over 2}-q^{-{1\over 2}})^{2g} {\rm e}^{-Q\cdot t}.\cr}}
In this formula $R_{\alpha},R_{\alpha}',R_{\alpha}''$ label
representations of the symmetric group
$S_\ell$, which can be labeled
by a Young tableau with a total of $\ell$ boxes.
$C_{R\,R'\,R''}$ are the Clebsch-Gordon coefficients
of the symmetric group, and the monomials $S_R (q)$ are defined as
follows. If $R$ is a hook representation
\eqn\hook{
\tableau{6 1 1 1 1}}
with $\ell$ boxes in total, and with $\ell-d$ boxes in the first row, then
\eqn\expsr{
S_R (q)=(-1)^d q^{ -{\ell -1 \over 2}+d} ,}
and it is zero otherwise.
Finally, $N_{(R_1, \cdots, R_L),g,Q}$ are {\it integers}
 associated to open string amplitudes. They compute the net number
of BPS domain walls of charge $Q$ and spin $g$ transforming
in the representations $R_{\alpha}$ of $U(M)$, where we are using
the fact that representations of $U(M)$ can also be labeled by Young
tableaux. It is also useful to introduce a generating functional for these
degeneracies as in \lmv:
\eqn\tildefrlinks{
{\widehat f}_{(R_1, \cdots, R_L)}(q, \lambda)=
\sum_{g \ge 0}\sum_Q N_{(R_1, \cdots, R_L),g,Q}
(q^{{1\over 2}}-q^{-{1 \over 2}})^{2g}{\rm e}^{-Q\cdot t}.}
We then have
the relation:
\eqn\relafslinks{
 f_{(R_1, \cdots, R_L)}(q, \lambda)=(q^{{1\over 2}}-q^{-{1 \over 2}})^{L-2}
\sum_{R_1', \cdots, R'_L}
M_{R_1, \cdots, R_L;R'_1, \cdots, R_L'}(q)
{\widehat f}_{(R_1, \cdots, R_L)}(q, \lambda),}
where the matrix $M_{R_1, \cdots, R_L;R'_1, \cdots, R_L'}(q)$ is given by
$$
M_{R_1, \cdots, R_L;R'_1, \cdots, R_L'}(q)=
 \prod_{\alpha=1}^L \sum_{R_{\alpha}''}
C_{R_{\alpha}\,R_{\alpha}'\,R_{\alpha}''}S_{R_{\alpha}''}(q)
$$
and it is invertible \lmv. Finally, it is also useful sometimes to
write BPS degeneracies in the winding number basis:
\eqn\kintslinks{
n_{({\vec k}^{(1)}, \cdots, {\vec k}^{(L)}),g,Q}=
\sum_{R_1, \cdots, R_L} \prod_{\alpha=1}^L
\chi_{R_{\alpha}} (C({\vec k}^{(\alpha)}))
N_{(R_1, \cdots, R_L),g,Q}.}
Notice that the BPS degeneracies $N_{(R_1, \cdots, R_L),g,Q}$ in the
representation basis are more fundamental than the degeneracies in the
winding number basis, as emphasized in \lmv.

The $f_{(R_1, \cdots, R_L)}$ introduced in \conj\ can be extracted
from $Z_{(R_1, \cdots, R_L)}$ through a procedure spelled out in detail in
\refs{\lm ,\lmv ,\newp}. One has, for example,
\eqn\exm{
f_{\tableau{1},\tableau{1}}=
Z_{\tableau{1},\tableau{1}}-Z_{\tableau{1},\cdot} Z_{\cdot,\tableau{1}}.}
It is also convenient to introduce the quantities
\eqn\zkbasis{
Z_{(\vec k^{(1)}, \cdots, \vec k^{(L)})}=\sum_{R_{\alpha}} \prod_{\alpha=1}^L
\chi_{R_{\alpha}} (\vec k^{(\alpha)}) Z_{(R_1, \cdots, R_L)}, }
in such a way that
\eqn\partfk{
Z_{\rm open}(V_1, \cdots, V_L)=\sum_{\vec k^{(1)}, \cdots, \vec k^{(L)}}
Z_{(\vec
k^{(1)}, \cdots, \vec k^{(L)})}
\prod_{\alpha=1}^L {1 \over z_{\vec k^{(\alpha)}}}
\Upsilon_{\vec k^{(\alpha)}}(V_{\alpha}).}

We can now write the total free energy as:
\eqn\totaln{
F(V)=\sum_{g=0}^{\infty} \sum_{\vec k^{(\alpha)}}g_s^{2g-2+h}
 F_{g, \vec k^{(\alpha)}} (t) \prod_{\alpha=1}^L
\Upsilon_{\vec k^{(\alpha)}} (V_{\alpha}),}
where $h=\sum_{\alpha} h_{\alpha}$ is the total number of holes. We have then
that,
\eqn\freez{
 \sum_{g=0}^{\infty} F_{g, \vec k^{(\alpha)}} (t) g_s ^{2g-2 + h}
={1 \over \prod_{\alpha=1}^L z_{\vec k^{(\alpha)}}}
Z_{\vec k^{(\alpha)}}^{(c)},}
where $(c)$ denotes the connected piece.

\newsec{The vertex amplitude}

In this section we use the apparatus developed above to calculate
some values of the vertex amplitude. This will provide highly
non-trivial checks that the vertex amplitude derived in section 6
using large N-dualities is in fact the correct expression.

As discussed in the previous section, we can extract the
free energy of the vertex amplitude with fixed winding numbers
as the connected part of $C_{ \vec{k}_1\vec{k}_2\vec{k}_3}$:
$$
F^{(n_1, n_2, n_3)}_{\vec k^{(1)}, \vec k^{(2)}, \vec k^{(3)}}(g_s)=
{1 \over \prod_{\alpha=1}^3 z_{\vec k^{(\alpha)}}}
(C^{(c)})^{(n_1, n_2, n_3)}_{\vec k^{(1)}, \vec k^{(2)}, \vec k^{(3)}}.
$$
Consider the part of this amplitude corresponding to a single hole
on each of the three stacks of D-branes. Since only the winding numbers
remain to be specified, we can simply denote this by
$F_{k,l,m}$, corresponding to $\vec k^{(i)}$, $i=1,2,3$
with a single nonzero entry in positions $k$, $l$, $m$, respectively.
Then, one has the following formulae:
$$
\eqalign{
F_{k,0,0}^{(n,0,0)}&={1 \over k} {[k+n k-1]! \over [k]![n k]!}\cr
F_{k,l,0}^{(0,0,0)}&={(-1)^{l+1} \over k l}{[k l]\over [k]}
\Bigg[ {k \atop l} \Bigg],\cr}
$$
where the $q$-number is $[x]= q^{x\over 2} - q^{-{x\over 2}}$,
the $q$-factorial is given by
\eqn\qfact{
[x]!=[x][x-1]\cdots [1], }
and finally the $q$-combinatorial number is defined as
\eqn\qcomb{
\Bigg[ {x \atop y} \Bigg]={[x]!\over [x-y]! [y]!}.}

Note that the leading $g_s$ terms $F_{k,0,0}$ and $F_{k,m,0}$ are
$$
\eqalign{
F_{g=0; k,0,0}^{(n,0,0)}&={1 \over g_s  k} {(k+n k-1)! \over (k)!(n k)!}\cr
F_{g=0; k,l,0}^{(0,0,0)}&={(-1)^{l+1} \over k}
\Bigg( {k \atop l} \Bigg),\cr}
$$
and these agree with \onep\ and \twop\ respectively up to a choice of
coordinate, $Tr V_i^{m}\rightarrow(-1)^{m}  Tr V_i^{m}$
and the over-all sign of the free
energy!

Let us now look at some explicit values of the vertex \wfb\ which
we can easily compute using the explicit expressions for $W_{R_1
R_2}$, that we gave in the previous section.  Using explicit
evaluation of $C$ one can verify that at least for small number of
boxes the highly non-trivial symmetry prediction \syma\ and \symb\
are indeed satisfied.

We give here a list of values for the trivalent vertex up to
five boxes in total. For the sake of space, we mostly list values which
are not related by symmetries, although we have included some to make manifest
the properties that we derived in section 3. The dot $\cdot$
stands for the trivial representation.

$$
\eqalign{
C_{\tableau{1} \cdot \cdot}&={1 \over q^{1\over 2} -q ^{-{1\over 2}}},\cr
C_{\tableau{1} \tableau{1} \cdot}&={q^2 -q+1 \over (q-1)^2},\qquad
C_{\tableau{2} \cdot \cdot}={q^2 \over (q-1)(q^2-1)},\cr
C_{\tableau{1 1} \cdot \cdot}&={q \over (q-1)(q^2-1)},}
$$
$$
\eqalign{
C_{\tableau{1} \tableau{1} \tableau{1}}&= { q^4 -q^3 + q^2-q+1 \over q^{1\over 2}(q-1)^3}, \qquad
C_{\tableau{2} \tableau{1} \cdot}={q^{3\over 2} (q^3-q^2 +1)\over  (q-1)^2 (q^2-1)}, \cr
C_{\tableau{1 1} \tableau{1} \cdot}&={ q^3-q^2 +1\over q^{1\over 2}(q-1)^2(q^2-1)}, \qquad
C_{\tableau{3}\cdot \cdot}={q^{9\over 2}\over (q-1) (q^2-1) (q^3-1)},\cr
C_{\tableau{2 1}\cdot \cdot}&={q^{5\over 2}\over  (q-1)^2 (q^3-1)},\qquad
C_{\tableau{1 1 1} \cdot \cdot}= {q^{3\over 2}\over (q-1)(q^2-1)(q^3-1) },}
$$
$$\eqalign{
C_{\tableau{2} \tableau{1} \tableau{1}}&={q^6-q^5 + q^3 -q+1\over (q-1)^3 (q^2-1)},\qquad
C_{\tableau{1 1} \tableau{1} \tableau{1}}={q^6-q^5 + q^3 -q+1\over
q(q-1)^3(q^2-1)},\cr
C_{\tableau{2} \tableau{2} \cdot }&={q^2 (q^4-q^2 +1) \over (q-1)^2 (q^2-1)^2},\qquad
C_{\tableau{2} \tableau{1 1} \cdot }={q (q^6-q^5 -q^4 + 2q^3 -q +1) \over
(q-1)^2 (q^2-1)^2},}
$$
$$
\eqalign{
C_{\tableau{1 1} \tableau{2} \cdot }&={q^6-q^5  + 2q^3 -q^2-q +1 \over q(q-1)^2 (q^2-1)^2}, \qquad
C_{\tableau{1 1} \tableau{1 1} \cdot }={q^4-q^2+1 \over (q-1)^2 (q^2-1)^2},\cr
C_{\tableau{3} \tableau{1} \cdot }&={q^4(q^4-q^3+1) \over (q-1)^2(q^2-1)(q^3-1)},\qquad
C_{\tableau{2 1} \tableau{1} \cdot }={q(q^4-q^3+q^2 -q+1) \over  (q-1)^3 (q^3-1)},\cr
C_{\tableau{1 1 1} \tableau{1} \cdot }&={q^4-q+1 \over  q (q-1)^2 (q^2-1) (q^3-1)},\cr
C_{\tableau{4} \cdot \cdot}&={q^8 \over (q-1)(q^2-1)(q^3-1) (q^4-1)},\qquad
C_{\tableau{3 1} \cdot \cdot}={q^5 \over (q-1)^2 (q^2-1) (q^4-1)},\cr
C_{\tableau{2 2} \cdot \cdot}&={q^4 \over (q-1)^2 (q^2-1)(q^4-1)},\qquad
C_{\tableau{2 1 1} \cdot \cdot}={q^3 \over (q-1)^2 (q^2-1) (q^4-1)},\cr
C_{\tableau{1 1 1 1} \cdot \cdot}&={q^2 \over (q-1)(q^2-1)(q^3-1)(q^4-1)},\cr
C_{\tableau{2} \tableau{2} \tableau{1}}&={q^{1\over 2}(q^8-q^7+q^5-q^4+q^3-q+1)\over
(q-1)^3 (q^2-1)^2},}
$$
$$
\eqalign{
C_{\tableau{2} \tableau{1 1} \tableau{1}}&={q^9-q^8-q^7+2q^6-q^4+q^3-q+1\over
q^{1\over 2}(q-1)^3(q^2-1)^2},\cr
C_{\tableau{1 1} \tableau{2} \tableau{1}}&={q^9-q^8+q^6-q^5+ 2q^3-q^2-q+1\over
q^{3\over 2} (q-1)^3(q^2-1)^2},\cr
C_{\tableau{1 1} \tableau{1 1} \tableau{1}}&={q^8-q^7+q^5-q^4+q^3-q+1\over
q^{3\over 2}(q-1)^3(q^2-1)^2},\cr
C_{\tableau{3} \tableau{1} \tableau{1}}&={q^{3\over 2}(q^8-q^7+q^4-q+1)\over
(q-1)^3 (q^2-1)(q^3-1)},\cr
C_{\tableau{2 1} \tableau{1} \tableau{1}}&={q^8-2q^7 + 3q^6 -3q^5+ 3q^4
-3 q^3 + 3q^2-2q +1 \over
q^{1\over 2} (q-1)^4(q^3-1)},\cr
C_{\tableau{1 1 1} \tableau{1} \tableau{1}}&={q^8-q^7+q^4-q+1\over
q^{3\over 2} (q-1)^3 (q^2-1)(q^3-1)},\cr
C_{\tableau{3} \tableau{2} \cdot}&={q^{9\over 2}(q^5-q^3+1)\over
(q-1)^2 (q^2-1)^2 (q^3-1)},\cr
C_{\tableau{3} \tableau{1 1} \cdot}&={q^{7 \over 2} (q^8-q^7-q^6 +q^5 +q^4 -q^2 +1)\over
(q-1)^2 (q^2-1)^2(q^3-1)},\cr
C_{\tableau{2 1} \tableau{2} \cdot}&={q^{1 \over 2}(q^7-q^6 +q^4 -q +1) \over
(q-1)^3 (q^2-1)(q^3-1)},\cr
C_{\tableau{2 1} \tableau{1 1} \cdot}&={q^{1 \over 2}(q^7-q^6 +q^3 -q +1) \over
(q-1)^3 (q^2-1)(q^3-1)},\cr
C_{\tableau{1 1 1} \tableau{2} \cdot}&={q^8-q^6 +q^4+q^3 -q^2-q +1 \over
q^{5 \over 2} (q-1)^2 (q^2-1)^2(q^3-1)},\cr
C_{\tableau{1 1 1} \tableau{1 1} \cdot}&={q^5-q^2 + 1 \over
q^{1 \over 2} (q-1)^2 (q^2-1)^2(q^3-1)},\cr
C_{\tableau{4} \tableau{1} \cdot}&={q^{15 \over 2} (q^5-q^4 +1) \over
(q-1)^2 (q^2-1)(q^3-1) (q^4-1)},\cr
C_{\tableau{3 1} \tableau{1} \cdot}&={q^{7\over 2}(q^5 -q^4 +q^2 -q+1) \over
(q-1)^3 (q^2-1) (q^4-1)},\cr
C_{\tableau{2 2} \tableau{1} \cdot}&={q^{5 \over 2} (q^4 -q^2+1) \over
(q-1)^2 (q^2-1)^2 (q^3-1)},\cr
C_{\tableau{2 1 1} \tableau{1} \cdot}&={q^{1 \over 2}(q^5 -q^4 +q^3 -q+1) \over
(q-1)^3 (q^2 -1)(q^4-1)},\cr
C_{\tableau{1 1 1 1} \tableau{1} \cdot}&={q^5  -q+1 \over
q^{3 \over 2} (q-1)^2 (q^2 -1)(q^3-1)(q^4-1)},\cr
C_{\tableau{5} \cdot \cdot}&={q^{25\over 2} \over
(q-1)(q^2-1)(q^3-1)(q^4-1)(q^5-1)},}
$$
$$
\eqalign{
C_{\tableau{4 1} \cdot \cdot}&={q^{17\over 2} \over
(q-1)^2 (q^2-1)(q^3-1)(q^5-1)},\cr
C_{\tableau{3 2} \cdot \cdot}&={q^{13\over 2} \over
(q-1)^2 (q^2-1) (q^3-1)(q^4-1)},\cr
C_{\tableau{3 1 1} \cdot \cdot}&={q^{11\over 2} \over
(q-1)^2 (q^2-1)^2 (q^5-1)},\cr
C_{\tableau{2 2 1} \cdot \cdot}&={q^{9\over 2} \over
(q-1)^2 (q^2-1)(q^3-1)(q^4-1)},\cr
C_{\tableau{2 1 1 1} \cdot \cdot}&={q^{7\over 2} \over
(q-1)^2 (q^2-1) (q^3-1) (q^5-1)},\cr
C_{\tableau{1 1 1 1 1} \cdot \cdot}&={q^{5\over 2} \over
(q-1)  (q^2-1) (q^3-1)(q^4-1) (q^5-1)}.\cr}
$$

Note for example that:
$$C_{\tableau{3} \tableau{1} \tableau{1}}=q^{3} C_{\tableau{1 1 1}
\tableau{1} \tableau{1}},$$
while on the other hand $k_{\tableau{3}}=6$, so that this precisely agrees
with \symb .

The vertex amplitude, being an open string amplitude, has to satisfy
strong integrality requirements that we have reviewed in section 6.
In order to check this,
we can compute the generating functionals for BPS states
${\widehat f}_{R_1 R_2 R_3}$ for arbitrary framings in the legs. If we denote
$z=(q^{1 \over2} -q^{-{1\over 2}})^2$, one finds, for example:
$$
\eqalign{
{\widehat f}_{\tableau{1}, \tableau{1}, \tableau{1}}=&
(-1)^{n_1 + n_2 + n_3} , \cr
{\widehat f}_{\tableau{2}, \tableau{1}, \tableau{1}}=&-(-1)^{n_2 + n_3}
\biggl({ q^{n_1\over 2} -q^{-{n_1\over 2}}
\over q^{1 \over 2} -q ^{-{1 \over 2}}}\biggr)^2 \cr
&=-(-1)^{n_2 + n_3} n_1^2  -{1 \over 12}
(-1)^{n_2 + n_3} n_1^2 (n_1^2 -1)z + \cdots, \cr
{\widehat f}_{\tableau{1 1}, \tableau{1}, \tableau{1}}=&-(-1)^{n_2 + n_3}
\biggl({q^{n_1+1\over 2} -q^{-{n_1+1\over 2}}
\over q^{1 \over 2} -q ^{-{1 \over 2}}}\biggr)^2 \cr
&=-(-1)^{n_2 + n_3}(1+ n_1)^2 -{1 \over 12}
(-1)^{n_2 + n_3} n_1 (n_1 +1)^2 (n_1 +2)z +\cdots,}
$$
$$
\eqalign{
{\widehat f}_{\tableau{2},\tableau{2},\tableau{1}}=&
{(-1)^{n_3} \over 4} \Bigl( 2n_2 (n_2-1) + n_1 (n_2^2 -3 n_2 +2)
+ n_1^2 (3 n_2^2- n_2 +2 )\Bigr)+ \cdots,\cr
{\widehat f}_{\tableau{2},\tableau{1 1},\tableau{1}}=&
{(-1)^{n_3} \over 4} \Bigl( n_1n_2 (n_2-1) +2 n_2 (1+n_2)
+ n_1^2 (3 n_2^2+5 n_2 +4 )\Bigr)+ \cdots,\cr
{\widehat f}_{\tableau{1 1},\tableau{2},\tableau{1}}=&
{(-1)^{n_3} \over 4} \Bigl(4 -6 n_2 + 6n_2^2
+ n_1 ( 7n_2^2 - 5n_2 +6) + n_1^2 ( 3n_2^2 - n_2 +2)
\Bigr)+ \cdots,\cr
{\widehat f}_{\tableau{1 1},\tableau{1 1},\tableau{1}}=&
{(-1)^{n_3} \over 4} \Bigl(4 +6 n_2 + 6n_2^2
+ n_1 ( 7n_2^2 + 9 n_2 +8) + n_1^2 ( 3n_2^2 + 5 n_2 +4)
\Bigr)+ \cdots,\cr
}
$$
This passes the integrality check.

\newsec{Examples of open and closed string amplitudes from the vertex.}

In this section we compute various closed and open string amplitudes using the
trivalent vertex. In the examples below we have made many
checks of the vertex against
amplitudes of closed and open string calculations using other means.

In the closed string case the vertex can be checked, 
in principle to all genera, by comparison with mirror
B-model calculations using
holomorphic anomaly \BCOV , 
see \refs{\ckyz ,\Hosono ,\AganagicWV}, as well as against A-model localisation
calculations using the techniques of \refs{\kon ,\gz ,\kz}.
More directly one can compare the vertex with open string amplitudes.
B-model calculations for the open string were so far only available for
the disk following \refs{\AV ,\AKV}.
A-model localization calculations for all genus
open string amplitudes with a single stack of D-branes 
have been introduced in \refs{\kl,\gz}. 
At the operational level this is a minor modification of the localization
procedure of the closed string case, and we provide a computer
program which computes this for general toric configurations\foot{This
program can be distributed on request. It requires the evaluation of
2d gravity correlations functions, which where implemented in Maple
by Carel Faber, see also \faber.}. 
Due to the extended combinatorics of the graphs indexing the fixed points
of the torus action on the moduli space of stable maps \refs{\kon ,
\gz ,\kz}, the computer
calculation is very slow compared with the
techniques developed in the present paper, 
in particular for amplitudes with higher genus and larger
degree (w.r.t the compact K\"ahler classes).
A-model calculations
provide, on the other hand, expressions which describe all windings.
While the results of the calculations
have been checked for many cases, 
see in particular \gz , the procedure has
not been established rigorously and leaves interesting conceptual issues
to be developed, in particular in regards to multiple stacks of branes.
It should be possible to derive within the localization
approach general expressions for the vertex for general windings
and framings. Some attmepts in this directions have
already been made \digr .

To begin with, it is useful to see how the vertex works in a few
simple examples where the complete amplitudes are known.

\subsec{Example I}
\ifig\exampleone{The left hand side is the graph $\Gamma$ of a 
geometry containing a
chain of ${\bf P^1}$'s with
five independent classes in $H_2$. The right hand side depicts a
decomposition of this graph in terms of three-point vertices. All the
vertices are obvious repetitions of the first two, moreover, all the
vertex amplitudes equal $C_{R_1 R_2R_3}$ with representation ordered
cyclically, counter-clockwise. Some of the representations are set to
be trivial, as corresponding legs are non-compact. The small figure in the
upper left corner is the corresponding graph $\hat \Gamma$.}
{\epsfxsize 5.0truein\epsfbox{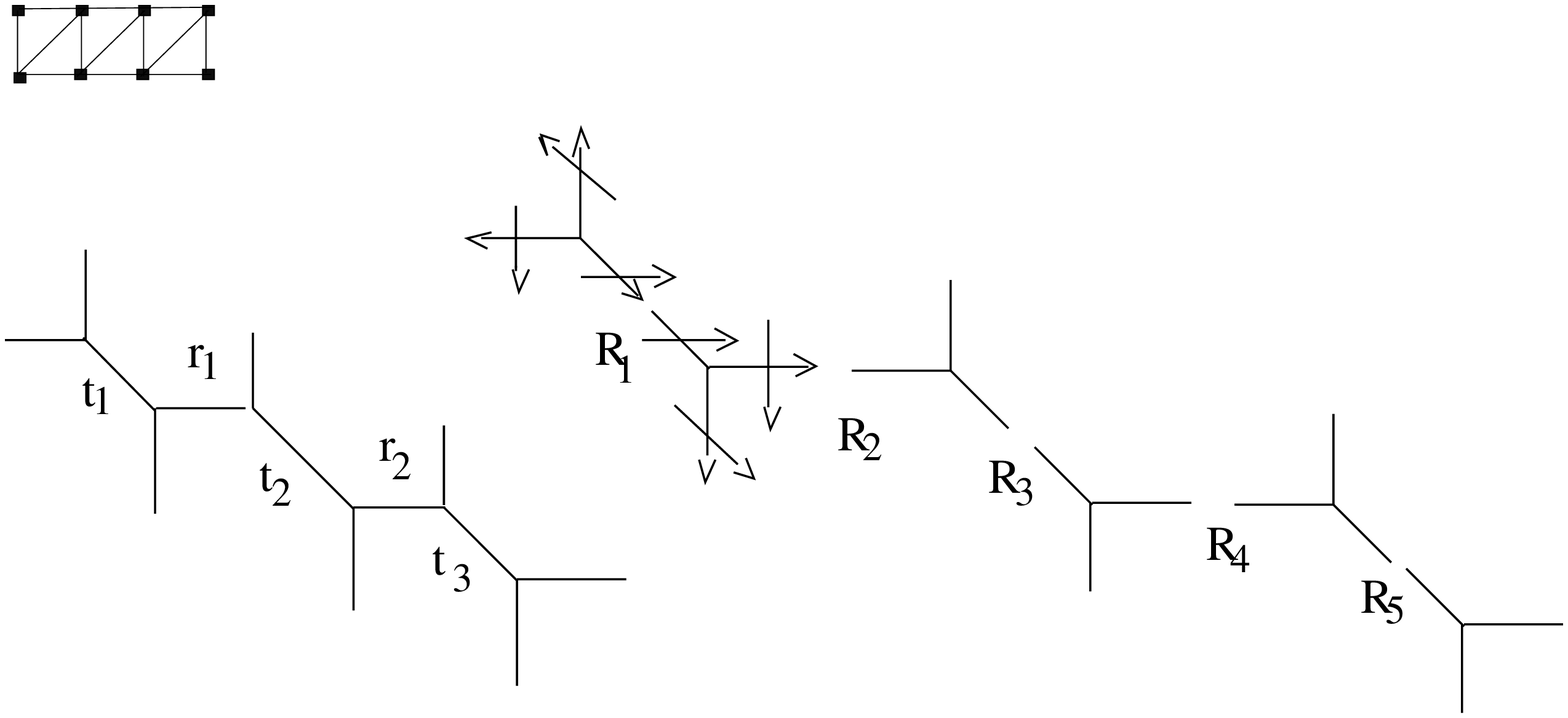}}

The closed string amplitude in \exampleone\ can be written in
terms of six trivalent vertices glued together. Two of them are of
the kind we have already discussed. Using an $SL(2,{\bf Z})$
transformation, we find the differently oriented trivalent vertex
corresponding to \exampleone.

{}From this, and using the gluing rules above, we find that the
closed string amplitude can be written as
\eqn\ea{\eqalign{Z =\sum_{R_1,\cdots, R_5}\;(-1)^{\sum_i \ell(R_i)}
\;C_{\cdot \cdot R_1}\; e^{-\ell(R_1) t_1}
C_{\cdot R_1^t R_2}\; e^{-\ell(R_2) r_1}\; C_{R_2^t \cdot R_3}\cr
\times e^{-\ell(R_3) t_2}\;
C_{\cdot R_3^t R_4}\; e^{-\ell(R_4) r_2}\;
C_{R_4^t \cdot R_5}\;e^{-\ell(R_5) t_3}
C_{\cdot R_5^t \cdot},}}
Note that to get the all genus answer up to degree $n$ in any one of
the five classes, we only need to perform the sum over the corresponding
representation with up to $n$ boxes.
It is not difficult to check that this is the correct A-model amplitude
on $X$, which is known to all genera. For example, one way to calculate
the amplitude is to use mirror symmetry to calculate the genus zero amplitude,
and integrality to fix the full free energy. This is possible as
there are no curves in $X$ with genus higher than zero, so we find
$$Z = \exp\biggl[ \sum_n {N_{\vec Q} \; e^{-n \vec{Q} \cdot\vec{t}}
\over n (2 \sinh(n g_s/2))^2}\biggr]$$
where $N_{\vec{Q}}$ are the degeneracies of BPS states corresponding to
a ${\bf P^1}$ in class ${\vec{Q}}$ in $X$ whose values are as follows.
First, there can be no BPS states
corresponding to ${\bf P^1}$ 's which are disconnected.
For the chains of connected ${\bf P^1}$'s we have that:
$N_{\vec{Q}} = -1$ if ${\vec{Q}}$ corresponds to the class of an odd
number of connected ${\bf P^1}$'s in $X$ (e.g. BPS states with
masses $r_1$, $r_1+t_2+r_2$ and $t_1+r_1+t_2+r_2+t_3$ all have
$N_{\vec{Q}}=-1$), $N_{\vec{Q}} = 1$ if $\vec{Q}$  corresponds to a class of an
even number of connected ${\bf P^1}$'s (e.g. $r_2+t_2$ and
$r_1+t_2+_2+t_3$).
\subsec{Example II}

It is also easy to see that our rules reproduce
the ${\cal O}(-3) \rightarrow {\bf P^2}$ amplitudes computed in \amv. In
\amv, the all genus amplitude was computed using a quiver-type
Chern-Simons theory with three nodes $G=U(N_1)\times U(N_2)\times
U(N_3)$ and bifundamental matter-fields, in the $N_i\rightarrow
\infty$ limit.

\ifig\exampletwo{The trivalent vertices for
gluing to  ${\cal O}(-3)\rightarrow {\bf P^2}$
amplitude (with D-branes on external legs). The superscript on the
representations give the framing $n$ in the corresponding propagator.}
{\epsfxsize 2.0truein\epsfbox{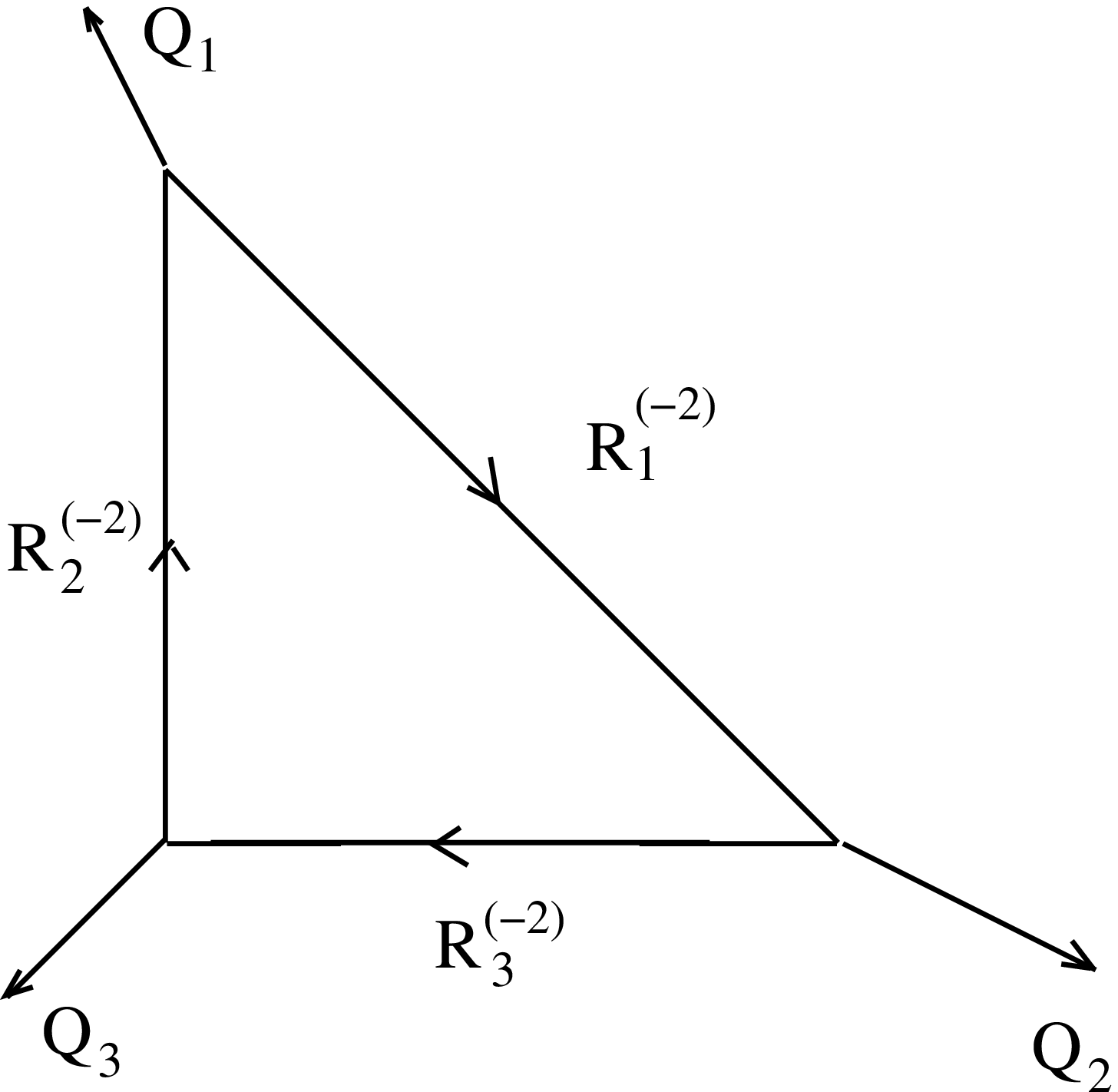}}

Using $SL(2,{\bf Z})$ transformations and adjusting framings
appropriately, it is easy to see that the amplitude corresponding
to the graph in \exampletwo, when there are no branes in the outer
legs, can be written as
\eqn\ptwo{Z_{\bf P^2}= \sum_{R_1,R_2,R_3}
(-1)^{\sum_i \ell(R_i)}e^{-\sum_i \ell(R_i) t}
q^{\sum_i \kappa_{R_i}} C_{\cdot R_2 R_3^t} C_{\cdot R_1 R_2^t}
C_{\cdot R_3 R_1^t},}
where $t$ is the K\"ahler parameter of ${\cal O}(-3) \rightarrow {\bf P^2}$.
Using the fact that, e.g.
$$C_{\cdot R_2 R_3^t}= W_{R_2 R_3}q^{-\kappa_{R_3}/2}$$
the amplitude \ptwo\ precisely equals the amplitude obtained in \amv\
by related, but different methods.

\subsec{Example III}

\ifig\gluing{This shows ${\cal O}(-1) + {\cal O}(-1) \rightarrow {\bf P^1}$
 with
D-branes on external legs}
{\epsfxsize 2.0truein\epsfbox{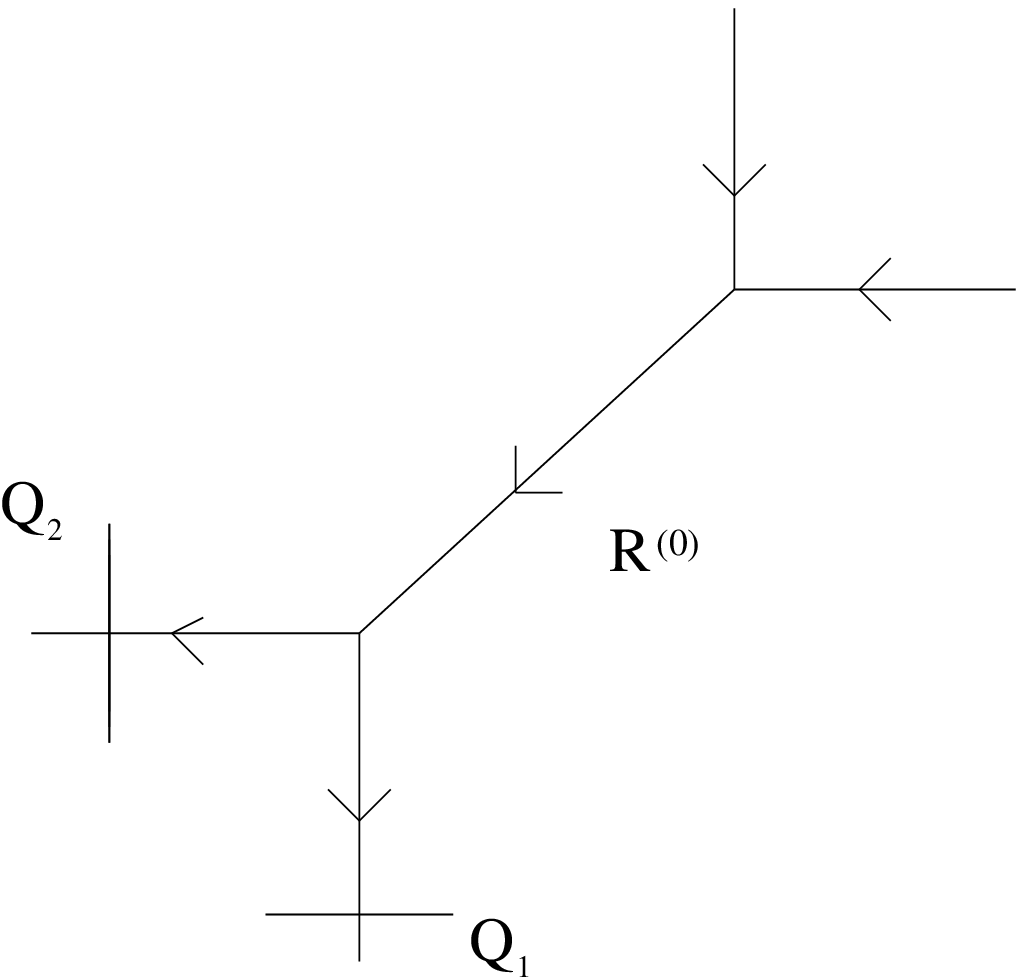}}

Another example where we can use the full vertex amplitude is shown
in \gluing: two D-branes on the outer legs of
${\cal O}(-1) + {\cal O}(-1) \rightarrow {\bf P^1}$.
According to the gluing rules we have
\eqn\glu{
Z(V_1, V_2)=\sum_{R, Q_1, Q_2} C_{Q_1 Q_2 R^t} (-1)^{\ell(R)}
e^{-\ell(R)  t} C_{R \cdot \cdot} \,
{\rm Tr}_{Q_1} V_1\;{\rm Tr}_{Q_2} V_2.}
On the other hand, this amplitude corresponds to a
Hopf link inside ${\bf S}^3$ with linking number $-1$,
therefore we should have
\eqn\equal{
Z(V_1, V_2)= \sum_{Q_1, Q_2}\lambda^{-{\ell_{Q_1} + \ell_{Q_2}\over 2}}
S_{Q_1 \overline Q_2} {\rm Tr}_{Q_1} V_1 {\rm Tr}_{Q_2} V_2.}
where $\lambda = e^{-t}$.
One can check that indeed \glu\ and \equal\ agree.
Namely, we have that
$$S_{Q_1 \overline Q_2} = {\cal W}_{Q_1 Q_2}(q,\lambda) S_{00}(\lambda)$$
where ${\cal W}_{Q_1 Q_2}$ is the invariant calculated in \mlfor, and
$S_{00}$ is the partition function of CS on ${\bf S}^3$,
which can be written as\foot{There are pieces in the free energy which
are finite polynomials in $t$ which encode certain topological data
in the compact case.  In the non-compact case at hand
they are ambiguous and in our vertex amplitudes we have naturally
set them to zero.} \gv\
\eqn\seriess{
S_{00}({\rm e}^{-t})=\exp \biggl( -\sum_{k=1}^{\infty}  { {\rm e}^{-kt}
\over k \bigl(q^{k\over 2} - q^{-{k \over 2}}\bigr)^2} \biggr).}
For example, for
$Q_1=R=\tableau{1}$,
agreement requires that
\eqn\tripar{
C_{\tableau{1} Q \tableau{1}}=W_{Q \tableau{1}} W_{\tableau{1}}-{W^{(1)}_{Q
\tableau{1}}\over W_{\tableau{1}}} , }
where $W^{(1)}_{Q_1 Q_2}$ is defined by the expansion
\eqn\expa{
{\cal W}_{Q_1 Q_2}(q,\lambda)=
\lambda^{\ell_{Q_1} + \ell_{Q_2}\over 2}W_{Q_1 Q_2}(q) +
\lambda^{{\ell_{Q_1} + \ell_{Q_2}\over 2}-1}W^{(1)}_{Q_1 Q_2}(q)+\cdots.}
Using the explicit formula \mlfor, we find that
\eqn\specval{
 C_{\tableau{1} Q \tableau{1}}= q^{\kappa_Q /2} {
W_{\tableau{1}Q^t} W_{\tableau{1} Q} \over W_Q} + W_Q,}
in agreement with \wfb.

\subsec{Example IV}

Another non-trivial configuration involving the full trivalent vertex
for which we have an immediate
prediction is ${\cal O}(-3) \rightarrow {\bf  P^2}$ with
``outer'' D-branes on the external legs.
This corresponds to the amplitude \ptwo\ but where we allow
non-trivial external representations on the trivalent vertices:
\eqn\ptwod{
\eqalign{
&Z_{\bf P^2}(V_1, V_2, V_3)=\cr
& \sum_{R_i, Q_i}
 C_{Q_3 R_2 R_3^t} C_{Q_1 R_1 R_2^t}
C_{Q_2 R_3 R_1^t}(-1)^{\sum_i \ell(R_i)}e^{-\sum_i \ell(R_i) t}
q^{\sum_i \kappa_{R_i}} {\rm Tr}_{Q_1}V_1 \, {\rm Tr}_{Q_2} V_2 \,
{\rm Tr}_{Q_3} V_3.}
}
This amplitude is the product of the closed string amplitude $Z_{\bf P^2}$
given in \ptwo, and the open string amplitude properly speaking, so we will
write
$$
Z_{\rm open}(V_1, V_2, V_3)={Z_{\bf ^2}(V_1, V_2, V_3)\over
Z_{\bf P^2}}= \sum_{Q_i} Z_{(Q_1,Q_2, Q_3)}(q, e^{-t})
{\rm Tr}_{Q_1}V_1 \, {\rm Tr}_{Q_2} V_2 \,
{\rm Tr}_{Q_3} V_3.
$$
Notice that the amplitudes are completely symmetric in $Q_1, Q_2, Q_3$, as
they should be by the symmetry of the geometry.

The generating functions
${\widehat f}_{(Q_1, Q_2, Q_3)}$ are computed from
$Z_{(Q_1,Q_2, Q_3)}$. Let us denote $z=(q^{1 \over 2} -q ^{-{1 \over 2}})^2$,
$y=e^{-t}$, so that
$${\widehat f}_{(Q_1, Q_2, Q_3)}=\sum_{g,Q}
N_{(R_1, R_2, R_3), g, Q}
z^g y^Q,$$
where $N_{(R_1, R_2, R_3), g, Q}$ are the degeneracies of BPS states with
the corresponding charges.
We find, for the first few representations and up to degree five in the
K\"ahler parameter, the
following results:
\eqn\resouter{
\eqalign{
{\widehat f}_{\tableau{1},\cdot, \cdot}=&\,  1 -2 \,y + 5\, y^2
 -(32 + 9\,z)\,y^3
+(286 + 288 \, z + 108\, z^2 + 14 \, z^3) \, y^4  \cr
&-(3038 + 6984 \, z + 7506 \, z^2 + 4519 \, z^3 + 1542\, z^4 + 276 \, z^5 + 20
\, z^6) \, y^5 + \cdots, \cr
{\widehat f}_{\tableau{1},\tableau{1}, \cdot}=&\, -y + 4\, y^2 -
(35 + 8 \, z) \,y^3 + (400 + 344 \, z + 112 \, z^2 + 13 \, z^3)\, y^4 \cr
& -( 5187 + 10504 \, z + 10036\, z^2 + 5434\, z^3 + 1691\, z^4 + 280 \, z ^5
+ 19 \, z^6) \, y^5 + \cdots, \cr
{\widehat f}_{\tableau{2},\cdot, \cdot}=&\, 7 \, y^3 -
(110 + 68 \, z + 12 \, z^2)\, y^4 \cr & + (1651 + 2938 \, z + 2353 \, z^2 +
992 \, z^3 + 212 \, z^4 + 18 \, z^5) \, y^5 + \cdots \cr
{\widehat f}_{\tableau{1 1},\cdot, \cdot}=&\, y - 4\, y^2 +
(28 + 8 \, z^2)\, y^3 -
(290 + 276 \, z + 100 \, z^2 + 13 \, z^3)\, y^4 \cr
& + (3536 + 7566 \, z + 7683 \, z^2 + 4442 \, z^3 + 1479\, z^4 + 262\, z^5 +
19 \, z^6)\, y^5 +  \cdots, \cr
{\widehat f}_{\tableau{1},\tableau{1},\tableau{1}}= &\,3 \, y^2
 -(36 + 7 \, z)\, y^3 + (531 + 396 \, z + 114 \, z^2 + 12 \, z^3) \, y^4 \cr
&-(8472 + 15210\, z + 13026\, z^2 + 6399\, z^3 + 1830 \, z^4 + 282 \, z^5
+ 18\, z^6)\, y^5 + \cdots,}
}
and so on. For representations involving only one nontrivial representation,
the degeneracies obtained above agree with the ones obtained in B-model
computations \AKV\ (see also \bmodel) and in A-model
computations through localization \refs{\gz ,\mayr}.

One can also compute the amplitudes in nontrivial framings, just by framing
the trivalent vertex in the appropriate way. For a single nontrivial
representation with framing $n$, we find $N_{(\tableau{1},\cdot,\cdot),
g,d}(n)=(-1)^n N_{(\tableau{1},\cdot,\cdot),g,d}(0)$, and
\eqn\framedf{
\eqalign{
{\widehat f}_{\tableau{2},\cdot, \cdot}(n)=& \,
{1\over 8} (1-(-1)^n -2\, n^2) \, +
{1 \over 96}(-3 + 3(-1)^n + 8\, n^2 -2\, n^4)\, z + \cdots \cr
& +  \Bigl(n^2 +  {1 \over 12} n^2 (n^2 -1)\, z + \cdots\Bigr) \, y
\cr &           + \Bigl( {1 \over 4}(-1 + (-1) ^n - 14\, n^2) +
{1 \over 48}( 3 - 3 (-1) ^n + 8 n^2 - 14 n^4)\, z + \cdots\Bigr)y^2 + \cdots,
\cr
{\widehat f}_{\tableau{1 1},\cdot, \cdot}(n)=& \,
{1\over 8}(-1+(-1)^n-4 n -2\, n^2) +
{1 \over 96}(3 - 3\, (-1)^n +8 \, n- 4\, n^2 -8 \, n^3- 2\,n^4)\, z +
 \cdots \cr & +
\Bigl(  (n+1)^2 + {1 \over 12} n (1+n)^2 (n+2) \, z + \cdots\Bigr) y
\cr & +  \Bigl( {1 \over 4}(-15 -  (-1) ^n - 28\,n - 14\, n^2 )\cr &
+ {1 \over 48}(-3 + 3\,  (-1)^n - 40\,n - 76\, n^2 -
56\,n^3 - 14\,n^4 )\, z+ \cdots \Bigr)y^2 + \cdots, \cr}}

\subsec{Example V}

\ifig\examplesix{The ${\cal O}(-3) \rightarrow {\bf P^2}$
with an outer brane and another brane in an inner edge.}
{\epsfxsize 3.0truein\epsfbox{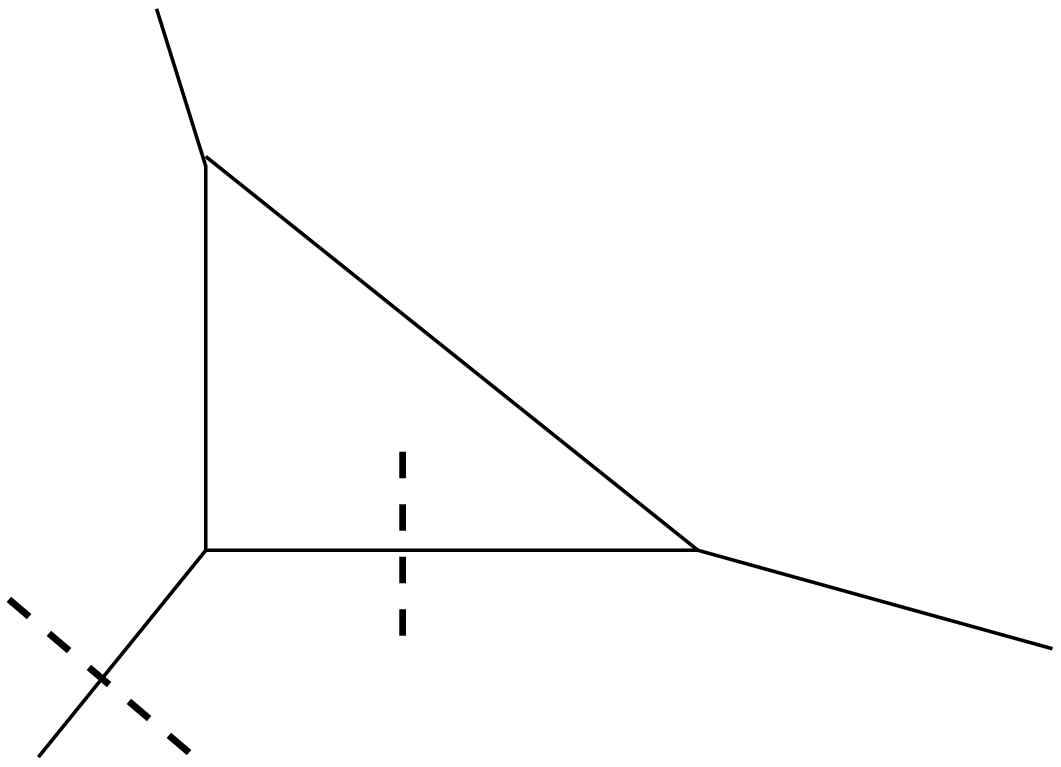}}

So far we have considered open string amplitudes where D-branes
were sitting on outer edges, but our formalism also allows to compute
 amplitudes
with branes in inner edges, as we saw in \oneinner. A simple example of such
a situation is local ${\bf P^2}$ with an inner and an outer brane, as
 depicted in
\examplesix. The framings are as in \exampletwo. The prediction for this
amplitude is
\vskip 0.3cm
$$
\eqalign{
Z(V_1, V_2)=
\sum_{R_i, Q_i}&(-1)^{s(R_i,Q_i)}\;
q^{f(R_i,Q_i)}\;
e^{-L(R_i,Q_i)}\;
C_{Q_1 R_2 R_3 \otimes Q_2}
\cr
& C_{\cdot R_3^t\otimes Q_3 R_1^t}\; C_{\cdot R_1 R_2^t}\;
\Tr_{Q_1} V_1\; \Tr_{Q_2}V_2 \; \Tr_{Q_3}V_2^{-1}.}
$$
\vskip 0.3cm
where
$$L(R_i,Q_i) = \sum_i \ell(R_i)t+
+\ell(Q_1)r_1 + \ell(Q_2)r_2+ \ell(Q_3)(t-r_2)$$
$$f(R_i,Q_i) = \kappa_{R_1} +
\kappa_{R_2}+ n \kappa_{Q_1}/2 +
p\kappa_{R_3 \otimes Q_2}/2 + (p+2)
\kappa_{R_3^t \otimes Q_3}/2,$$
$$s(R_i,Q_i) = \sum_i \ell(R_i) + n\ell(Q_1)+
p\ell (R_3 \otimes Q_2) + (p+2) \ell(R_3^t \otimes Q_3)$$
The integers $p$ and $n$ correspond to the framing of the inner brane and
the outer branes, respectively.  $p$ is related
to the framing $p'$ in the B-model of \AKV\ by $p=-1-p'$.
We can again compute the generating functionals $\hat f$
for degeneracies of BPS states
for different representations.
We present some results
corresponding to  $p=-1$, and $n=0$
(for the inner brane, this is the zero framing of \AKV ) where
we absorb $e^{-r_i}$ in $V_i$:
\eqn\resinner{
\eqalign{
{\widehat f}_{\cdot,\tableau{1}, \cdot}=&\,  -1 + \,y - (5 + z)\, y^2
 +(40 + 31 \, z + 9\,z^2 + z^3)\,y^3 \cr
&\, -(399 + 743 \, z + 648\, z^2 + 322 \, z^3 + 94 \,z^4 + 15 \, z^5 + z^6)
 \, y^4  \cr
&+(4524 + 16146 \, z + 29256 \, z^2 + 33523 \, z^3 + 26079\, z^4 + 14151 \,
 z^5 + 5364
\, z^6 \cr
& \,\,\,\ + 1390\, z^7 + 234 \, z^8 + 23\, z^9 + z^{10}) \, y^5 + \cdots, \cr
{\widehat f}_{\cdot,\cdot, \tableau{1}}=&\,  -1 + 2\, y -(12 + 3\, z)\, y
+ (104 + 96\, z + 33 \, z^2 + 4\, z^3)\, y^3 \cr
&-(1085 + 2328 \, z + 2334 \, z^2 + 1315 \, z^3 + 423 \, z^4 + 72 \, z^5 + 5
\, z^6)\, y^4 \cr
&-(12660 + 50874 \, z + 103683 \, z^2 + 133002\, z^3 + 114732 \, z^4 + 68040
 \, z^5
+ 27711 \, z^6 \cr
&\,\,\,\, + 7590 \, z^7 + 1332 \, z^8 + 135\, z^9 + 6 \, z^{10})\, y^5 +
 \cdots,\cr
{\widehat f}_{\tableau{1},\tableau{1}, \cdot}=&\, -1+y - (6 + z)\, y^2 +
(59 + 39 \, z + 10\, z^2 +z^3) \,y^3 \cr
&- (706 +1152 \, z +895 \, z^2 + 403 \, z^3 + 108\, z^4 + 16\, z^5 + z^6)\,
 y^4 \cr
& +( 9372 + 29927 \, z +  48964 \, z^2 + 51169\, z^3 + 36663 \, z^4 + 18485 \,
 z ^5
+ 6561 \, z^6 + 1603 \, z^7\cr  & \,\,\,\, + 256 \, z^8 + 24 \, z^9 + z^{10})
 \, y^5 + \cdots, \cr
{\widehat f}_{\cdot,\tableau{1}, \tableau{1}}=&\, 2 \, y^2 -
( 46 + 30 \, z + 5 \, z^2)\, y^3 +
(852 + 1682 \, z + 1285 \, z^2 + 536 \, z^3 + 111\,z^4 + 9\, z^5 )\, y^4 \cr
& - (14848 + 55104 \, z + 101054 \, z^2 + 113629 \, z^3 +83274 \, z^4 + 40375
 \, z^5 +
12800 \, z^6 \cr
& \,\,\,\, +2544\, z^7 + 287\, z^8 + 14\, z^9 )\, y^5 +  \cdots, \cr
}}
and so on. The results for $\widehat f_{\cdot, R, \cdot}$ and $\widehat
 f_{\cdot, \cdot, R}$
correspond to inner branes with positive and negative winding numbers,
 respectively, and they
agree with the B-model results of \AKV\ in the case of disc amplitudes.
 For higher genus and/or
number of holes, our results agree with those obtained through localization
 in \mayr. The
amplitudes with two nontrivial representations can be also obtained through
 localization, and in
all cases we have found perfect agreement with the above results.

\subsec{Example VI}
\ifig\examplefive{There are three K\"ahler parameters in the problem.
The size of leg
corresponding to $R_0$ is $s$, and the sizes of
 $R_3$ and $R_2$ correspond to
$t_1 $, $t_2$, respectively.}
{\epsfxsize 4.0truein\epsfbox{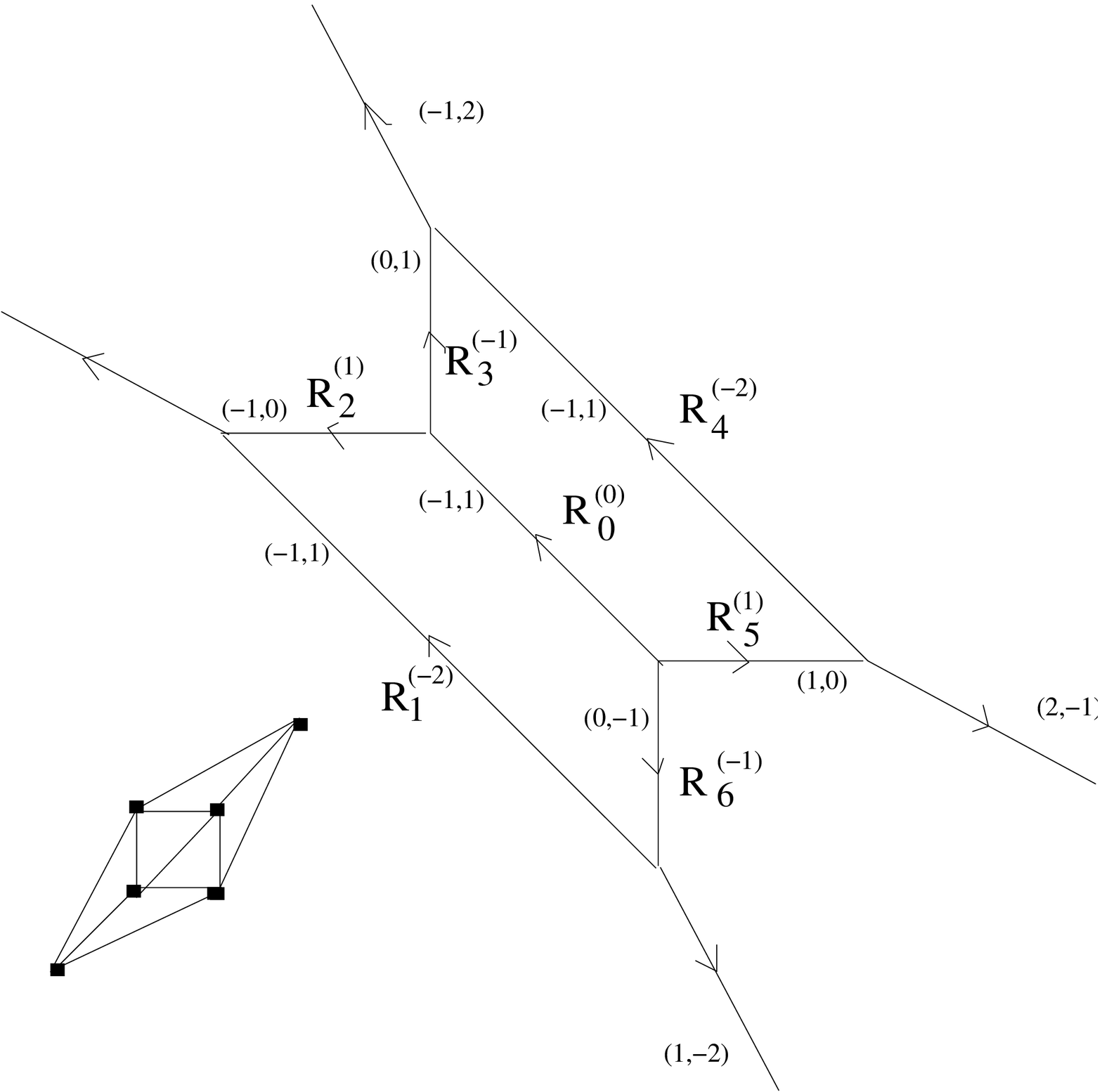}}

We now consider more complicated examples of closed string amplitudes.
Consider for example the toric diagram in \examplefive. There are
three K\"ahler parameters involved, $s$, $t_1$ and $t_2$, as indicated
in the figure. The amplitude is symmetric in $t_1$, $t_2$. When, say,
$t_2$ is taken to infinity, the resulting geometry is that of a
local Hirzebruch surface ${\bf F}_1$, where $s$ corresponds to
the K\"ahler parameter of the base in local ${\bf F}_1$, while
$t_1$ corresponds to the fiber. The amplitude for \examplefive\ can be
computed by using our rules in section 3, and the result is:
\eqn\dx{\eqalign{Z(X)&= \sum_{R_{0\ldots 6}} (-1)^{\ell(R_0) + \ell(R_1)
+ \ell(R_4)}q^{\kappa_{R_1}}
C_{R_1^t \cdot R_2^t} q^{-\kappa_{R_2}/2}
C_{R_2 R_3 R_0^t}q^{\kappa_{R_3}/2}C_{R_3^t \cdot R_4^t}\cr
&q^{-\kappa_{R_4}}
C_{R_4 \cdot  R_5^t}q^{-\kappa_{R_5}/2}
C_{R_5R_6R_0} q^{\kappa_{R_6}/2}C_{R_6^t \cdot  R_1}
e^{-L(R_i)}}}
where we wrote
\eqn\mess{
L(R_i)= (\ell(R_0)+\ell(R_1)+\ell(R_4))s
+(\ell(R_2)+\ell(R_6))t_1
+(\ell(R_3)+\ell(R_5))t_2.}
Notice that when $t_2 \rightarrow \infty$ \dx\ becomes the
amplitude for local ${\bf F}_1$, which was computed from Chern-Simons theory
in \refs{\amer,\ik}\ with the techniques of \amv.
We will write the answer in terms of a generating functional ${\cal F}_g$
of BPS degeneracies at genus $g$,
as
$$
{F}_g (s, t_1, t_2)=
\sum_{\ell=0}^{\infty} {\rm e}^{-\ell s}{\cal F}_\ell^g
(t_1, t_2)
$$
where
$$
{\cal F}_\ell^g(t_1, t_2)=\sum_{d_1, d_2}
n^g_{\ell, d_1, d_2} q_1^{d_1} q_2^{d_2} ,
$$
and we have written $q_i={\rm e}^{-t_i}$ (these shouldn't be confused
with the Chern-Simons variable introduced before). We then find,
up to order four in $q_i$, the following results (symmetrization w.r.t.
$q_1$, $q_2$ is understood):
$$
\eqalign{
{\cal F}_0^0(t_1, t_2)&= -2\, q_1 -  2\, q_1 q_2 + \cdots, \cr
{\cal F}_1^0(t_1, t_2)&=-1 - 3\, q_1 -5\, q_1^2 - 7\, q_1^3  -9 \, q_1^4
-4\, q_1 q_2 - 8\,
q_1 q_2^2 -12\,  q_1  q_2^3 -16\,  q_1  q_2^4- 9 \,   q_1^2   q_2^2 \cr
& -15 \, q_1^2   q_2^3 -21\, q_1^2   q_2^4 -16\, q_1^3   q_2^3
-24 \, q_1^3   q_2^4 -25 \, q_1^4   q_2^4  + \cdots, \cr
{\cal F}_2^0(t_1, t_2)&=-6\, q_1^2 - 32\, q_1^3  -110  \, q_1^4
- 10\,
q_1 q_2^2 -70\,  q_1  q_2^3 - 270 \,  q_1  q_2^4 -32 \,   q_1^2   q_2^2
-126 \, q_1^2   q_2^3  \cr
& - 456 \, q_1^2   q_2^4- 300\, q_1^3   q_2^3
- 784 \, q_1^3   q_2^4- 1584 \, q_1^4   q_2^4  + \cdots, \cr
{\cal F}_2^1(t_1, t_2)&=9 \, q_1^3  +68  \, q_1^4
+16\,  q_1  q_2^3 + 144 \,  q_1  q_2^4 +
21 \, q_1^2   q_2^3 +204 \, q_1^2   q_2^4+ 59\, q_1^3   q_2^3  \cr
& + 297 \, q_1^3   q_2^4+ 684 \, q_1^4   q_2^4  + \cdots,\cr
{\cal F}_2^2(t_1, t_2)&=-12  \, q_1^4
- 22 \,  q_1  q_2^4  -30 \, q_1^2   q_2^4-36 \, q_1^3   q_2^4-
94 \, q_1^4   q_2^4  + \cdots, \cr
{\cal F}_3^0(t_1, t_2)&=27\, q_1^3  +286 \, q_1^4
+64\,  q_1  q_2^3 +800 \,  q_1  q_2^4 +25 \,   q_1^2   q_2^2
+266 \, q_1^2   q_2^3 +1998 \, q_1^2   q_2^4 \cr
& +1332\, q_1^3   q_2^3
+6260\, q_1^3   q_2^4+21070 \, q_1^4   q_2^4  + \cdots,}
$$
$$
\eqalign{
{\cal F}_3^1(t_1, t_2)&=-10 \, q_1^3  -288  \, q_1^4
-18\,  q_1  q_2^3 -688 \,  q_1  q_2^4  \cr
& -64 \, q_1^2   q_2^3 -1404 \, q_1^2   q_2^4-516\, q_1^3   q_2^3
-4372 \, q_1^3   q_2^4-18498 \, q_1^4   q_2^4  + \cdots, \cr
{\cal F}_3^2(t_1, t_2)&=108  \, q_1^4
+224\,  q_1  q_2^4 +375 \, q_1^2   q_2^4+49\, q_1^3   q_2^3
+1168 \, q_1^3   q_2^4+6837 \, q_1^4   q_2^4  + \cdots,\cr
{\cal F}_3^3(t_1, t_2)&=-14  \, q_1^4
-26\,  q_1  q_2^4 -36 \, q_1^2   q_2^4
-114 \, q_1^3   q_2^4-1196 \, q_1^4   q_2^4  + \cdots, \cr
{\cal F}_3^4(t_1, t_2)&=81\, q_1^4   q_2^4  + \cdots, \cr}
$$
These numbers agree with the results obtained with
localization\foot{In order to compare with the results using localization
 and mirror symmetry, we have redefined $g_s \rightarrow i g_s$ and
therefore $n_Q^g \rightarrow (-1)^{g-1} n_Q^g$.}.
Notice that, at genus $0$ and for $t_2 \rightarrow \infty$,
the above results coincide with the results for local ${\bf F}_1$
presented for example in \ckyz.

One can find a similar result for the $A_2$ fibration over ${\bf P^1}$.
 In this case,
the amplitude reads
\eqn\dx{\eqalign{Z(X)&= \sum_{R_{0\ldots 6}}q^{5\kappa_{R_1}/2}
C_{R_1^t \cdot R_2^t} q^{-\kappa_{R_2}/2}
C_{R_2 R_3 R_0^t}q^{\kappa_{R_3}/2+ 3\kappa_{R_0}/2 }C_{R_3^t \cdot R_4^t}\cr
&q^{\kappa_{R_4}/2}
C_{R_4 \cdot  R_5^t}q^{-\kappa_{R_5}/2}
C_{R_5R_6R_0} q^{\kappa_{R_6}/2}C_{R_6^t \cdot  R_1}
e^{-L(R_i)},\cr}}
where now
$$
\eqalign{
L(R_i) &= (\ell(R_0)+\ell(R_1)+\ell(R_4))s \cr
&+ (4 \ell (R_1) + \ell(R_2)+\ell(R_6))t_1+ (2 \ell(R_0) + 2 \ell(R_1) +
\ell(R_3)+\ell(R_5))t_2.}
$$
Here, $s$ corresponds to the K\"ahler parameter of the base of the fibration,
and $t_1$, $t_2$ correspond to the K\"ahler parameters of the fibers.
 Denoting the
generating functional as before, we obtain
in this case,
$$
\eqalign{
{\cal F}_0^0(t_1, t_2)&=-2 \, q_1 - 2\, q_2 - 2 \, q_1 q_2 , \cr
{\cal F}_1^0(t_1, t_2)&=-1 -2 \, q_2 -4 \, q_2^2 - 6 \, q_2^3  - 8  \, q_2^4
- 2\, q_1 q_2 - 6 \,
q_1 q_2^2 - 10 \,  q_1  q_2^3 - 14 \,  q_1  q_2^4- 6 \,   q_1^2   q_2^2 \cr
& -12 \, q_1^2   q_2^3- 18 \, q_1^2   q_2^4 - 4 \, q_1^3 q_2 ^2 -12\, q_1^3
   q_2^3
- 20  \, q_1^3   q_2^4 - 6\, q_1^4 q_2^2 - 10 q_1^4 q_2^3 -
20 \, q_1^4   q_2^4  + \cdots, \cr
{\cal F}_2^0(t_1, t_2)&=-6\, q_2^3  -32  \, q_2^4
- 10\,  q_1  q_2^3 - 70 \,  q_1  q_2^4 -12 \,   q_1^2   q_2^3-
96 \, q_1^2   q_2^4- 12\, q_1^3   q_2^3 -110 \, q_1^3   q_2^4  \cr
& - 10 \, q_1^4  q_2^3 - 112 \, q_1^4  q_2^4 + \cdots, \cr
{\cal F}_2^1(t_1, t_2)&=9  \, q_1^4
+16 \,  q_1  q_2^4  + 21\, q_1^2   q_2^4+
24 \, q_1^3   q_2^4+ 25  \,q_1^4  q_2^4 + \cdots, \cr}
$$
and so on, again in agreement with the results for
 genus zero in \ckyz. We have also checked some of these
results at higher genus with localization techniques.

\bigskip
\leftline{\bf
   Acknowledgments}

We would like to thank D.-E.Diaconescu, R. Dijkgraaf, J. Gomis, A. Grassi,
A. Iqbal, A. Kapustin, S. Katz, V. Kazakov, I. Kostov, C-C. Liu,
H. Ooguri, J. Schwarz, S. Shenker and E. Zaslow for valuable
discussions (and the cap!).
The research of MA and CV was supported in part by NSF grants
PHY-9802709 and DMS-0074329.  In
addition, CV thanks the hospitality of the theory group at Caltech,
where he is a Gordon Moore Distinguished Scholar. M.A. is grateful to the
Caltech theory group for hospitality during part of this work.
A.K. is supported in part by the DFG
grant KL-1070/2-1.

\listrefs
\end
\bye